\documentclass[letterpaper,11pt]{article}
\pdfoutput=1
\pdfoutput=1
\pdfoutput=1
\pdfoutput=1

\usepackage{jheppub}
\usepackage{multirow}

\usepackage{subfig}
\usepackage{xspace}
\usepackage[countmax]{subfloat}

\usepackage{amssymb}
\usepackage{amsmath}
\usepackage{cancel}
\usepackage{tabu,booktabs}
\usepackage{color}
\usepackage{braket}
\usepackage{graphicx}
\usepackage{multirow}
\usepackage{verbatim}
\usepackage{amsthm}
\usepackage{slashed}
\usepackage{wasysym}
\usepackage{simplewick}
\usepackage{mathtools}
\usepackage{soul}
\usepackage{xspace}

\definecolor{dancomment}{RGB}{0,159,0}

\def\cB{\mathcal{B}}
\def\cC{\mathcal{C}}

\def\cL{\mathcal{L}}
\def\cM{\mathcal{M}}

\def\cO{\mathcal{O}}

\def\cP{\mathcal{P}}

\def\cY{\mathcal{Y}}

\def\tr{{\rm tr}}

\def\tO{\tilde{O}}

\def\nn{{\nonumber}}

\newcommand{\hard}{\mathrm{hard}}
\newcommand{\dyn}{\mathrm{dyn}}

\newcommand{\BPS}{\mathrm{BPS}}

\newcommand{\Eq}[1]{Equation~\eqref{#1}}

\DeclareRobustCommand{\Sec}[1]{Sec.~\ref{#1}}

\DeclareRobustCommand{\App}[1]{App.~\ref{#1}}
\DeclareRobustCommand{\Tab}[1]{Table~\ref{#1}}

\DeclareRobustCommand{\Eq}[1]{Eq.~(\ref{#1})}
\DeclareRobustCommand{\Eqs}[2]{Eqs.~(\ref{#1}) and (\ref{#2})}

\DeclareRobustCommand{\Ref}[1]{Ref.~\cite{#1}}
\DeclareRobustCommand{\Refs}[1]{Refs.~\cite{#1}}

\def\be{\begin{equation}}
\def\ee{\end{equation}}

\newcommand{\SCETi}{\mbox{${\rm SCET}_{\rm I}$}\xspace}

\def\l{\langle}
\def\r{\rangle}

\def\bt{\beta}

\newcommand{\lotsdots}{{
+\cdot\cdot:\cdot\cdot 
(\cdot\cdot:\cdot\cdot\ldots\cdot\cdot:\cdot\cdot)
[\cdot\cdot:\cdot \cdot-]
}}

\newcommand{\Sl}[1]{\slashed{#1}}

\renewcommand{\arraystretch}{1.05}
\allowdisplaybreaks[3]

\newcommand{\eq}[1]{Eq.~\eqref{eq:#1}}
\newcommand{\eqs}[2]{Eqs.~\eqref{eq:#1} and \eqref{eq:#2}}
\newcommand{\eqss}[3]{Eqs.~\eqref{eq:#1}, \eqref{eq:#2} , and \eqref{eq:#3}}

\newcommand{\app}[1]{App.~\ref{app:#1}}

\newcommand{\ord}[1]{\mathcal{O}(#1)}

\newcommand{\mae}[3]{\langle#1\rvert#2\rvert#3\rangle}

\newcommand{\df}{\mathrm{d}}

\newcommand{\sdt}{\!\cdot\!}

\newcommand{\al}{\alpha}

\newcommand\bn{{\bar n}}
\newcommand{\ga}{\gamma}

\newcommand{\de}{\delta}

\newcommand{\ve}{\varepsilon}
\newcommand{\la}{\lambda}

\newcommand{\w}{\omega}
\newcommand{\balpha}{{\bar \alpha}}
\newcommand{\bbeta}{{\bar \beta}}
\newcommand{\bgamma}{{\bar \gamma}}
\newcommand{\bdelta}{{\bar \delta}}

\newcommand{\fd}[2]{\parbox{#1}{\includegraphics[width=#1]{#2}}}

\newcommand{\hH}{\widehat{H}}
\newcommand{\hS}{\widehat{S}}

\newcommand{\vT}{\bar{T}}

\newcommand{\vC}{\vec{C}}
\newcommand{\vO}{\vec O}

\newcommand{\lp}{\tilde p}        

\newcommand{\bnP}{\overline {\mathcal P}}

\newcommand{\id}{\mathbf{1}}

  \newcount\hour \newcount\minute
  \hour=\time \divide \hour by 60 \minute=\time
  \newcommand{\todaytime}{\today \ -- \number\hour :\ifnum \minute<10 0\fi\number\minute}

\preprint{MIT-CTP 4934}

\title{A Subleading Power Operator Basis for the Scalar Quark Current}

\author{Cyuan-Han Chang,}
\author{Iain W. Stewart,}
\author{Gherardo Vita}

\affiliation{Center for Theoretical Physics, Massachusetts Institute of Technology, Cambridge, MA 02139, USA}

\emailAdd{cyuanhan@mit.edu}
\emailAdd{iains@mit.edu}
\emailAdd{vita@mit.edu}

\abstract{Factorization theorems play a crucial role in our understanding of the strong interaction. For collider processes they are typically formulated at leading power and much less is known about power corrections in the $\lambda\ll 1$ expansion. Here we present a complete basis of power suppressed operators for a scalar quark current at $\mathcal{O}(\lambda^2)$ in the amplitude level power expansion in the Soft Collinear Effective Theory, demonstrating that helicity selection rules significantly simplify the construction. This basis applies for the production of any color singlet scalar in $q\bar{q}$ annihilation (such as $b \bar b \to H$). We also classify all operators which contribute to the cross section at $\mathcal{O}(\lambda^2)$ and perform matching calculations to determine their tree level Wilson coefficients.  These results can be exploited to study  power corrections in both resummed and fixed order perturbation theory, and for analyzing the factorization properties of gauge theory amplitudes and cross sections at subleading power.}


\begin{document} 

\maketitle

\section{Introduction}\label{sec:intro}

Studying the behavior of observables at all orders in perturbation theory is an important goal towards the understanding of the theory of strong interactions. 
In Quantum Chromo Dynamics (QCD) factorization theorems are typically formulated at leading power~\cite{Collins:1989gx}, whereas the structure of power corrections has received much less attention. 
A formalism for studying factorization in QCD is the Soft Collinear Effective Theory (SCET) \cite{Bauer:2000ew, Bauer:2000yr, Bauer:2001ct, Bauer:2001yt, Bauer:2002nz}, an effective field theory describing the soft and collinear limits of QCD. SCET allows for a systematic power expansion in $\lambda \ll 1$ at the level of the Lagrangian, and simplifies many aspects of factorization. 
Since subleading power corrections are of significant theoretical and practical interest, SCET has been used both to study power corrections at the level of the amplitude \cite{Larkoski:2014bxa} and to derive factorization theorems at subleading power for $B$ decays \cite{Lee:2004ja,Beneke:2004in,Hill:2004if,Bosch:2004cb,Beneke:2004rc,Paz:2009ut,Benzke:2010js}.
More recently,  progress has been made using SCET towards  understanding subleading power corrections for event shape observables \cite{Freedman:2013vya,Freedman:2014uta,Kolodrubetz:2016uim,Moult:2016fqy,Feige:2017zci,Goerke:2017lei} and Higgs production in gluon fusion \cite{Moult:2017rpl}. Approaches to power corrections calculations in frameworks different from SCET can be in \cite{Laenen:2008gt,Laenen:2008ux,Bonocore:2015esa} for Drell-Yan in the next-to-soft threshold limit.

In this paper, we focus on the power suppressed hard scattering operators describing the quark antiquark initiated production (or decay to exclusive jets) of a color singlet scalar. 
We present a complete operator basis to $\cO(\lambda^2)$ in the SCET power expansion using operators of definite helicity \cite{Moult:2015aoa,Kolodrubetz:2016uim,Feige:2017zci}, and discuss how helicity selection rules simplify the structure of the basis. 
We also classify all operators which can contribute at the cross section level at $\cO(\lambda^2)$, and discuss the structure of interference terms between different operators in the squared matrix element. 
We then perform the tree level matching onto our operators. These results can be used to study subleading power corrections either in fixed order, or resummed perturbation theory, and are intended to compliment recent analyses for the case of vector quark currents~\cite{Feige:2017zci} and color singlet scalar production in gluon fusion~\cite{Moult:2017rpl}.

We consider the case of the Yukawa interaction after electroweak symmetry breaking, yielding couplings in the mass basis
\begin{align}\label{eq:yukawaint}
\cL_m=-m_d^i\bar d^i_Ld^i_R\frac{h}{v}-m_u^i\bar u^i_Lu^i_R\frac{h}{v}\,,
\end{align}
where $i=1,2,3$ is the flavor index, $h$ is the Higgs field, and $v=(\sqrt{2} G_F)^{-1/2}=246$ GeV is the Higgs vacuum expectation value.  
The corresponding hard scattering operators in SCET describe the quark antiquark initiated production of a color singlet scalar, which we will take for concreteness to be the Higgs, and can be used to study the underlying hard Born process
\begin{equation} \label{eq:interaction}
q(p_1)\, \bar q(p_2) \to h(k)
\,,\end{equation}
where $q \bar{q}$ denote the colliding quark-antiquark pair, and $h$ the outgoing Higgs particle. 
For the purpose of constructing a subleading power basis we will treat the dynamics of the incoming quarks as if they were massless, and hence effectively organize our analysis as an expansion near the massless limit. 
We are also interested in exclusive jet processes ($pp \to H+\text{0-jets}$ and $H\to \text{quark dijets}$) where it is meaningful to organize the hard scattering operators at the amplitude level.

It is possible to write a factorization formula for the active-parton exclusive jet cross section corresponding to \eq{interaction} for a variety of jet resolution variables. 
For concreteness, we consider the case of beam thrust, $\tau_B$. The factorization formula at leading power for the beam thrust cross section, can be written schematically using quark and antiquark beam functions $B_{q \bar{q}}$, in the form \cite{Stewart:2009yx,Stewart:2010tn}
\begin{align} \label{eq:sigma}
\frac{\df\sigma^{(0)}}{\df \tau_B} &=
\int\!\df x_a\, \df x_b\, \df \Phi(q_1 \!+ q_2; k)\, \widehat\cM(\{k\})\
 \hH^{(0)}(\{q_i\})\: 
\Bigl[ B_q^{(0)} B_{q}^{(0)}  \Bigr]\otimes \hS^{(0)}  
\,,\end{align}
where $\widehat\cM(\{k\})$ denotes the measurement made on the color singlet final state, the $x_{a,b}$ are the momentum fractions of the incoming partons and $\df \Phi$ denotes the Lorentz-invariant phase space for the Born process in \eq{interaction}.
\footnote{By referring to active-parton factorization we mean that this formula ignores contributions that occur through the Glauber Lagrangian of Ref.~\cite{Rothstein:2016bsq} like proton spectator interactions~\cite{Gaunt:2014ska}.} 
The hard function $\hH(\{q_i\})$ encodes the dependence on the underlying hard interaction and the trace is over color.
The beam functions $B_i$ describe energetic initial-state radiation along the beam directions \cite{Stewart:2009yx,Fleming:2006cd}, while the soft function $\hS$ describes soft radiation. 
Factorization formulas allow towers of logarithms of $\tau_B$ to be resummed to all orders through the renormalization group evolution of these hard, beam and soft functions. For the process $H\to q\bar q$ we can similarly consider a measurement of the classic thrust, and obtain an analogous factorization formula to \eq{sigma} with the beam functions $B_q^{(0)}$ replaced by jet functions $J_q^{(0)}$.

The factorization formula in \Eq{eq:sigma}, being at leading power, describes only the terms in the cross section proportional to $\tau_B^{-1}$, including delta function terms. 
The full QCD beam thrust cross section $\frac{\df\sigma}{\df\tau_B}$ can be expanded in powers of $\tau_B$ as,
\begin{align}\label{eq:cross_expand}
\frac{\df\sigma}{\df\tau_B} &=\frac{\df\sigma^{(0)}}{\df\tau_B} +\frac{\df\sigma^{(1)}}{\df\tau_B} +\frac{\df\sigma^{(2)}}{\df\tau_B}+\frac{\df\sigma^{(3)}}{\df\tau_B} +{\cal O}(\tau)\,.
\end{align}
and it might be expected that the power corrections in \Eq{eq:cross_expand} obey a factorization formula similar to that of \Eq{eq:sigma}. Schematically, 
\begin{align} \label{eq:sigma_sub}
&\hspace{-0.25cm}\frac{\df\sigma^{(n)}}{\df\tau_B} =
\int\!\df x_a\, \df x_b\, \df \Phi(q_1 \!+ q_2; k)\,M(\{k\})\
\sum_{j}   H^{(n_{Hj})}_{j} \otimes 
  \Big[ B^{(n_{Bj})}_{j}  B^{(n'_{Bj})}_{j}\Big] \otimes S_j^{(n_{Sj})}  
,\end{align}
where $j$ sums over the different terms that contribute at each order, $n_{Hj}+n_{Bj}+n_{Bj}'+n_{Sj}=n$, and $\otimes$ denotes a set of color contractions and convolutions, whose detailed structure and definition is beyond the scope of this paper, but it is known to be more complicated than the typical leading power factorization theorems.
Deriving a factorization theorem of the form of \eq{sigma_sub} would allow the resummation of subleading power logarithms. As a matter of fact by solving the renormalization group evolution of the different functions appearing in \Eq{eq:sigma_sub}, it is possible to resum subleading power logarithms allowing for an all orders understanding of power corrections to the soft and collinear limits.

To derive a factorization theorem in SCET the procedure is to match QCD onto SCET, which consists of a Lagrangian $\cL_\hard$ describing the hard scattering process and a Lagrangian $\cL_\dyn$ describing the dynamics of soft and collinear radiation
\begin{align} 
\cL_{\text{SCET}}=\cL_\hard+\cL_\dyn \,.
\end{align}
The dynamical Lagrangian can be divided into two parts
\begin{align}
\cL_\dyn=\cL_{\text{fact}}+\cL_{G}^{(0)} \,,
\end{align}
and the hard scattering Lagrangian consist of hard scattering operators multiplied by Wilson coefficients
\begin{align}
\cL_\hard=\sum_{i} \cC_{i} O_{i} \,.
\end{align}
In the dynamical Lagragian, $\cL_{G}^{(0)}$ is the leading power Glauber Lagrangian, which was derived in Ref.~\cite{Rothstein:2016bsq}. $\cL_{G}^{(0)}$ couples together soft and collinear fields in an apriori non-factorizable manner, while $\cL_{\text{fact}}$ includes both the leading interactions which can be factorized into independent soft and collinear Lagrangians, and subleading power interactions which are factorizable, via an order by order insertion procedure, as products of soft and collinear fields.
Our focus here is on determining the subleading power $\cL_\hard$ for $q\bar{q}\to H$. For our analysis $\cL_\dyn$ only plays a minor role when we carry out explicit matching calculations, and $\cL_{G}^{(0)}$ does not appear for these tree level matching calculations. 

The Wilson coefficients are obtained through the matching of the full theory diagrams onto SCET, hence they are process dependent, and \Sec{sec:matching} provides matching results for the process $q \bar{q} \to h$. 
The hard scattering operators are more universal since they depend only on the color charged states of the underlying hard Born process and on the spin of the non-QCD interacting fields. 
Therefore the basis of hard scattering operators presented in \Sec{sec:basis} is valid for all quark antiquark initiated production or decay with coupling to any number of color singlet scalars. 
The Lagrangian $\cL_\dyn$ is universal and the relevant terms for our analysis are known in SCET to $\cO(\lambda^2)$ in the power expansion \cite{Manohar:2002fd,Chay:2002vy,Beneke:2002ni,Beneke:2002ph,Pirjol:2002km,Bauer:2003mga}.

In order to decouple the leading power soft and collinear interactions in $\cL_{\text{fact}}$ a BPS field redefinition \cite{Bauer:2002nz} can be performed in the effective theory. 
If the leading power glauber lagrangian, $\cL_{G}^{(0)}$, is proven to be irrelevant, then the Hilbert spaces for the soft and collinear degrees of freedom are factorized, and the cross section can be written as a product of squared matrix elements, each involving only collinear fields or soft fields after a series of algebraic manipulations, such as the application of color and dirac fierz identities.
This procedure it is used to define each of the functions appearing in \Eq{eq:sigma_sub} in terms of hard scattering operators and Lagrangian insertions in SCET. In the case of \Eq{eq:sigma_sub}, $B_q, B_{\bar q}$ are the squared matrix elements containing only collinear fields and the soft function $\hS$ is a squared matrix element of only soft operators.
Given that the Lagrangian insertions are process independent and therefore universal, the remaining ingredient necessary to derive a subleading power factorization theorem for the $ q \bar{q}\to H$ process is a complete basis of subleading power hard scattering operators.
The derivation of a basis, which is the goal of this paper, provides the groundwork for future systematic studies of power corrections for color singlet production through quark antiquark annihilation. 

Recently, there has been considerable work focused on the use of event shape observables for performing NNLO fixed order calculations. An event shape observable can be used to compute the NNLO subtractions using the $q_T$ \cite{Catani:2007vq} or $N$-jettiness \cite{Boughezal:2015aha,Gaunt:2015pea} subtraction schemes.
Therefore, an important application of the results presented in this paper is the calculation of subleading power corrections to event shape observables for $q\bar{q}\to H$, such as $0$-jettiness \cite{Stewart:2010tn}. 
The use of event shape observables for performing NNLO subtractions has been already applied to color singlet production \cite{Catani:2009sm,Ferrera:2011bk,Catani:2011qz,Grazzini:2013bna,Cascioli:2014yka,Ferrera:2014lca,Gehrmann:2014fva,Grazzini:2015nwa,Grazzini:2015hta,Campbell:2016yrh,Boughezal:2016wmq}, to the production of a single jet in association with a color singlet particle \cite{Boughezal:2015aha,Boughezal:2015dva,Boughezal:2016isb,Boughezal:2016dtm}, to inclusive photon production \cite{Campbell:2016lzl} and to vector-boson pair production~\cite{Grazzini:2017mhc}. 
It is possible to improve the stability and numerical accuracy of the subtraction by analytically computing the power corrections for it. This was shown explicitly in two recent works where the SCET based analytic calculation of the leading power corrections for $0$-jettiness has been carried out both for $q\bar q$ initiated Drell Yan like production of a color singlet vector boson~\cite{Moult:2016fqy} (see also~\cite{Boughezal:2016zws}) and for Higgs production in gluon fusion~\cite{Moult:2017jsg}. It would be interesting to extend this calculation to $q\bar{q}\to H$.

The paper is organized in the following way. In \Sec{sec:review} we provide a brief review of SCET focusing on the relevant elements such as helicity building blocks that are needed in the rest of the paper. 
In \Sec{sec:basis} we present a complete basis of operators to $\cO(\lambda^2)$ for the quark antiquark initiated production (or decay to exclusive jets) of a color singlet, and carefully classify which operators can contribute to the cross section at $\cO(\lambda^2)$. 
To simplify the presentation we will always refer to the scalar quark current for the process $b\bar b\to H$ or $H\to b\bar b$, so that we can identify the quark flavor in the current, and distinguish cases where additional quarks are of the same or different flavor. In \Sec{sec:matching} we perform the tree level matching to the relevant operators. 
Conclusions are given in \Sec{sec:conclusions}. 
Some extensions are included in the appendices, including enumerating operators with an additional Lagrangian mass insertion that causes a helicity flip.

\section{SCET and Helicity Operators}\label{sec:review}

In this section we briefly review salient features of SCET~\cite{Bauer:2000ew, Bauer:2000yr, Bauer:2001ct, Bauer:2001yt, Bauer:2002nz} needed for our analysis (see also \Refs{iain_notes,Becher:2014oda}).  
We will also review the use of helicity operators in SCET following the construction of  \Refs{Moult:2015aoa,Kolodrubetz:2016uim,Feige:2017zci}, to which we refer the reader for further details.
SCET is an effective field theory of QCD describing the interactions of soft and collinear particles in the presence of a hard interaction. 
Soft particles are characterized by small momenta with homogenous scaling in all its components, while collinear particles carry a larger momentum along a particular light-like direction. 
For each such direction $\hat n_i$ present in the problem we define two light-like reference vectors $n_i = (1,\hat n_i)$ and $\bn_i=(1,-\hat n_i)$ such that $n_i^2 = \bn_i^2 = 0$ and $n_i\cdot\bn_i = 2$. 
Any four-momentum $p$ can then be decomposed with these basis vectors as
\begin{equation} \label{eq:lightcone_dec}
p^\mu = \bn_i\sdt p\,\frac{n_i^\mu}{2} + n_i\sdt p\,\frac{\bn_i^\mu}{2} + p^\mu_{n_i\perp}\
\,.\end{equation}
A particle with momentum $p$ close to the $\hat n_i$ direction is called $n_i$-collinear and has momentum components scaling as $(n_i\!\cdot\! p, \bn_i \!\cdot\! p, p_{n_i\perp}) \sim \bn_i\cdot p$ $\,(\la^2,1,\la)$. 
Here $\la \ll 1$ is a formal power counting parameter determined by  measurements or kinematic restrictions made on the QCD radiation.
The choice of reference vectors is not unique, and selecting any two reference vectors, $n_i$ and $n_i'$, with $n_i\cdot n_i' \sim \ord{\lambda^2}$ will describe the same physics.
The freedom in the choice of $n_i$ and the auxiliary $\bn_i$ induces a symmetry in the effective theory known as reparametrization invariance (RPI) \cite{Chay:2002vy,Manohar:2002fd}. 
More explicitly, there are three classes of RPI transformations under which the EFT is invariant
\begin{alignat}{3}\label{eq:RPI_def}
&\text{RPI-I} &\qquad &  \text{RPI-II}   &\qquad &  \text{RPI-III} \nn \\
&n_{i \mu} \to n_{i \mu} +\Delta_\mu^\perp &\qquad &  n_{i \mu} \to n_{i \mu}   &\qquad & n_{i \mu} \to e^\alpha n_{i \mu} \nn \\
&\bar n_{i \mu} \to \bar n_{i \mu}  &\qquad &  \bar n_{i \mu} \to \bar n_{i \mu} +\epsilon_\mu^\perp  &\qquad & \bar n_{i \mu} \to e^{-\alpha} \bar n_{i \mu}\,,
\end{alignat}
where the transformation parameters have a power counting $\Delta^\perp \sim \lambda$, $\epsilon^\perp \sim \lambda^0$, and $\alpha\sim \lambda^0$, and satisfy $n_i\cdot \Delta^\perp=\bar n_i\cdot \Delta^\perp=n_i \cdot \epsilon^\perp=\bar n_i \cdot \epsilon^\perp=0$.
Additionally, while $\alpha$ here corresponds with a finite transformation, the parameters $\Delta^\perp$ and $\epsilon^\perp$ were chosen as infinitesimal (this choice is for convenience, and independent of the power counting). 
RPI symmetries can be used to relate the Wilson coefficients of operators at different orders in the power expansion, and we will exploit this property in our analysis. 
The Wilson coefficients must also satisfy the rescaling symmetries of RPI-III, and at 
tree level are simply rational functions of the large momentum components of the fields appearing in the operator.

To facilitate manifest power counting in  SCET it is useful to decompose momenta into label and residual components
\begin{equation} \label{eq:label_dec}
p^\mu = \lp^\mu + k^\mu = \bn_i \sdt\lp\, \frac{n_i^\mu}{2} + \lp_{n_i\perp}^\mu + k^\mu\,.
\,\end{equation}
The momenta $\bn_i \cdot\lp \sim Q$ and $\lp_{n_i\perp} \sim \la Q$ are referred to as the label components, where $Q$ is a typical scale of the hard interaction, while fluctuations about the label momentum are described by a small residual momentum $k\sim \la^2 Q$.
From this decomposition we can obtain fields with momenta of definite scaling by performing a multipole expansion. 
The effective theory consists of collinear quark fields $\xi_{n_i,\lp}(x)$ and collinear gluon fields $A^\mu_{n_i,\lp}(x)$ for each direction $n_i$, as well as soft quark and gluon fields, $q_{us}(x)$ and $A_{us}(x)$ respectively. 
In this paper we will restrict ourselves to the SCET$_\text{I}$ theory where the soft degrees of freedom are referred to as ultrasoft so as to distinguish them from the soft modes of SCET$_\text{II}$ \cite{Bauer:2002aj} (see \Ref{Feige:2017zci} for a discussion of SCET$_\text{II}$ in the context of subleading power helicity operators). 
Independent gauge symmetries are enforced for each set of fields, which have support for the corresponding momenta carried by that field \cite{Bauer:2003mga}. 
The leading power gauge symmetry is exact, and is not corrected at subleading powers. 
The fields for $n_i$-collinear quarks and gluons are labeled by their collinear direction $n_i$ and their large momentum $\lp$. 
They  are in a mixed representation, with position space for the residual momenta in all components, and momentum space for the large momentum components. 
While the label momentum operator $\cP^\mu$ gives the label momentum component, derivatives acting on collinear fields give the residual momentum dependence, which scales as $i \partial^\mu \sim k \sim \la^2 Q$. 
It acts on a collinear field as $\cP^\mu\, \xi_{n_i,\lp} = \lp^\mu\, \xi_{n_i,\lp}$. 
Note that we do not need an explicit $n_i$ label on the label momentum operator, since it is implied by the field that the label momentum operator is acting on, and we often use the shorthand notation $\bnP = \bn_i\sdt\cP$. 
We also typically suppress the momentum labels on the collinear fields, keeping only the label of the collinear sector, ${n_i}$. 
The ultrasoft fields carry residual momenta, $i \partial^\mu \sim \la^2Q$, and not label momenta, and their quanta can exchange residual momenta between distinct collinear sectors.

SCET is constructed such that at every stage of a calculation manifest power counting in the expansion parameter $\la$ is preserved.
All fields have a definite power counting as discussed in~\cite{Bauer:2001ct}, and the SCET Lagrangian is expanded as a power series in $\lambda$
\begin{align} \label{eq:SCETLagExpand}
\cL_{\text{SCET}}=\cL_\hard+\cL_\dyn= \sum_{i\geq0} \cL_\hard^{(i)}+ 
 {\cal L}_G^{(0)} + \sum_{i\geq0} \cL^{(i)} \,.
\end{align}
Here $(i)$ denotes objects at ${\cal O}(\lambda^i)$ in the power counting. 
The Lagrangians $ \cL_\hard^{(i)}$ contain the hard scattering operators $O^{(i)}$. The hard scattering operators encode all process dependence and are determined by an explicit matching calculation.
On the other hand the $\cL^{(i)}$ describe the dynamics of ultrasoft and collinear modes in the effective theory, and are universal.
The terms we need from $ \cL_\hard^{(i)}$ and $\cL^{(i)}$ are explicitly known to $\mathcal{O}(\lambda^2)$, and can be found in a summarized form in \cite{iain_notes}.
Finally, ${\cal L}_G^{(0)} $ is the leading power Glauber Lagrangian \cite{Rothstein:2016bsq}, which describes the leading power coupling of soft and collinear degrees of freedom in the forward scattering limit.

In this paper we will be interested in subleading power hard scattering operators, in particular, $\cL_\hard^{(1)}$ and $\cL_\hard^{(2)}$. 

Hard scattering operators are constructed out of collinear building blocks that are gauge invariant products of fields and Wilson lines~\cite{Bauer:2000yr,Bauer:2001ct}. 
These building blocks include quark fields $\chi_{n_i}\sim \lambda$, gluon fields ${\cal B}_{n_i\perp}^\mu\sim \lambda$, and derivatives $\cP_\perp^\mu\sim\lambda$, where we have also indicated their power counting in $\lambda$. 
Here 
\begin{align} \label{eq:chiB}
\chi_{{n_i},\w}(x) &= \Bigl[\delta(\w - \bnP_{n_i})\, W_{n_i}^\dagger(x)\, \xi_{n_i}(x) \Bigr]
\,,\\
\cB_{{n_i}\perp,\w}^\mu(x)
&= \frac{1}{g}\Bigl[\delta(\w + \bnP_{n_i})\, W_{n_i}^\dagger(x)\,i  D_{{n_i}\perp}^\mu W_{n_i}(x)\Bigr]
 \,, \nn
\end{align}
where the collinear covariant derivative in \eq{chiB} is given by
\begin{equation}
i  D_{{n_i}\perp}^\mu = \cP^\mu_{{n_i}\perp} + g A^\mu_{{n_i}\perp}\,,
\end{equation}
and the collinear Wilson line satisfies $\bn_i\cdot D_{n_i} W_{n_i}=0$, has $W_{n_i}\sim \lambda^0$, and is defined as
\begin{equation} \label{eq:Wn}
W_{n_i}(x) = \biggl[~\sum_\text{perms} \exp\Bigl(-\frac{g}{\bnP_{n_i}}\,\bn\sdt A_{n_i}(x)\Bigr)~\biggr]\,.
\end{equation}
The square brackets indicate that the label momentum operators act only on the fields in the Wilson line.
The operators in \eq{chiB} are localized with respect to the residual position $x$, and behave as local quark and gluon fields from the perspective of the ultrasoft degrees of freedom.
Collinear fields transform under ultrasoft gauge transformations as background fields of the appropriate representation. 
Dependence on the ultrasoft degrees of freedom enters the operators through the ultrasoft quark field $q_{us}\sim \lambda^3$, and the ultrasoft covariant derivative $D_{us}\sim \lambda^2$, 
\begin{equation}
i  D_{us}^\mu = i  \partial^\mu + g A_{us}^\mu\,,
\end{equation}
which can be used to construct other operators like the ultrasoft gluon field strength. 
All other field and derivative combinations can be reduced to this set by the use of equations of motion and operator relations~\cite{Marcantonini:2008qn}. 

The hard effective Lagrangian at each power is given by a product of hard scattering operators $\vec O^{(j)}$ constructed from building block fields, and Wilson coefficients $\vec C^{(j)}$,
\begin{align} \label{eq:Leff_sub_explicit}
\cL^{(j)}_{\text{hard}} = \sum_{\{n_i\}} \sum_{A,\cdot\cdot} 
  \bigg[ \prod_{i=1}^{\ell_A} \int \! \! \df \omega_i \bigg] \,
& \vO^{(j)\dagger}_{A\,\lotsdots}\big(\{n_i\};
   \omega_1,\ldots,\omega_{\ell_A}\big) \nn\\
& \times
\vC^{(j)}_{A\,\lotsdots}\big(\{n_i\};\omega_1,\ldots,\omega_{\ell_A} \big)
\,.
\end{align}
The collinear sectors $\{n_i\}$ are determined by the directions found in the collinear states of the hard process being considered.
If there is a direction $n_1'$ in the state then we sum over the cases where each of $n_1$, $\ldots$, $n_4$ is set equal to this $n_1'$.\footnote{The $n_i$ in $\{n_i\}$ are really representatives of an equivalence class determined by demanding that distinct classes $\{n_i\}$ and $\{n_j\}$ have $n_i\cdot n_j\gg \lambda^2$.} 
For most jet processes only a single collinear field appears in each sector at leading power, while subleading power operators can involve multiple collinear fields in the same collinear sector, as well as $\cP_\perp$ insertions.
The scaling of an operator is simply obtained by adding up the powers for the building blocks it contains.
The sum over $A,\cdot\cdot$ in \eq{Leff_sub_explicit} runs over the full basis of operators that appear at this order, which are specified by either helicity labels $\cdot\cdot$ and/or explicit labels $A$ on the operators and coefficients.
A complete basis is necessary to guarantee that the renormalization group evolution of operators will close and that the operators will fully reproduce the IR structure of QCD in this limit.
Moreover, the $\vC^{(j)}_{A}$ are also vectors in the color subspace in which the $\mathcal{O}(\lambda^j)$ hard scattering operators $\vec O_A^{(j)\dagger}$ are decomposed. Explicitly, in terms of color indices, we use the notation of \Ref{Moult:2015aoa} and have
\begin{align} \label{eq:Opm_color}
\vO^\dagger_\lotsdots  &= O_\lotsdots^{a_1\dotsb \alpha_n}\, \vT^{\, a_1\dotsb \alpha_n}
 \,, \nn\\
C_{\lotsdots}^{a_1\dotsb\alpha_n}
 &= \sum_k C_{\lotsdots}^k T_k^{a_1\dotsb\alpha_n}
\equiv \vT^{ a_1\dotsb\alpha_n} \vC_{\lotsdots}
\,.\end{align}
Here $\vT^{\, a_1\dotsb\alpha_n}$ is a row vector of color structures that spans the color conserving subspace. 
The $\alpha_i$ are fundamental indices and the $a_i$ are adjoint indices.
While the color structures do not necessarily have to be independent, they must be complete. 

An efficient approach to simplify operator bases in SCET is to use operators of definite helicity \cite{Moult:2015aoa,Kolodrubetz:2016uim,Feige:2017zci}, which has already been anticipated by labeling our operators $\vec O_{A\,\lotsdots}$ with subscripts $\pm$ for these helicities, following the notation of Ref.~\cite{Feige:2017zci}.
This general philosophy is commonly used in the study of on-shell scattering amplitudes, where it leads to compact expressions, makes symmetries manifest, and removes gauge redundancies.
The use of helicities is also natural in SCET since the effective theory is formulated as an expansion about identified directions $\hat n_i$ which are natural for defining helicities.  

SCET helicity operators were introduced in \cite{Moult:2015aoa} where they were used to study leading power processes with high multiplicities and extended to subleading power in \cite{Kolodrubetz:2016uim} where it was shown that the use of helicity operators is also convenient when multiple fields appear in the same collinear sector.
In \Tab{tab:helicityBB} we give a summary of the complete set of operators that we will use.
We define collinear gluon and collinear quark fields of definite helicity as
\begin{subequations}
	\label{eq:cBpm_quarkhel_def}
\begin{align} 
\label{eq:cBpm_def}
\cB^a_{i\pm} &= -\ve_{\mp\mu}(n_i, \bn_i)\,\cB^{a\mu}_{n_i\perp,\w_i}
\,, \\
\label{eq:quarkhel_def}
 \chi_{i \pm}^\alpha &= \frac{1\,\pm\, \gamma_5}{2} \chi_{n_i, - \omega_i}^\alpha
\,,\qquad\quad
\bar{\chi}_{i \pm}^\balpha =  \bar{\chi}_{n_i,  \omega_i}^\balpha \frac{1\,\mp\, \gamma_5}{2}\,.
\end{align}
\end{subequations}
Here $\alpha$, $\balpha$, and $a$ are $3$, $\bar 3$, and adjoint color indices respectively, and the $\omega_i$ labels on both the gluon and quark building blocks are taken to be outgoing, which is also used for our helicity convention.
We use the standard spinor helicity notation, following for example \cite{Dixon:1996wi},
\begin{align} \label{eq:braket_def}
|p\rangle\equiv \ket{p+} &= \frac{1 + \ga_5}{2}\, u(p)
  \,,
 & |p] & \equiv \ket{p-} = \frac{1 - \ga_5}{2}\, u(p)
  \,, \\
\bra{p} \equiv \bra{p-} &= \mathrm{sgn}(p^0)\, \bar{u}(p)\,\frac{1 + \ga_5}{2}
  \,, 
 & [p| & \equiv \bra{p+} = \mathrm{sgn}(p^0)\, \bar{u}(p)\,\frac{1 - \ga_5}{2}
  \,, \nn 
\end{align}
with $p$ lightlike. With this notation, the polarization vector of an outgoing gluon with momentum $p$ is
\begin{equation}
 \ve_+^\mu(p,k) = \frac{\mae{p+}{\ga^\mu}{k+}}{\sqrt{2} \langle kp \rangle}
\,,\qquad
 \ve_-^\mu(p,k) = - \frac{\mae{p-}{\ga^\mu}{k-}}{\sqrt{2} [kp]}
\,,\end{equation}
where $k\neq p$ is an arbitrary light-like reference vector. In \eq{cBpm_def} it is chosen to be $\bn_i$.

Since fermions always arise in pairs, we can define fermion currents with definite helicities.
Here we will restrict to the case of two back to back directions, $n$ and $\bar n$ which is relevant for our analysis.
We define helicity currents where the quarks are in opposite collinear sectors,
 \begin{align} \label{eq:jpm_back_to_bacjdef}
 & h=\pm 1:
 & J_{n \bn \pm}^{\balpha\beta}
 & = \mp\, \sqrt{\frac{2}{\omega_n\, \omega_\bn}}\, \frac{   \ve_\mp^\mu(n, \bn) }{\langle \bn \mp | n \pm\rangle}   \, \bar{\chi}^\balpha_{n\pm}\, \gamma_\mu \chi^\beta_{\bn \pm}
 \,, \\
 & h=0:
 & J_{n \bn 0}^{\balpha\beta}
 & =\frac{2}{\sqrt{\vphantom{2} \omega_n \,\omega_\bn}\,  [n \bn] } \bar \chi^\balpha_{n+}\chi^\beta_{\bn-}
 \,, \qquad
 (J^\dagger)_{n \bn 0}^{\balpha\beta}=\frac{2}{\sqrt{ \vphantom{2} \omega_n \, \omega_\bn}  \langle n  \bn \rangle  } \bar \chi^\balpha_{n-}\chi^\beta_{\bn+}
 \,, \nn
 \end{align}
 while helicity currents where the quarks are in the same collinear sector are defined as,
\begin{align}\label{eq:coll_subl}
 & h=0:
 & J_{i0}^{\balpha \beta} 
  &= \frac{1}{2 \sqrt{\vphantom{2} \omega_{\bar \chi} \, \omega_\chi}}
  \: \bar \chi^\balpha_{i+}\, \Sl{\bar n}_i\, \chi^\beta_{i+}
   \,,\qquad
   J_{i\bar 0}^{\balpha \beta} 
  = \frac{1}{2 \sqrt{\vphantom{2} \omega_{\bar \chi} \, \omega_\chi}}
  \: \bar \chi^\balpha_{i-}\, \Sl {\bar n}_i\, \chi^\beta_{i-}
 \,, \\[5pt]
  & h=\pm 1:
 & J_{i\pm}^{\balpha \beta}
  &= \mp  \sqrt{\frac{2}{ \omega_{\bar \chi} \, \omega_\chi}}  \frac{\epsilon_{\mp}^{\mu}(n_i,\bar n_i)}{ \big(\l n_i \mp | \bar{n}_i \pm \r \big)^2}\: 
   \bar \chi_{i\pm}^\balpha\, \gamma_\mu \Sl{\bar n}_i\, \chi_{i\mp}^\beta
 \,. \nn
\end{align}
Here $i$ can be either $n$ or $\bar n$.
The Feynman rules for these currents can be found in~\cite{Feige:2017zci}.
Note that the operators $J_{n \bn 0}^{\balpha\beta}$, $(J^\dagger)_{n \bn 0}^{\balpha\beta}$, and $J_{i\pm}^{\balpha \beta}$ have quarks of the opposite chirality, and hence are the ones that will be generated by coupling to a scalar.  

\begin{table}
 \begin{center}
  \begin{tabular}{|c|c|cc|ccc|c|ccc|}
	\hline \phantom{x} & \phantom{x} & \phantom{x} 
	& \phantom{x} & \phantom{x} & \phantom{x} & \phantom{x} 
	& \phantom{x} & \phantom{x} & \phantom{x} & \phantom{x} 
	\\[-13pt]                      
 Field: & 
    $\cB_{i\pm}^a$ & $J_{ij\pm}^{\balpha\beta}$ & $J_{ij0}^{\balpha\beta}$ 
    & $J_{i\pm}^{\balpha \beta}$ 
	& $J_{i0}^{\balpha \beta}$ & $J_{i\bar 0}^{\balpha \beta}$  
    & $\cP^{\perp}_{\pm}$ 
	& $\partial_{us(i)\pm}$ & $\partial_{us(i)0}$ & $\partial_{us(i)\bar{0}}$
	\\[3pt] 
 Power counting: &	
    $\lambda$ &  $\lambda^2$ &  $\lambda^2$
	& $\lambda^2$ & $\lambda^2$& $\lambda^2$ & $\lambda$ 
    & $\lambda^2$ & $\lambda^2$  & $\lambda^2$
	\\
 Equation: & 
   (\ref{eq:cBpm_def}) & \multicolumn{2}{c|}{(\ref{eq:jpm_back_to_bacjdef})} 
     & \multicolumn{3}{c|}{(\ref{eq:coll_subl})} & (\ref{eq:Pperppm}) 
     & \multicolumn{3}{c|}{(\ref{eq:partialus})}
    \\
  \hline  
  \end{tabular}\\
\vspace{.3cm} 
  \begin{tabular}{|c|cc|}
	\hline  \phantom{x} &  \phantom{x} & \phantom{x} 
	\\[-13pt]                        
 Field: & 
 	$\cB^a_{us(i)\pm}$ & 
    \!\!$\cB^a_{us(i)0}$  
	\\[3pt] 
 Power counting: &
 	$\lambda^2$ & $\lambda^2$ 
	\\ 
 Equation: & 
     \multicolumn{2}{c|}{(\ref{eq:Bus})}  
    \\
	\hline
  \end{tabular}
 \end{center}
\vspace{-0.3cm}
\caption{The helicity building blocks in $\text{SCET}_\text{I}$ together with their power counting order in the $\lambda$-expansion, and the equation numbers where their definitions may be found. The building blocks also include the conjugate currents $J^\dagger$ in cases where they are distinct from the ones shown.
} 
\label{tab:helicityBB}
\end{table}

At subleading power one must also consider insertions of the $\cP_{\perp}^\mu$ operator which acts on the perpendicular subspace defined by the vectors $n_i, \bar n_i$.
It is therefore natural to define
\begin{align} \label{eq:Pperppm}
\cP_{+}^{\perp}(n_i,\bar n_i)=-\epsilon^-(n_i,\bar n_i) \cdot \cP_{\perp}\,, \qquad \cP_{-}^{\perp}(n_i,\bar n_i)=-\epsilon^+(n_i,\bar n_i) \cdot \cP_{\perp}\,.
\end{align} 
The $\cP^\perp_\pm$ operator carry helicity $h=\pm 1$. 
We use square brackets to denote which fields are acted upon, for example
$ \left [ \cP^{\perp}_{+}  \cB_{i-}  \right]  \cB_{i-} \cB_{i+}$,
indicates that the $\cP^{\perp}_{+}$ operator acts only on the first field.
For currents, we use a curly bracket notation
\begin{align}\label{eq:p_perp_notation}
  \big\{ \cP^{\perp}_\lambda J_{i 0 }^{\balpha \beta} \big\}  
  & = \frac{1}{2 \sqrt{\vphantom{2}\omega_{\bar \chi} \, \omega_\chi }} \:
   \Big[  \cP^{\perp}_{\lambda}  \bar \chi^\balpha_{i +}\Big] \Sl {\bar n}_i \chi^\beta_{i+}
  \,, \\
 \big\{ J_{i0 }^{\balpha \beta} (\cP^{\perp}_{\lambda})^\dagger \big\}
  &=  \frac{1}{2\sqrt{\vphantom{2}\omega_{\bar \chi} \, \omega_\chi}} \:
  \bar \chi^\balpha_{i+} \Sl {\bar n}_i \Big[   \chi^\beta_{i+} (\cP^{\perp}_{\lambda})^\dagger \Big]
  \,, \nn
\end{align}
to indicate which of the fields is acted on.

To work with gauge invariant ultrasoft gluon fields, we construct our basis post BPS field redefinition.
The BPS field redefinition is~\cite{Bauer:2002nz}
\be \label{eq:BPSfieldredefinition}
\cB^{a\mu}_{n\perp}\to \cY_n^{ab} \cB^{b\mu}_{n\perp} , \qquad \chi_n^\alpha \to Y_n^{\alpha \bbeta} \chi_n^\beta,
\ee
and is performed in each collinear sector.
Here $Y_n$, $\cY_n$ are fundamental and adjoint ultrasoft Wilson lines. 
For a generic representation, $(r)$, the ultrasoft Wilson line is defined by
\be
Y^{(r)}_n(x)=\bold{P} \exp \left [ ig \int\limits_{-\infty}^0 ds\, n\cdot A^a_{us}(x+sn)  T_{(r)}^{a}\right]\,,
\ee
where $\bold P$ denotes path ordering.
The BPS field redefinition decouples the ultrasoft and collinear degrees of freedom at leading power, and accounts for the full physical path of ultrasoft Wilson lines~\cite{Chay:2004zn,Arnesen:2005nk}.

The ultrasoft Wilson lines introduced by the BPS field redefinition can be arranged with the ultrasoft fields to define ultrasoft gauge invariant building blocks.
Particularly, in an arbitrary representation, r, the gauge covariant derivative can be sandwiched by Wilson lines and decomposed in the following way:
\begin{align}\label{eq:soft_gluon}
Y^{(r)\,\dagger}_{n_i} i D^{(r)\,\mu}_{us} Y^{(r)}_{n_i }=i \partial^\mu_{us} + [Y_{n_i}^{(r)\,\dagger} i D^{(r)\,\mu}_{us} Y^{(r)}_{n_i}]=i\partial^\mu_{us}+T_{(r)}^{a} g \cB^{a\mu}_{us(i)}\,.
\end{align}
Here, the ultrasoft gauge invariant gluon field is defined by
\begin{align} \label{eq:softgluondef}
g \cB^{a\mu}_{us(i)}= \left [   \frac{1}{in_i\cdot \partial_{us}} n_{i\nu} i G_{us}^{b\nu \mu} \cY^{ba}_{n_i}  \right] \,.
\end{align}
From \eq{softgluondef} we have  $n_i\cdot \cB^{a}_{us(i)}= 0$. 
The Wilson lines which remain after this procedure can be absorbed into a generalized color structure, $\vT_{\BPS}$. For details about this procedure see \cite{Kolodrubetz:2016uim}.
Determining these color structures is straightforward, see for example~\cite{Feige:2017zci}.
We can next define ultrasoft gauge invariant gluon helicity fields and derivative operators which are analogs of their collinear counterparts.
For the ultrasoft gluon helicity fields we have three building blocks
\begin{equation} \label{eq:Bus}
\cB^a_{us(i)\pm} = -\ve_{\mp\mu}(n_i, \bn_i)\,\cB^{a\mu}_{us(i)},\qquad  \cB^a_{us(i)0} =\bar n_\mu  \cB^{a \mu}_{us(i)}   
\,,\end{equation}
and similarly, for the ultrasoft derivative operators we have
\begin{equation}  \label{eq:partialus}
\partial_{us(i)\pm} = -\ve_{\mp\mu}(n_i, \bn_i)\,\partial^{\mu}_{us},\qquad   \partial_{us(i)0} =\bar n_{i\mu} \partial^{\mu}_{us}, \qquad \partial_{us(i)\bar 0} = n_{i \mu} \partial^{\mu}_{us}
\,.\end{equation}
Note that for the ultrasoft gauge invariant gluon field we use three building block fields to describe the two physical degrees of freedom, unlike for the gauge invariant collinear gluon fields where only two are needed.
This is a consequence of the fact that the ultrasoft gluons are homogenous in their components and are not fundamentally associated with any direction.
Generically, their polarization vectors do not lie in the perpendicular space of any fixed external reference vector.
As we have done for the $\cP_\perp$ operators in \Eq{eq:p_perp_notation}, we will use the same curly bracket notation when inserting ultrasoft derivatives into operators.

At subleading powers, gauge invariant ultrasoft quark fields can also appear explicitly in operator bases, but are not needed here.
Subleading power helicity operators involving ultrasoft quarks are discussed in~\cite{Feige:2017zci}.
For jet collider processes they are not relevant for determining the $\cO(\lambda^2)$ operator basis since they power count as $\cO(\lambda^3)$.
Although ultrasoft quarks do not appear in the hard scattering operators at $\cO(\lambda^2)$ they do appear in the calculation of cross sections or amplitudes at $\cO(\lambda^2)$ through subleading power Lagrangian insertions.
These ultrasoft quark dependent Lagrangians were important for the subleading power perturbative SCET calculation of \Ref{Moult:2016fqy}.

Finally, the helicity operator basis discussed here only provides a complete basis in $d=4$, and we have not discussed evanescent operators \cite{Buras:1989xd,Dugan:1990df,Herrlich:1994kh}.
In general additional building block fields would be introduced, for example an $\epsilon$ scalar gluon $\cB^a_{\epsilon}$ to encode the $(-2\epsilon)$ transverse degrees of freedom of the gluon.
The extension of our basis to include evanescent operators is best done in the context of explicit calculations of for example the one-loop Wilson coefficients.
Since we do not perform a one-loop matching to our operators, we leave the treatment of evanescent operators to future work.

\section{Operator Basis}\label{sec:basis}

{
\renewcommand{\arraystretch}{1.4}
\begin{table}[h]
\hspace{-0.15cm} \scalebox{0.9}{
\begin{tabular}{| c | l | l | c | c | c | }
	\hline 
Order & $\!$Category &  Operators (equation number) 
 & \#$\!$  helicity & \#$\!$ of
 & $\!\sigma_{2j}^{\cO(\lambda^2)}\!\! \ne\! 0\!$
 \\[-8pt]
 & & & configs & \! color\! & 
 \\ \hline 
$\mathcal{O}(\lambda^0)$  
   & $\! H b\bar b$ & $O_{(\lambda_1)}^{(0)ab}=J_{n\bar n\lambda_1}^{\balpha\bt}H$ \,(\ref{eq:Hbb})
   & 2  & 1 & $\checkmark$
    \\ \hline
$\mathcal{O}(\lambda^1)$ 
   & $\! H b \bar{b} g$  
    & $O_{\cB \lambda_1 (\lambda_2)}^{(1)a\,\balpha\bt}
	=  \cB_{\bar n\lambda_1}^a\, J_{n\, \lambda_2}^{\balpha\bt}\,H$ \,(\ref{eq:H1_basis}) 
    & 2 & 1 &   $\checkmark$ 
    \\ \hline
$\mathcal{O}(\lambda^2)$
   & $\! H b \bar b g g$  & $O_{\cB1\lambda_1 \lambda_2(\lambda_3)}^{(2)ab\, \balpha\bt}
	=  \cB_{n \lambda_1}^a \cB_{\bar n \lambda_2}^b \, J_{n\bar n\,{\lambda_3} }^{\balpha\bt}   \,H $ \,(\ref{eq:Hbbgg_basis1}) 
    & 4  & 3 & 
    \\
	& & $O_{\cB2\lambda_1 \lambda_2(\lambda_3)}^{(2)ab\, \balpha\bt}
	=  \cB_{n \lambda_1}^a \cB_{n \lambda_2}^b \, J_{n\bar n\,{\lambda_3} }^{\balpha\bt}   \,H $ \,(\ref{eq:Hbbgg_basis2}) 
    & 2  & 3 & $\checkmark$
    \\
        & & $O_{\cB3\lambda_1 \lambda_2(\lambda_3)}^{(2)ab\, \balpha\bt}
	=  \cB_{\bar n \lambda_1}^a \cB_{\bar n \lambda_2}^b \, J_{n\bar n\,{\lambda_3} }^{\balpha\bt}   \,H $ \,(\ref{eq:Hbbgg_basis3}) 
    & 2  & 3 & $\checkmark$
    \\ \cline{2-6}
   & $\! H b \bar b q \bar q$  & $O_{bq1(\lambda_1;\lambda_2)}^{(2)\balpha\bt\bgamma\delta}
	= J_{(b)n {\lambda_1}\, }^{\balpha\bt}\, J_{(q) n\bar n {\lambda_2}\, }^{\bgamma\delta}\,H$ \,(\ref{eq:Z2_basis_bq_1}) 
    & 2 & 2 &
    \\
	& & $O_{bq2(\lambda_1;\lambda_2)}^{(2)\balpha\bt\bgamma\delta}
	= J_{(b) \bar n  \lambda_1\, }^{\balpha\bt}\, J_{(q) n\bar{n}  \, \lambda_2\, }^{\bgamma\delta}\,H $ \,(\ref{eq:Z2_basis_bq_2})
    & 2  & 2 &
    \\
    & & $O_{bq3(\lambda_1;\lambda_2)}^{(2)\balpha\bt\bgamma\delta}
	= J_{(b) n \bar n \lambda_1\, }^{\balpha\bt}\, J_{(q) n \lambda_2\, }^{\bgamma\delta}\,H$ \,(\ref{eq:Z2_basis_bq_3})
    & 4 &  2 & $\checkmark$
    \\
    & & $O_{bq4(\lambda_1;\lambda_2)}^{(2)\balpha\bt\bgamma\delta}
	= J_{(b) n \bar n \lambda_1\, }^{\balpha\bt}\, J_{(q) \bar n \lambda_2\, }^{\bgamma\delta}\,H$ \,(\ref{eq:Z2_basis_bq_4})
    & 4 & 2  & $\checkmark$
    \\ \cline{2-6}
   & $\! H b \bar b b \bar b$  & $O_{bb1(\lambda_1;\lambda_2)}^{(2)\balpha\bt\bgamma\delta}
	= S\,J_{(b)n\bar n {\lambda_1}\, }^{\balpha\bt}\, J_{(b)n {\lambda_2}\, }^{\bgamma\delta}\,H$ \,(\ref{eq:Z2_basis_bb_1}) 
    & 4 & 2 &  $\checkmark$
    \\
	& & $O_{bb2(\lambda_1;\lambda_2)}^{(2)\balpha\bt\bgamma\delta}
	=S\, J_{(b) n \bar n \lambda_1\, }^{\balpha\bt}\, J_{(b) \bar n \lambda_2\, }^{\bgamma\delta}\,H$ \,(\ref{eq:Z2_basis_bb_2})
    & 4 & 2 & $\checkmark$
    \\ \cline{2-6}
   &  $\cP_\perp$   & $O_{\cP\chi 1\lambda_1 (\lambda_2)[\lambda_{\cP}]}^{(2)a\,\balpha\bt}
	= \cB_{n\lambda_1}^a \, \{\cP_{\perp}^{\lambda_{\cP}}J_{n\bar n\, {\lambda_2}}^{\balpha\bt}\}\,H$  \,(\ref{eq:Hbbgpperp_basis1})  
    & 4 & 1 & $\checkmark$
    \\
	& & $O_{\cP\chi 2 \lambda_1 (\lambda_2)[\lambda_{\cP}]}^{(2)a\balpha\bt}
	= \cB_{\bar n\lambda_1}^a\, \{J_{n\bar n\, {\lambda_2}}^{\balpha\bt}(\cP_{\perp}^{\lambda_{\cP}})^\dagger\}\, H$  \,(\ref{eq:Hbbgpperp_basis2}) 
    & 4  & 1 & $\checkmark$ 
    \\ \cline{2-6}
	&  	$\!$Ultrasoft $\!\!\!$    & $O_{\cB(us(n))0:(\lambda_1)}^{(2)a\,\balpha\bt} 
	= \cB_{us(n)0}^a \, J_{n\bar n\,\lambda_1}^{\balpha\bt}\,H $ \,(\ref{eq:soft_insert_basis1}) 
    & 2 & 1 & $\checkmark$ 
    \\
	&   & $O_{\cB(us(\bar n))0:(\lambda_1)}^{(2)a\,\balpha\bt} 
	= \cB_{us(\bar n)0}^a \, J_{n\bar n\,\lambda_1}^{\balpha\bt}\,H $ \,(\ref{eq:soft_insert_basis2}) 
    & 2 & 1 & $\checkmark$ 
    \\
	& & $O_{\partial(us(n))0:(\lambda_1)}^{(2)\,\balpha\bt}
	= \{\partial_{us(n)0} \, J_{n\bar n\,\lambda_1}^{\balpha\bt}\}\,H $ \,(\ref{eq:soft_derivative_basis1}) 
    & 2 & 1 & $\checkmark$ 
    \\
    	& & $O_{\partial(us(\bar n))0:(\lambda_1)}^{(2)\,\balpha\bt}
	= \{J_{n\bar n\,\lambda_1}^{\balpha\bt}\,(\partial_{us(\bar n)0})^\dagger\}\,H $ \,(\ref{eq:soft_derivative_basis2}) 
    & 2 & 1 & $\checkmark$ 
    \\ \hline
\end{tabular}}
\vspace{0.1cm}
\caption{Basis of hard scattering operators for $H\to b\bar b$ or $b\bar b\to H$ up to ${\cal O}(\lambda^2)$. The $\lambda_i$ denote helicities, $S$ represents a symmetry factor present for some cases, and detailed lists of operators can be found in the indicated equation.  In the fourth column, we summarize the number of allowed helicity configurations. The final column indicates which operators contribute to the cross section up to $\mathcal{O}(\lambda^2)$ in the power expansion, as discussed in detail in Sec. \ref{sec:discussion}. Counting the helicity configurations there are a total of 48 operators, of which only 40 contribute to the cross section at $\mathcal{O}(\lambda^2)$.
Of those 40 operators, only 16 of them have non zero Wilson coefficients at tree level. These numbers do not include the number of distinct color configurations which are indicated in the 5th column.
}
\label{tab:summary}
\end{table}
}

In this section, we construct the $\cO(\lambda)$ and $\cO(\lambda^2)$ basis of the power suppressed hard scattering operators for the $H\to b\bar b$ process (where other than this coupling the $b$ quark is treated as massless). When we write the operator basis using the helicity operators, the basis will be greatly simplified by the symmetries that arise from the helicity conservation. The helicity-operator approach is particularly powerful in this case due to the fact that the Higgs is spinless. We summarize the complete basis of field structures in \Tab{tab:summary}, and we will show which operators contribute to the cross section at $\cO(\lambda^2)$ in \Sec{sec:discussion}. These operators are indicated with a check mark in the table.

From \eq{Leff_sub_explicit} we can see that the hard Lagrangian in SCET it is written as a sum over label momenta of the hard operators.  If we take the special case of two back-to-back collinear sectors this reduces to
\begin{align}\label{eq:sum_dir}
\cL_{\text{hard}}^{(j)} = \sum_{n} \sum_{A,\cdot\cdot}  
\bigg[ \prod_{i=1}^{\ell_A} \int \! \! \df \omega_i \bigg] \,
& \vO^{(j)\dagger}_{A\,\lotsdots}\big(n,\bn;
   \omega_1,\ldots,\omega_{\ell_A}\big) \nn\\
& \times
\vC^{(j)}_{A\,\lotsdots}\big(n,\bn;\omega_1,\ldots,\omega_{\ell_A} \big)
\,.
\end{align}
Therefore, we do not need to include twice those operators which are identical up to the swap of $n\leftrightarrow \bar n$ when writing the basis. This implies that when we consider an operator with different field structures in the two collinear sectors we have the freedom to make an arbitrary choice for which is labeled $n$ and which $\bar n$, and this choice can be made independently for each operator. All possible interferences are properly incorporated by the sum over directions in \Eq{eq:sum_dir} when squaring matrix elements.

\subsection{Leading Power}\label{sec:basis_lp}

The leading power operators for $b\bar b\to H$ or $H\to b\bar b$ in the Higgs effective theory are well known. Due to the fact that the Higgs is spin zero and the quark-antiquark pair from the Yukawa interaction have opposite chirality, the only two operators are
\begin{align}
 \boldsymbol{b_n \bar b_{\bn}:}   {\vcenter{\includegraphics[width=0.18\columnwidth]{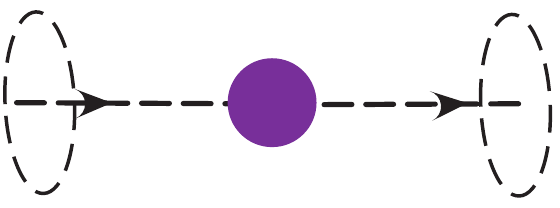}}} \nn
\end{align}
\vspace{-0.4cm}
\begin{alignat}{2}\label{eq:Hbb}
 &O_{0}^{(0)\balpha\bt}
=J_{n\bar n\,0}^{\balpha\bt}\, H
\,, \qquad &&O_{0^\dagger}^{(0)\balpha\bt}
=(J^\dagger)_{n\bar n\,0}^{\balpha\bt}\, H\,.
\end{alignat}
Here the purple circled denotes that this is a hard scattering operator in the effective theory, while the dashed circles indicate which fields are in each collinear sector.
Note that here we have opted not to include a symmetry factor at the level of the operator. We will include symmetry factors in the operator only when there is an exchange symmetry within a given collinear sector. We assume that overall symmetry factors which involve exchanging particles from different collinear sectors are taken into account at the phase space level. Following the notation in \eq{Opm_color}, we see that the color basis here is one-dimensional, with
\begin{align} \label{eq:leading_color}
 \vT^{\alpha\bar \bt} = \de_{\alpha\bar \bt}\,, \qquad 
 \vT_{\BPS}^{\alpha\bar \bt} = \bigl( Y_{n}^\dagger Y_{\bar n} \bigr)_{\alpha\bar \bt}  
\,.
\end{align}
Here $\vT_{\BPS}^{\alpha\bar\bt}$ combines the color basis with the ultrasoft Wilson lines obtained from the BPS field redefintion in \eq{BPSfieldredefinition}.

\subsection{Subleading Power}\label{sec:nlp}

To simplify our operator basis, we will work in the center of mass frame. Furthermore, we will choose the $n$ and $\bar n$ axes such that the total label $\perp$ momentum of each collinear sector vanishes. Such choice is allowed in an SCET$_{\text{I}}$ theory since the ultrasoft sector does not have label momentum. Therefore, we do not need to consider operators where the $\cP_\perp$ operator acts on a sector with a single collinear field (We consider the generalization away from this choice in \app{gen_pt}.). After excluding such operators, the suppression of the operators at the $\cO(\lambda)$ order must, therefore, come from an explicit collinear field. 

It turns out that the $\cO(\lambda)$ operators are highly constrained due to the spin-0 nature of the Higgs. From the above discussions, we see that there are two possibilities for the collinear field content of the operators, either three collinear gluon fields, or two collinear quark fields and a collinear gluon field. Surprisingly, one can immediately see that the operators involving three collinear gluon fields are not possible since they cannot sum to a state with zero helicity. As a result, only the operators with two collinear quark fields and a collinear gluon field are in the operator basis.  To satisfy the helicity constraints, the collinear quark current in these operators must have helicity $\pm1$ in order to cancel the spin of the collinear gluon field. Furthermore, the quark-antiquark pair arises from the Yukawa interaction, and therefore must have opposite chirality. Together this implies that the quarks are described by the current $J_{n\,\pm}^{\balpha\bt}$, where we choose the convention that the quark is in the $n$-collinear sector. By label momentum conservation, the remaining gluon is in the $\bn$-collinear section. The only two operators in the basis at  $\cO(\lambda)$ are therefore
\begin{align}
\boldsymbol{(b\bar b)_n (g)_{\bn}:}   {\vcenter{\includegraphics[width=0.18\columnwidth]{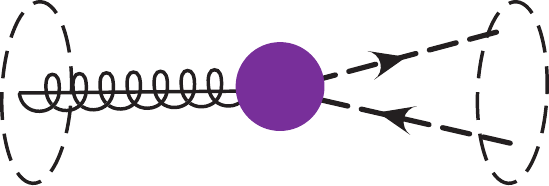}}} \nn
\end{align}
\vspace{-0.4cm}
\begin{alignat}{2} \label{eq:H1_basis}
&O_{\cB\bar n+(+)}^{(1)a\,\balpha\bt}
= \cB_{\bar n+}^a\, J_{n\,+}^{\balpha\bt} \,H
\,,\qquad &
&O_{\cB\bar n-(-)}^{(1)a\,\balpha\bt}
= \cB_{\bar n-}^a \, J_{n\,-}^{\balpha\bt}\,H
\,,
\end{alignat}
The color basis is one-dimensional $\vT^{a\, \al\bbeta} = T^a_{\al\bbeta}$. After the BPS field redefinition we have
\begin{align} \label{eq:nlp_color}
 \vT_{\BPS}^{ a \al\bbeta} 
    &= \left (\cY_{\bar n}^{ba} \cY_{n}^{bc} T^c   \right )_{\alpha \bar \beta}
    \,,
\end{align}
for \eq{H1_basis}.

\subsection{Subsubleading Power}\label{sec:nnlp}

At the subsubleading power, the allowed operators can include either only collinear field insertions, insertions of one collinear field and one $\cP_\perp$ operator, or ultrasoft field insertions. We will discuss each of these cases separately.

\subsubsection{Collinear Field Insertions}\label{sec:nnlp_collinear}

We start with operators involving only collinear field insertions, which can have four collinear fields at $\cO(\lambda^2)$. Moreover, the bottom quark and bottom antiquark from the Yukawa interaction carry opposite chirality, but the gluon splitting interaction does not change the chirality of the quark. Therefore, the allowed operators must at least include two quark fields, and they can be composed purely of collinear quark fields or of two collinear gluon fields and a collinear quark current. In each of these cases, the possible helicity combinations of the operators will be restricted by helicity selection rules.

\vspace{0.4cm}
\noindent{\bf{Two Quark-Two Gluon Operators:}}

We start with operators involving two collinear quark fields and two collinear gluon fields. The possible helicity combination of these operators are again severely constrained by the helicity selection rules. We notice that the total helicity of two gluon fields is either $0$ or $2$, and the total helicity of two quark fields is either $0$ or $1$. Therefore, to achieve a total spin zero, both the gluon fields and the quark fields must be in helicity zero configurations. Furthermore, since the quark fields arise from the Yukawa interaction they must have opposite chirality. This implies that all operators must involve only the currents $J_{n\,\bar n\,0}^{\balpha\bt}$ or $(J^\dagger)_{n\,\bar n\,0}^{\balpha\bt}$, where we have taken without loss of generality that the bottom quark is in the $n$-collinear sector, as per the discussion below \Eq{eq:sum_dir}. The two gluon fields can then either be in opposite collinear sectors, or in the same collinear sector. The color basis before BPS field redefinition is identical for the two cases. It is three dimensional, and we take as a basis
\begin{equation} \label{eq:bbggll_color}
\vT^{\, ab \alpha\bbeta}
= \Bigl(
(T^a T^b)_{\alpha\bbeta}\,,\, (T^b T^a)_{\alpha\bbeta} \,,\, \tr[T^a T^b]\, \delta_{\alpha\bbeta}
\Bigr)
\,.\end{equation} 

In the case that the two collinear gluons are in opposite collinear sectors, a basis of helicity operators is given by
\begin{align}
 \boldsymbol{(bg)_n (\bar bg)_{\bn}:}  {\vcenter{\includegraphics[width=0.18\columnwidth]{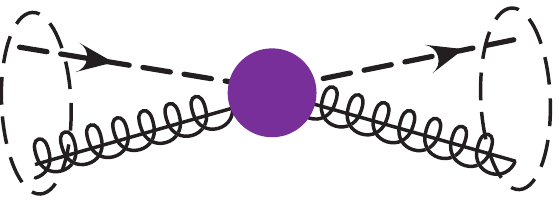}}} \nn
\end{align}
\vspace{-0.4cm}
\begin{alignat}{2} \label{eq:Hbbgg_basis1}
&O_{\cB1++(0)}^{(2)ab\, \balpha\bt}
=   \cB_{n+}^a\, \cB_{\bar n+}^b \, J_{n\bar n\,{0} }^{\balpha\bt}  \,H\, , \qquad 
&&O_{\cB1++(0^\dagger)}^{(2)ab\, \balpha\bt}
=\cB_{n+}^a\, \cB_{\bar n+}^b  \, (J^\dagger)_{n\bar n\,0 }^{\balpha\bt}   \,H\, ,  \\
&O_{\cB1--(0)}^{(2)ab\, \balpha\bt}
=  \cB_{n-}^a\, \cB_{\bar n-}^b \, J_{n\bar n\,{0} }^{\balpha\bt}    \,H\, , \qquad 
&&O_{\cB1--(0^\dagger)}^{(2)ab\, \balpha\bt}
= \cB_{n-}^a\, \cB_{\bar n-}^b  \, (J^\dagger)_{n\bar n\,0 }^{\balpha\bt}   \,H\, . \nn
\end{alignat}
The color basis after BPS field redefinition is given by
\begin{equation}
\vT_{\BPS}^{\, ab \alpha\bbeta}
= \Bigl(
(T^aY_n^\dagger Y_{\bar n} T^b)_{\alpha\bbeta} \,,\, 
(Y_n^\dagger T^d\cY_{\bar n}^{db} T^c \cY_{n}^{ca} Y_{\bn})_{\alpha\bbeta} \,,\, 
T_F (\cY_n^T \cY_\bn)^{ab} \, \bigl(Y_n^\dagger Y_{\bar n}\bigr)_{\alpha\bbeta}
\Bigr)
\,,
\end{equation}
where the identity $\tr[T^a T^b] = T_F \delta^{ab}$ has been used.

The two gluons can also be in the same collinear sector. In the case that they are in $n$-collinear sector a basis of helicity operators is given by:
\begin{align}
\boldsymbol{(bgg)_n (\bar b)_{\bn}:} {\vcenter{\includegraphics[width=0.18\columnwidth]{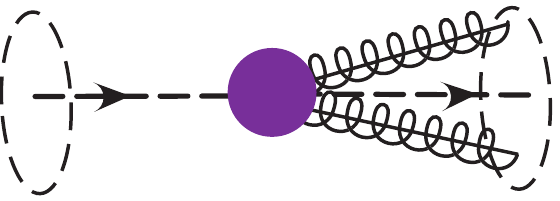}}} \nn
\end{align}
\begin{alignat}{2}\label{eq:Hbbgg_basis2}
&O_{\cB2+-(0)}^{(2)ab\, \balpha\bt}
=  \cB_{n+}^a\, \cB_{n-}^b \, J_{n\bar n\,{0} }^{\balpha\bt}    \,H\, , \qquad 
&O_{\cB2+-(0^\dagger)}^{(2)ab\, \balpha\bt}
= \cB_{n+}^a\, \cB_{n-}^b  \, (J^\dagger)_{n\bar n\,0 }^{\balpha\bt}   \,H\,.
\end{alignat}
After BPS field redefinition, the color basis is
\begin{equation}
\vT_{\BPS}^{\, ab \alpha\bbeta}
= \Bigl(
(T^aT^bY_n^\dagger Y_{\bar n})_{\alpha\bbeta}
  \,,\, 
(T^bT^aY_n^\dagger Y_{\bar n})_{\alpha\bbeta}
   \,,\, 
T_F \delta^{ab} \, \bigl(Y_n^\dagger Y_{\bar n}\bigr)_{\alpha\bbeta} \Bigr)
\,.
\end{equation}

In the case that the two gluons are in $\bn$-collinear sector a basis of helicity operators is given by:
\begin{align}
\boldsymbol{(b)_n (\bar bgg)_{\bn}:} {\vcenter{\includegraphics[width=0.18\columnwidth]{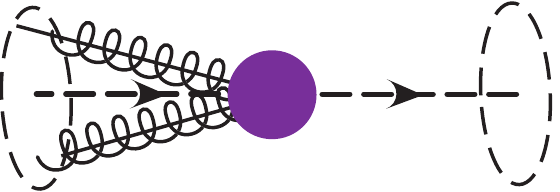}}}\nn
\end{align}
\begin{alignat}{2}\label{eq:Hbbgg_basis3}
&O_{\cB3+-(0)}^{(2)ab\, \balpha\bt}
=  \cB_{\bar n+}^a\, \cB_{\bar n-}^b \, J_{n\bar n\,{0} }^{\balpha\bt}    \,H\, , \qquad 
&O_{\cB3+-(0^\dagger)}^{(2)ab\, \balpha\bt}
= \cB_{\bar n+}^a\, \cB_{\bar n-}^b  \, (J^\dagger)_{n\bar n\,0 }^{\balpha\bt}   \,H\,.
\end{alignat}
After BPS field redefinition, the color basis is
\begin{equation}
\vT_{\BPS}^{\, ab \alpha\bbeta}
= \Bigl(
(Y_n^\dagger Y_{\bar n}T^aT^b)_{\alpha\bbeta}
  \,,\, 
(Y_n^\dagger Y_{\bar n}T^bT^a)_{\alpha\bbeta}
   \,,\, 
T_F \delta^{ab} \, \bigl(Y_n^\dagger Y_{\bar n}\bigr)_{\alpha\bbeta} \Bigr)
\,.
\end{equation}

\vspace{0.4cm}
\noindent{\bf{Four Quark Operators:}}

We now consider the operators involving four collinear quark fields. In this case, we first notice that one quark-antiquark pair is produced from the Yukawa interaction and the other quark-antiquark pair is produced from a gluon splitting, and therefore we have one quark-antiquark pair with opposite chirality and the other pair with the same chirality.

When constructing the operator basis, we must also consider separately the case of identical quark flavors $H b \bar b b\bar b$ and distinct quark flavors $H  b \bar b q\bar q $. For the case of distinct quark flavors $H b \bar b q\bar q $ the two quarks of flavor $b$ are of opposite chirality, and the two quarks of flavor $q$ are of the same chirality. We choose the quarks of the same flavor to appear in the same current, and the current will be labeled by the flavor $(b)$ or $(q)$. For all these cases, the color basis is
\begin{equation} \label{eq:qqqq_color}
\vT^{\, \al\bbeta\ga\bdelta} =
\Bigl(
\de_{\al\bdelta}\, \de_{\ga\bbeta}\,,\, \delta_{\al\bbeta}\, \de_{\ga\bdelta}
\Bigr)
\,.\end{equation}
We will give results for the corresponding $\bar T_{\rm BPS}^{\, \al\bbeta\ga\bdelta}$ basis after BPS field redefinition when we consider each case below.

We first consider the case of operators with distinct quark flavors $H b \bar b q\bar q $. Due to the chirality constraint of the two quark pairs, the current formed by the bottom quark pair has helicity $\pm1$ if the bottom quarks are in the same collinear sector, and has helicity $0$ if the bottom quarks are in different collinear sectors. For the other flavor, the current has helicity $0$ if the quarks are in the same collinear sector and has helicity $\pm1$ if the quarks are in different collinear sectors. Therefore, in order for the state to have total helicity $0$, we must have three quarks or antiquarks in one collinear sector and one quark or antiquark in the other collinear sector. There are four different cases, corresponding to the cases where $b$, $\bar b$, $q$, $\bar q$ are in a collinear sector alone, respectively. In the case that $\bar q$ alone is in a collinear sector, the bottom quark current must have helicity $\pm1$, and the other current must have helicity $\mp 1$. Then, the operator basis satisfying the chirality and angular momentum constraint is given by:
\vspace{0.3cm}
\begin{align}
 \boldsymbol{(b \bar bq)_n(\bar q)_\bn:} {\vcenter{\includegraphics[width=0.18\columnwidth]{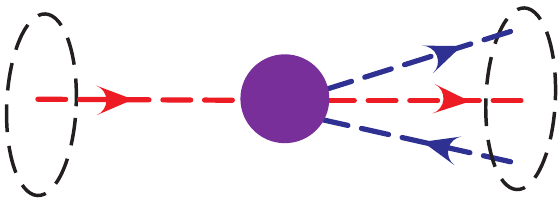}}} \nn
\end{align}
\vspace{-0.4cm}
\begin{alignat}{2}\label{eq:Z2_basis_bq_1}
&O_{bq1(+;-)}^{(2)\balpha\bt\bgamma\delta}
= \,  J_{(b)n+\, }^{\balpha\bt}\, J_{(q)n\bar n-\, }^{\bgamma\delta} H
\,,\qquad &
&O_{bq1(-;+)}^{(2)\balpha\bt\bgamma\delta}
=\, J_{(b)n-\, }^{\balpha\bt}\, J_{(q)n\bar n+\, }^{\bgamma\delta} H
\,,
\end{alignat}
For the operators in \eq{Z2_basis_bq_1} the color basis after BPS field redefinition is
\begin{align}  \label{eq:TBPS_Obq1}
\vT_{\BPS}^{ \al\bbeta\ga\bdelta} &=
\left(\bigl(Y_n^\dagger\,Y_{\bar n}\bigr)_{\alpha\bar \delta}\,\delta_{\gamma\bar \bt} \,,\, \delta_{\al\bbeta}\,\bigl(Y_n^\dagger\,Y_{\bar n}\bigr)_{\gamma\bar \delta} \right)
\,.
\end{align}
In the case that $q$ alone is in a collinear sector, the operator basis is given by:
\begin{align}
 \boldsymbol{(q)_n(b \bar b\bar q)_\bn:} {\vcenter{\includegraphics[width=0.18\columnwidth]{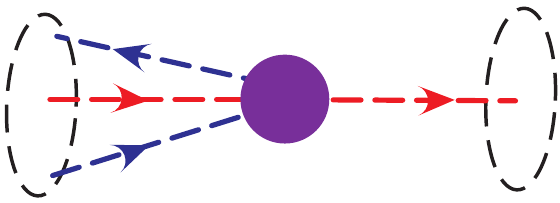}}} \nn
\end{align}
\vspace{-0.4cm}
\begin{alignat}{2}\label{eq:Z2_basis_bq_2}
&O_{bq2(+;+)}^{(2)\balpha\bt\bgamma\delta}
= \,  J_{(b)\bar n+\, }^{\balpha\bt}\, J_{(q)n\bar n+\, }^{\bgamma\delta} H
\,,\qquad &
&O_{bq2(-;-)}^{(2)\balpha\bt\bgamma\delta}
=\, J_{(b)\bar n-\, }^{\balpha\bt}\, J_{(q)n\bar n-\, }^{\bgamma\delta} H
\,,
\end{alignat}
For the operators in \eq{Z2_basis_bq_2} the color basis after BPS field redefinition is
\begin{align}  \label{eq:TBPS_Obq2}
\vT_{\BPS}^{ \al\bbeta\ga\bdelta} &=
\left(\delta_{\alpha\bar \delta}\,\bigl(Y_n^\dagger\,Y_{\bar n}\bigr)_{\gamma\bar \bt} \,,\, \delta_{\al\bbeta}\,\bigl(Y_n^\dagger\,Y_{\bar n}\bigr)_{\gamma\bar \delta} \right)
\,.
\end{align}
When the bottom antiquark $\bar b$ alone is in a collinear sector, both currents have helicity $0$ due to the chirality and angular momentum conservation. The operator basis in this case is then given by:
\begin{align}
\boldsymbol{(bq\bar q)_n(\bar b)_\bn:} {\vcenter{\includegraphics[width=0.18\columnwidth]{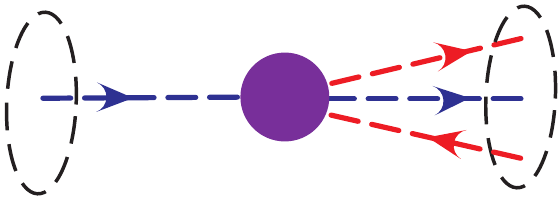}}} \nn
\end{align}
\vspace{-0.4cm}
\begin{align}\label{eq:Z2_basis_bq_3}
&O_{bq3(0;0)}^{(2)\balpha\bt\bgamma\delta}
=\, J_{(b) n \bar n 0\, }^{\balpha\bt}\, J_{(q) n0\,}^{\bgamma\delta} H
\,,\qquad &
&O_{bq3(0;\bar 0)}^{(2)\balpha\bt\bgamma\delta}
= \, J_{(b)n \bar n 0\, }^{\balpha\bt}\, J_{(q) n\bar 0\, }^{\bgamma\delta} H
\,,\\
&O_{bq3(0^\dagger;0)}^{(2)\balpha\bt\bgamma\delta}
=\, (J^\dagger)_{(b) n \bar n 0\, }^{\balpha\bt}\, J_{(q) n0\, }^{\bgamma\delta} H
\,,\qquad &
&O_{bq3(0^\dagger;\bar 0)}^{(2)\balpha\bt\bgamma\delta}
= \, (J^\dagger)_{(b)n \bar n 0\, }^{\balpha\bt}\, J_{(q) n\bar 0\, }^{\bgamma\delta} H
\,. \nn
\end{align}
For the operators in \eq{Z2_basis_bq_3} the color basis after BPS field redefinition is
\begin{align}  \label{eq:TBPS_Obq3}
\vT_{\BPS}^{ \al\bbeta\ga\bdelta} &=
\left(\delta_{\alpha\bar \delta}\,\bigl(Y_n^\dagger\,Y_{\bar n}\bigr)_{\gamma\bar \bt} \,,\, \bigl(Y_n^\dagger\,Y_{\bar n}\bigr)_{\al\bbeta}\,\delta_{\gamma\bar \delta} \right)
\,.
\end{align}
Finally, when the bottom quark $b$ is in a collinear sector alone, the operator basis is:
\begin{align}
\boldsymbol{(b)_n(\bar bq\bar q)_\bn:} {\vcenter{\includegraphics[width=0.18\columnwidth]{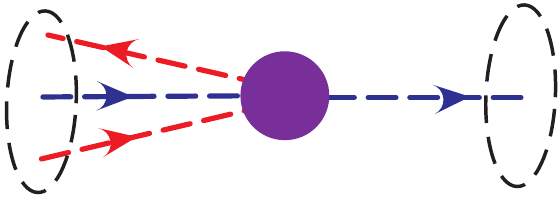}}} \nn
\end{align}
\vspace{-0.4cm}
\begin{align}\label{eq:Z2_basis_bq_4}
&O_{bq4(0;0)}^{(2)\balpha\bt\bgamma\delta}
=\, J_{(b) n \bar n 0\, }^{\balpha\bt}\, J_{(q)\bar n0\,}^{\bgamma\delta} H
\,,\qquad &
&O_{bq4(0;\bar 0)}^{(2)\balpha\bt\bgamma\delta}
= \, J_{(b)n \bar n 0\, }^{\balpha\bt}\, J_{(q)\bar n\bar 0\, }^{\bgamma\delta} H
\,,\\
&O_{bq4(0^\dagger;0)}^{(2)\balpha\bt\bgamma\delta}
=\, (J^\dagger)_{(b) n \bar n 0\, }^{\balpha\bt}\, J_{(q)\bar n0\, }^{\bgamma\delta} H
\,,\qquad &
&O_{bq4(0^\dagger;\bar 0)}^{(2)\balpha\bt\bgamma\delta}
= \, (J^\dagger)_{(b)n \bar n 0\, }^{\balpha\bt}\, J_{(q)\bar n\bar 0\, }^{\bgamma\delta} H
\,. \nn
\end{align}
For the operators in \eq{Z2_basis_bq_4} the color basis after BPS field redefinition is
\begin{align}  \label{eq:TBPS_Obq4}
\vT_{\BPS}^{ \al\bbeta\ga\bdelta} &=
\left(\bigl(Y_n^\dagger\,Y_{\bar n}\bigr)_{\alpha\bar \delta}\,\delta_{\gamma\bar \bt} \,,\, \bigl(Y_n^\dagger\,Y_{\bar n}\bigr)_{\al\bbeta}\,\delta_{\gamma\bar \delta} \right)
\,.
\end{align}
In the above four cases, we choose the collinear sector directions such that the quark of the pair that are in different collinear sectors is in the $n$-collinear sector. To implement this we took the current to be $J^{\balpha\bt}_{n\bar n\lambda}$ rather than $J^{\balpha\bt}_{\bar n n\lambda}$.

For identical quark flavors, the operators are similar to those in the distinct flavors case. However, the operators in \eq{Z2_basis_bq_1} are equivalent to the two operators in \eq{Z2_basis_bq_3} if the quark flavors are identical. Similarly, the operators in \eq{Z2_basis_bq_2} and \eq{Z2_basis_bq_4} are also equivalent. Therefore, when a bottom antiquark $\bar b$ is in a collinear sector alone, the operator basis is reduced to:
\begin{align}
\boldsymbol{(b\bar b b)_n(\bar b)_\bn:} {\vcenter{\includegraphics[width=0.18\columnwidth]{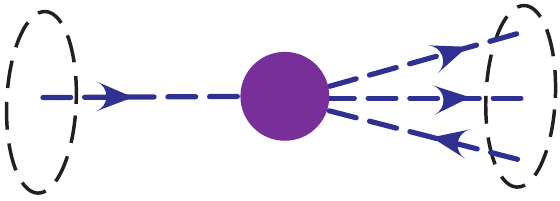}}}
\nn
\end{align}
\vspace{-0.4cm}
\begin{alignat}{2} \label{eq:Z2_basis_bb_1}
&O_{bb1(0;0)}^{(2)\balpha\bt\bgamma\delta}
= \frac{1}{2} \,  J_{(b)n\bar n0\, }^{\balpha\bt}\, J_{(b)n0\, }^{\bgamma\delta} H
\,,\qquad &
&O_{bb1(0;\bar 0)}^{(2)\balpha\bt\bgamma\delta}
=\, J_{(b) n\bar n0\, }^{\balpha\bt}\, J_{(b)n\bar 0\, }^{\bgamma\delta} H
\,, \\
&O_{bb1(0^\dagger;0)}^{(2)\balpha\bt\bgamma\delta}
= \,  (J^\dagger)_{(b)n\bar n0\, }^{\balpha\bt}\, J_{(b)n0\, }^{\bgamma\delta} H
\,,\qquad &
&O_{bb1(0^\dagger;\bar 0)}^{(2)\balpha\bt\bgamma\delta}
=\frac{1}{2} \, (J^\dagger)_{(b) n\bar n0\, }^{\balpha\bt}\, J_{(b)n\bar 0\, }^{\bgamma\delta} H
\,.\nn
\end{alignat}
Similarly, in the case that a bottom quark $b$ is in a collinear sector alone, the operator basis is reduced to:
\begin{align}
\boldsymbol{(b)_n(b\bar b\bar b)_\bn:} {\vcenter{\includegraphics[width=0.18\columnwidth]{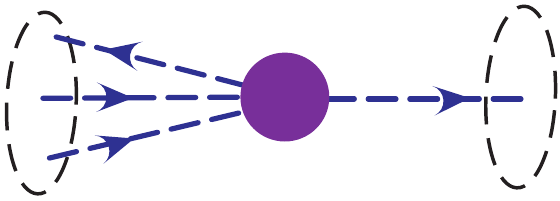}}} \nn
\end{align}
\vspace{-0.4cm}
\begin{alignat}{2} \label{eq:Z2_basis_bb_2}
&O_{bb2(0;0)}^{(2)\balpha\bt\bgamma\delta}
= \,  J_{(b)n\bar n0\, }^{\balpha\bt}\, J_{(b)\bar n0\, }^{\bgamma\delta} H
\,,\qquad &
&O_{bb2(0;\bar 0)}^{(2)\balpha\bt\bgamma\delta}
=\frac{1}{2} \, J_{(b) n\bar n0\, }^{\balpha\bt}\, J_{(b)\bar n\bar 0\, }^{\bgamma\delta} H
\,,\\
&O_{bb2(0^\dagger;0)}^{(2)\balpha\bt\bgamma\delta}
=\frac{1}{2} \,  (J^\dagger)_{(b)n\bar n0\, }^{\balpha\bt}\, J_{(b)\bar n0\, }^{\bgamma\delta} H
\,,\qquad &
&O_{bb2(0^\dagger;\bar 0)}^{(2)\balpha\bt\bgamma\delta}
=\, (J^\dagger)_{(b) n\bar n0\, }^{\balpha\bt}\, J_{(b)\bar n\bar 0\, }^{\bgamma\delta} H
\,.\nn
\end{alignat}
The symmetry factors $\frac{1}{2}$ in \eqs{Z2_basis_bb_1}{Z2_basis_bb_2} are due to the identical particles in the same collinear sector. We also have the same color bases as in \eq{TBPS_Obq3} for $O_{bb1}^{(2)}$ in \eq{Z2_basis_bb_1}, and the same color basis as in
\eq{TBPS_Obq4} for $O_{bb2}^{(2)}$ in \eq{Z2_basis_bb_2}.

\subsubsection{$\cP_\perp$ Insertions}\label{sec:nnlp_perp}

As discussed previously, we choose to work in a frame where the total $\perp$ momentum of each collinear sector vanishes. Therefore, operators involving $\cP_\perp$ insertions first appear at $\cO(\lambda^2)$, and the $\cP_\perp$ operator must act in a collinear sector composed of two or more fields. At $\cO(\lambda^2)$, the $\cP_\perp$ operator can be inserted into either an operator involving two quark fields and a gluon field, or an operator involving three gluon fields. However, due to the fact that the bottom quark pair have different chirality and the gluon splitting interaction preserves chirality, the three gluon case is not possible and is ruled out.

When the $\cP_\perp$ operator is inserted into an operator involving two quark fields and a gluon field, the helicity structure of the operator is highly constrained. In particular, since the $\cP_\perp$ operator and the gluon field both have helicity $\pm1$, the quark fields must be in a helicity zero configuration. Combined with the fact that they must have opposite chirality, this implies that all operators must involve only the currents $J_{n\bar n\,0}^{\balpha\bt}$ or $(J^\dagger)_{n\bar n\,0}^{\balpha\bt}$. Here we have again taken without loss of generality that the bottom quark is in the $n$-collinear sector. For the case that the gluon field and the $\cP_\perp$ operator are in the $n$-collinear sector, a basis of operators is then given by
\begin{align}
&   \boldsymbol{(bg\cP_\perp)_n (\bar b)_{\bn}:} {\vcenter{\includegraphics[width=0.18\columnwidth]{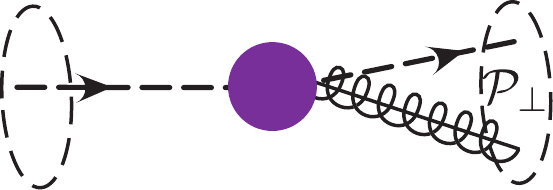}}}  \nn
\end{align}
\vspace{-0.4cm}
\begin{alignat}{2}\label{eq:Hbbgpperp_basis1}
&O_{\cP \chi1+ (0)[-]}^{(2)a\,\balpha\bt}
= \cB_{n+}^a\, \big\{ \cP_{\perp}^{-} J_{n\bar n\, 0}^{\balpha\bt} \big\}\,  H
\,,\qquad &
&O_{\cP\chi1- (0)[+]}^{(2)a\,\balpha\bt}
= \cB_{n-}^a\, \big\{ \cP_{\perp}^{+} J_{n\bar n\, 0}^{\balpha\bt} \big\} \, H 
\,,\\
&O_{\cP\chi1+ (0^\dagger)[-]}^{(2)a\,\balpha\bt}
=  \cB_{n+}^a\,  \big\{ \cP_{\perp}^{-} (J^\dagger)_{n\bar n\,0}^{\balpha\bt} \big\}\,  H
\,,\qquad &
&O_{\cP\chi1- (0^\dagger)[+]}^{(2)a\,\balpha\bt}
= \cB_{n-}^a \, \big\{ \cP_{\perp}^{+} (J^\dagger)_{n\bar n\,0}^{\balpha\bt} \big\}  \, H
\,.\nn
\end{alignat}
The color basis is one-dimensional
\begin{equation} \label{eq:nnlp_color_quark_perp}
\vT^{a\, \al\bbeta} = T^a_{\al\bbeta}\,.
\end{equation}
The structure after BPS redefinition is given by
\begin{align} \label{eq:nnlp_color_quark_perpBPS1}
\vT_{\BPS}^{ a \al\bbeta} 
=\left (T^aY^\dagger_{n}\,Y_{\bar n}\right)_{\alpha \bar \beta}
\,.
\end{align}

For the case that the gluon field and the $\cP_\perp$ operator are in the $\bar n$-collinear sector, the operator basis is given by:
\begin{align}
&   \boldsymbol{(b)_n (\bar bg\cP_\perp)_{\bn}:} {\vcenter{\includegraphics[width=0.18\columnwidth]{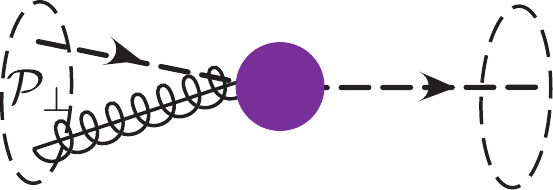}}}  \nn
\end{align}
\vspace{-0.4cm}
\begin{alignat}{2}\label{eq:Hbbgpperp_basis2}
&O_{\cP \chi2+ (0)[-]}^{(2)a\,\balpha\bt}
= \cB_{\bar n+}^a\, \big\{J_{n\bar n\, 0}^{\balpha\bt} (\cP_{\perp}^{-})^\dagger \big\}\,  H
\,,\qquad &
&O_{\cP\chi2- (0)[+]}^{(2)a\,\balpha\bt}
= \cB_{\bar n-}^a\, \big\{J_{n\bar n\, 0}^{\balpha\bt} (\cP_{\perp}^{+})^\dagger \big\} \, H 
\,,\\
&O_{\cP\chi2+ (0^\dagger)[-]}^{(2)a\,\balpha\bt}
=  \cB_{\bar n+}^a\,  \big\{(J^\dagger)_{n\bar n\,0}^{\balpha\bt} (\cP_{\perp}^{-})^\dagger\big\}\,  H
\,,\qquad &
&O_{\cP\chi2- (0^\dagger)[+]}^{(2)a\,\balpha\bt}
= \cB_{\bar n-}^a \, \big\{(J^\dagger)_{n\bar n\,0}^{\balpha\bt} (\cP_{\perp}^{+})^\dagger \big\}  \, H
\,.\nn
\end{alignat}
After BPS field redefinition the color structure is given by
\begin{align} \label{eq:nnlp_color_quark_perpBPS2}
\vT_{\BPS}^{ a \al\bbeta} 
=\left (Y^\dagger_{n}\,Y_{\bar n}T^a\right)_{\alpha \bar \beta}
\,.
\end{align}

Since we have assumed that the total $\cP_\perp$ in each collinear sector is zero, integration by parts can be used to make the $\cP_\perp$ operator act only on either the quark/antiquark, or the gluon field, which has been used in writing \eqs{Hbbgpperp_basis1}{Hbbgpperp_basis2} to avoid the need to consider cases where it acts on the gluon.

\subsubsection{Ultrasoft Insertions}\label{sec:nnlp_soft}

Operators with explicit ultrasoft insertions can also appear at $\cO(\lambda^2)$. Label momentum conservation implies that these operators must have a collinear field in each collinear sector. Moreover, we should have an ultrasoft insertion to two collinear quark fields for these operators. The operators with two collinear gluon fields and an ultrasoft insertion are impossible due to the chirality constraint of the Yukawa interaction.

The operator basis involving ultrasoft gluons is in general more complicated than a basis with only collinear operators because ultrasoft fields are not naturally associated with a given lightcone direction. There are therefore different choices that can be made when constructing the basis. Our choice will be to work in a basis where all ultrasoft derivatives acting on ultrasoft Wilson lines are absorbed into $\cB_{us}$ fields. Let us consider the following example involving two pre-BPS operators made of two collinear quark fields, and an ultrasoft derivative to understand why it is always possible to make this choice
\begin{align}
O^\mu_1=\bar \chi_{\bar n} (i D_{us}^\mu) \chi_n\,, \qquad O^\mu_2=\bar \chi_{\bar n} (-i \overleftarrow D_{us}^\mu) \chi_n\,,
\end{align}
where $(-i \overleftarrow D_{us}^\mu)=(i D_{us}^\mu)^\dagger$ and we have not made the $\mu$ index contraction explicit since it is not relevant to the current discussion. After performing the BPS field redefinition, we obtain 
\begin{align}
O^\mu_{1\text{BPS}}= i \bar \chi_{\bar n}Y_{\bar n}^\dagger D_{us}^\mu Y_n \chi_n\,, \qquad
 O^\mu_{2\text{BPS}}=-i \bar \chi_{\bar n} Y_{\bar n}^\dagger \overleftarrow D_{us}^\mu Y_n \chi_n
\end{align}
To absorb all derivatives acting on Wilson lines into $\cB_{us}$ fields, the Wilson lines in the operators must be organized as
\begin{align}
O^\mu_{1\text{BPS}}= i\bar \chi_{\bar n}Y_{\bar n}^\dagger Y_n (Y_n^\dagger  D_{us}^\mu Y_n) \chi_n\,, \qquad 
O^\mu_{2\text{BPS}}=-i \bar \chi_{\bar n} (Y_{\bar n}^\dagger \overleftarrow D_{us}^\mu Y_{\bar n}) Y_{\bar n}^\dagger  Y_n \chi_n
\end{align}
Using \Eq{eq:soft_gluon}, we see that $O^\mu_{1\text{BPS}}$ and $O^\mu_{2\text{BPS}}$ can be written entirely in terms of $\partial_{us}$ operators acting on collinear fields, and the two ultrasoft gauge invariant gluon fields $\cB_{us(n)}$ and $\cB_{us(\bar n)}$, respectively. However, it is good to notice that ultrasoft gluon fields defined with respect to both lightcone directions are required because of this choice. 
In principle it is possible to make another choice. For example, one can decide to work only with $\cB_{us(n)}$, but in this case we see that the ultrasoft derivative must also be allowed to act explicitly on pairs of ultrasoft Wilson lines, for example $[\partial_{us}^\mu (Y_n^\dagger Y_\bn)]$. We will choose to avoid this situation in constructing our complete basis, so that ultrasoft derivatives acting on soft Wilson lines will occur only within the explicit $\cB_{us}$ fields. Our choice also makes our basis more symmetric.

For the operators involving one ultrasoft gluon and two collinear quarks, we have the basis
\begin{align}
&  \boldsymbol{g_{us}(b)_n (\bar b)_{\bn}:}{\vcenter{\includegraphics[width=0.18\columnwidth]{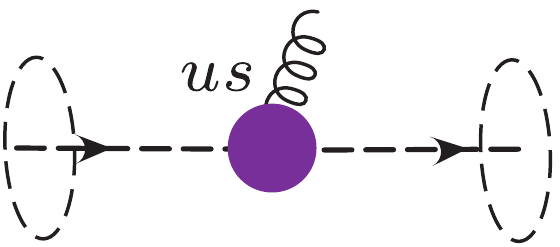}}}\nn
\end{align}
\vspace{-0.4cm}
\begin{alignat}{2} \label{eq:soft_insert_basis1}
&O_{\cB(us(n))0:(0)}^{(2)a\,\balpha\bt}
=  \cB_{us(n)0}^a\, J_{n\bar n\,0}^{\balpha\bt}\, H
\,,\qquad &
&O_{\cB(us(n))0:(0^\dagger)}^{(2)a\,\balpha\bt}
=\cB_{us(n)0}^a \, (J^\dagger)_{n\bar n\,0}^{\balpha\bt}\, H
\,, 
\end{alignat}
with the unique color structure
\begin{equation} 
\vT_{\BPS}^{\,a\, \al\bbeta} = \left ( T^a Y^\dagger_{n} Y_{\bn} \right )_{\alpha \bar\beta}
\,.\end{equation}
and
\begin{alignat}{2} \label{eq:soft_insert_basis2}
&O_{\cB(us(\bar n))0:(0)}^{(2)a\,\balpha\bt}
=  \cB_{us(\bar n)0}^a\, J_{n\bar n\,0}^{\balpha\bt}\, H
\,,\qquad &
&O_{\cB(us)(\bar n))0:(0^\dagger)}^{(2)a\,\balpha\bt}
=\cB_{us(\bar n)0}^a \, (J^\dagger)_{n\bar n\,0}^{\balpha\bt}\, H
\,, 
\end{alignat}
with the unique color structure
\begin{equation} 
\vT_{\BPS}^{\,a\, \al\bbeta}=\bigl(Y_n^\dagger Y_\bn T^a\bigr)_{\alpha\bbeta}
\,.\end{equation}
Note that the color structures associated with the two different projections of the $\cB_{us}$ field are distinct. All other helicity combinations vanish due to helicity selection rules.

Using RPI symmetry, the Wilson coefficients of the operators that include $\cB_{us(n)0}$ can be related to the Wilson coefficients of the leading power operators (see \cite{Larkoski:2014bxa}). In particular, we have 
\begin{align} \label{eq:usRPIrelation}
C^{(2)}_{\cB n(us)0:\lambda_1, \lambda_1}&=-\frac{\partial C^{(0)}_{\lambda_1, \lambda_1} }{\partial \omega_1}  
\,, 
\end{align}
where $C^{(0)}_{\lambda_1, \lambda_1}$ is the Wilson coefficient of the $\cO(\lambda^0)$ operator of \eq{Hbb}. We will verify this at the level of tree level matching in \Sec{sec:matching}, where this subleading power coefficient vanishes.

There are also operators involving a single ultrasoft derivative and two collinear quark fields, 
\begin{align}
&  \boldsymbol{\partial_{us} (b)_n (\bar b)_{\bn}:}{\vcenter{\includegraphics[width=0.18\columnwidth]{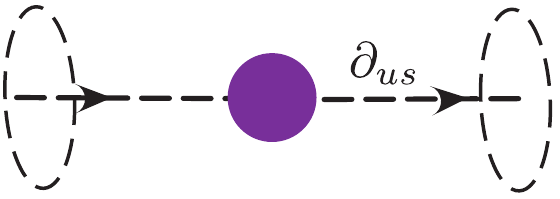}}}  \nn
\end{align}
\vspace{-0.4cm}
\begin{alignat}{2} \label{eq:soft_derivative_basis1}
&O_{\partial(us(n))0:(0)}^{(2)\,\balpha\bt}
=  \{\partial_{us(n)0}\, J_{n\bar n\,0}^{\balpha\bt}\}\, H
\,,\qquad &
&O_{\partial(us(n))0:(0^\dagger)}^{(2)\,\balpha\bt}
=  \{\partial_{us(n)0} \, (J^\dagger)_{n\bar n\,0}^{\balpha\bt}\}\, H\,,
\end{alignat}
with the color structures given before and after BPS field redefinition by
\begin{align} \label{eq:leading_color_deriv}
 \vT^{\al\bbeta} = (\de_{\al\bbeta})\,, \qquad 
 \vT_{\BPS}^{\al\bbeta} = \bigl(Y_n^\dagger Y_{\bar n}\bigr)_{\al\bbeta} 
\,,
\end{align}
and
\begin{alignat}{2} \label{eq:soft_derivative_basis2}
&O_{\partial^\dagger(us(\bar n))\bar 0:(0)}^{(2)\,\balpha\bt}
=  \{J_{n\bar n\,0}^{\balpha\bt}\,(\partial_{us(\bar n)0})^\dagger\}\, H
\,,~ &
&O_{\partial^\dagger(us(\bar n))\bar 0:(0^\dagger)}^{(2)\,\balpha\bt}
=  \{(J^\dagger)_{n\bar n\,0}^{\balpha\bt}\, (\partial_{us(\bar n)0})^\dagger\}\, H\,,
\end{alignat}
with the color structures given before and after BPS field redefinition by
\begin{align} \label{eq:leading_color_deriv2}
 \vT^{\al\bbeta} = (\de_{\al\bbeta})\,, \qquad 
 \vT_{\BPS}^{\al\bbeta} = \bigl(Y_n^\dagger Y_{\bar n}\bigr)_{\al\bbeta} \,.
\end{align}
Although the color structure is the same in both cases, we have separated them to highlight the different decompositions of the ultrasoft derivatives in the two cases. Note that the form of the ultrasoft derivatives which appear is constrained by the helicity constraints. The other ultrasoft derivatives allowed by the helicity constraints such as $\partial_{us(n)\bar 0}$ and $\partial_{us(\bar n)\bar 0}$ are not included since they can be removed by the equations of motion.

Using RPI symmetry, the Wilson coefficients of the operators involving $\partial_{us(n)0}$ can also be related to the Wilson coefficients of the leading power operators (see \cite{Larkoski:2014bxa}). In particular, we have 
\begin{align} \label{eq:usRPIrelationb}
C^{(2)}_{\partial(us)n:\lambda_1,\lambda_1}&=-\frac{\partial C^{(0)}_{\lambda_1,\lambda_1} }{\partial \omega_1}  
\,, 
\end{align}
where $C^{(0)}_{\lambda_1,\lambda_1}$ is the Wilson coefficient of the $\cO(\lambda^0)$ operator of \Eq{eq:Hbb}. We will discuss this further at the level of  tree level matching in \Sec{sec:matching}.

\section{Cross Section Contributions and Factorization}\label{sec:discussion}

The basis of operators presented in \Sec{sec:basis} generates the full set of operators that contribute at the amplitude level up to $\cO(\lambda^2)$ suppression with respect to the leading power. However, some of these operators will not contribute to a physical cross section at $\cO(\lambda^2)$. In this section, we briefly discuss the operators that can contribute to the cross section for an SCET$_\text{I}$ event shape observable, thrust $\tau$, measured on $H \to b\bar b$. Using helicity selection rules we determine which operators can contribute at $\cO(\lambda^2)=\cO(\tau)$. The results are summarized by the check marks in \Tab{tab:summary}.

{
\renewcommand{\arraystretch}{1.6}
	\begin{center}
\begin{table}[t]
	\begin{center}
		\begin{tabular}{| c | l | c | c | c | c | }
			\hline
			& Operators & Factorization  & Jet $n$ &  Jet $\bar n$ & Soft  \\
			\hline 
			$\!\!\mathbf{\mathcal{O}(\lambda^0)}\!\!$ 
			&$O^{(0)}  O^{(0)} $ 
			& $H_b^{(0)} J_{q}^{(0)} J_{q}^{(0)} S_q^{(0)}$
			& $\bar \chi_n \,\hat\delta\, \chi_n$ 
			& $\bar \chi_\bn \,\hat\delta\, \chi_\bn$
			& $Y_n^\dagger Y_\bn \widehat\cM^{(0)}\, Y_\bn^\dagger Y_n$ 
			\\
			\hline
			$\!\!\mathbf{\mathcal{O}(\lambda^2)}\!\!$ 
			&$O_{\cB}^{(1)}  O_{\cB}^{(1)}$
			& $H_{b1}^{(0)} J_{qq}^{(2)} J_{g}^{(0)} S_{g}^{(0)}$  
			& $\bar \chi_n \chi_n \,\hat\delta\,\bar \chi_n \chi_n $ 
			& $\cB_\bn \hat\delta\, \cB_\bn$   
			& $\cY_n^T \cY_\bn \widehat\cM^{(0)}\,   \cY_\bn^T \cY_n$   
			\\
			\cline{2-6}
			&$O^{(0)}  O_{\cB2}^{(2)} $
			& $H_{b2}^{(0)} J_{qgg}^{(2)} J_{q}^{(0)} S_{q}^{(0)}$  
			& $\!\!\bar \chi_n \cB_n  \cB_n \hat\delta\, \chi_n\!\!  $ 
			& $ \bar \chi_\bn \,\hat\delta\, \chi_\bn $   
			& $Y_n^\dagger Y_\bn \widehat\cM^{(0)}\, Y_\bn^\dagger Y_n$
			\\
			\cline{2-6}
			&$O^{(0)}  O_{bq3}^{(2)} $
			& $H_{b3}^{(0)} J_{q'\!q'\!q}^{(2)} J_{q}^{(0)} S_{q}^{(0)}$  
			& $\!\!\bar \chi_n \bar \chi_{(q')n} \chi_{(q')n} \hat\delta\, \chi_n\!\!  $ 
			& $ \bar \chi_\bn \,\hat\delta\, \chi_\bn $   
			& $Y_n^\dagger Y_\bn \widehat\cM^{(0)}\, Y_\bn^\dagger Y_n$  			
			\\
			\cline{2-6}
			&$O^{(0)}  O_{bb1}^{(2)} $
			& $H_{b4}^{(0)} J_{qqq}^{(2)} J_{q}^{(0)} S_{q}^{(0)}$  
			& $\!\!\bar \chi_n \bar \chi_{n} \chi_{n} \hat\delta\, \chi_n\!\!  $ 
			& $ \bar \chi_\bn \,\hat\delta\, \chi_\bn $   
			& $Y_n^\dagger Y_\bn \widehat\cM^{(0)}\, Y_\bn^\dagger Y_n$  			
			\\
			\cline{2-6}
			&$O^{(0)}  O_{\cP\chi1}^{(2)} $
			& $H_{b5}^{(0)} J_{qg P}^{(0)} J_{q}^{(2)} S_{q}^{(0)}$  
			& $\bar \chi_n [\cP_\perp \cB_n]\,\hat\delta\,\chi_n $ 
			& $\!\!\bar \chi_\bn \hat\delta\, \chi_\bn\!\!   $   
			& $Y_n^\dagger Y_\bn \widehat\cM^{(0)}\, Y_\bn^\dagger Y_n$   
			\\
			\cline{2-6}
			&$\!\! O^{(0)}  O_{\cB us(n)0}^{(2)} \!\!\!$
			& $H_{b6}^{(0)} J_q^{(0)} J_{q}^{(0)} S_{q\cB}^{(2)}$  
			&  $\bar \chi_n \,\hat\delta\,\chi_n   $	 
			&$\bar \chi_\bn \,\hat\delta\,\chi_\bn   $  
			& $\!\cB_{us(n) 0}\, Y_n^\dagger Y_\bn \widehat\cM^{(0)}\, Y_\bn^\dagger Y_n\!\!$     
			\\
			\cline{2-6}
			& $\!\!O^{(0)}  O_{\partial(us(n))0}^{(2)} \!\!\!$
			& $H_{b7}^{(0)} J_q^{(0)} J_{q}^{(0)} S_{q\partial}^{(2)}$  
			&  $\bar \chi_n \,\hat\delta\,\chi_n   $	 
			&$\bar \chi_\bn \,\hat\delta\,\chi_\bn   $  
			& $\!\partial_{us(n) 0}\, Y_n^\dagger Y_\bn \widehat\cM^{(0)}\, Y_\bn^\dagger Y_n\!\!$
			\\
			\hline
		\end{tabular}
		\caption{Beam and soft functions up to $\cO(\lambda^2)$ arising from products of hard scattering operators in the factorization of Higgs with a jet veto, and their field content. We have suppressed the helicity and color structures. We have not included products of operators whose beam and soft functions are identical to those shown by charge conjugation or $n\leftrightarrow \bn$.}		\label{tab:fact_func}
	\end{center}
\end{table}
	\end{center}
}

\subsection{Factorization beyond leading power and vanishing at $\cO(\lambda)$}

At subleading power, one has multiple contributions to the factorization theorem. The full subleading factorization theorem is composed by the sum of different factorized expressions. For example, a source of different contributions is the fact that for each subleading hard scattering operator contributing to the cross section at a given order in the power counting, there is a factorization theorem that allows us to express the contribution of this operator in terms of a hard function, i.e. the Wilson coefficient of the operator, a jet function for each collinear direction and a soft function at all orders in $\alpha_s$.
The subleading factorization theorem receives contributions also from subleading Lagrangian insertions, where the hard scattering operator is either leading or subleading power, and there is also a suppression is coming from T-products with subleading Lagrangians.
Finally, there can also be contributions coming from the expansion of the measurement to subleading power.
Here will not discuss the factorization of the cross section in details and the interested reader can find more details in \Refs{Feige:2017zci,Moult:2017rpl}.

Since our focus here is on subleading hard scattering operators, we restrict ourselves to determining the structure of the factorization theorem terms arising purely from these operators, written in terms of hard, jet and soft functions. A summary of these results is given in \Tab{tab:fact_func}.  In many cases the jet and soft functions which appear in the subleading power factorization formula are identical to those at leading power. \footnote{For gluon-gluon and quark-quark color channels the leading power soft functions are
\begin{align}\label{eq:soft_func_def}
S_g^{(0)}=\frac{1}{(N_c^2-1)}  \tr \big\langle 0 \big| \cY^T_{\bar n} \cY_n \widehat{\cM}^{(0)}\cY_n^T \cY_{\bar n} \big|0 \big\rangle\,, \qquad
S_q^{(0)}=\frac{1}{N_c}  \tr \big\langle 0 \big| Y^\dagger_{\bar n} Y_n \widehat{\cM}^{(0)}Y_n^\dagger Y_{\bar n} \big|0 \big\rangle\,, 
\end{align}
and depend on the kinematic variables probed by the measurement operator $\widehat{\cM}^{(0)}$. The leading power jet functions for quarks and gluons are
\begin{align} \label{eq:jet_func_def}
\delta^{\alpha\bar \beta} \Big( \frac{\Sl{n}}{2} \Big)^{\!ss'\!} J_q^{(0)}
  &= \int \!\!\frac{dx^-}{|\omega|}\: e^{\frac{i}{2} \ell^+ x^-} 
  \Big\langle 0\Big|\, \chi_{n}^{s\alpha} \big(x^- \text{\small $\frac{n}{2}$}\big) \,\hat{\delta}\, \bar \chi_{n,\omega}^{s'\bar\beta}(0) 
  \,\Big| 0 \Big\rangle
   \,, \\
\delta^{ab} g^{\mu\nu}_\perp J_g^{(0)}
  &=-\omega\!  \int \!\!\frac{dx^-}{|\omega|}\:  e^{\frac{i}{2} \ell^+ x^-}  \Big\langle 0 \Big|\, \cB^{\mu a}_{\perp} \big(x^- \text{\small $\frac{n}{2}$}\big)\, \hat{\delta}\, \cB^{\nu b}_{\perp,\omega}(0) \,\Big|0\Big\rangle
\,, \nn
\end{align}
where we take $\ell^+ \gg \Lambda_{\rm QCD}^2/\omega$.}
For the case of the soft functions this simplification arises due to color coherence, allowing a simplification to the Wilson lines in the soft functions that appear. 
For the beam functions, this simplification occurs since the power correction is often restricted to a single collinear sector. The other collinear sector is then described by the leading power beam functions (incoming jet functions) for gluons and quarks.
Here $\hat{\delta}$ appearing in these beam or soft functions is the leading power measurement function for the collinear and soft sectors. In general it depends on the factorization theorem being treated. 
Here we work in SCET$_{\text{I}}$ under the assumption that the measurement function does not fix the perpendicular momentum $\cP_\perp$ of the measured particle.\footnote{Measurements that fix the $\cP_\perp$, like broadening or $p_T$ spectrum, can be treated using a slightly different effective theory, SCET$_{\text{II}}$~\cite{Bauer:2002aj}.} This assumption has been explicitly used in writing the form of the jet functions in \Eq{eq:jet_func_def}, as well as in our construction of the operator basis. Since we treat the dynamics of the $b$ quarks in the scalar current as massless, we obtain the standard massless jet functions $J_q^{(0)}$ (or for incoming quarks, beam functions $B_q^{(0)}$) at leading power.

\paragraph{Vanishing at $\cO(\lambda)$.}
As for the case of gluon fusion~\cite{Moult:2017rpl} and Drell-Yan like processes~\cite{Feige:2017zci}, for color singlet scalar production and decay through quark antiquark annihilation, all the contributions to the factorization theorem at $\cO(\lambda) = \cO(\sqrt{\tau})$ vanish. The proof follows exactly the same steps as \Refs{Feige:2017zci,Moult:2017rpl}, namely:
\begin{itemize} 
	\item The $\cO(\lambda)$ hard scattering operators, \eq{H1_basis}, can't interfere with the leading power due to fermion number conservation in each collinear direction.
	\item There is no $\cO(\lambda)$ expansion of the measurement.
	\item There is no $\cO(\lambda)$ subleading Lagrangian that contributes to the process.
\end{itemize}
and we refer to those references for additional details.

\subsection{$\cO(\lambda^2)$ contributions}\label{sec:contribs}
Unlike the $\cO(\lambda)$ power corrections, the power corrections at $\cO(\lambda^2)=\cO(\tau)$ will not vanish. The cross section contributions at $\cO(\lambda^2)$ whose power suppression arises solely from hard scattering operators can be either a product of two $\cO(\lambda)$ operators or a product of an $\cO(\lambda^2)$ operator and an $\cO(\lambda^0)$ operator 
{\begin{small}
\begin{align}\label{eq:xsec_lam2}
&\frac{d\sigma}{d\tau}^{(2)} \supset N \sum_{X,i}  \tilde \delta^{(4)}_q  \bra{0} C_i^{(2)} \tO_i^{(2)}(0) \ket{X}\bra{X} C^{(0)} \tO^{(0)}(0) \ket{0} \delta\big( \tau - \tau^{(0)}(X) \big) +\text{h.c.}\nn \\
&+ N \sum_{X,i,j}  \tilde \delta^{(4)}_q  \bra{0} C_i^{(1)} \tO_i^{(1)}(0) \ket{X}\bra{X} C_j^{(1)} \tO_j^{(1)}(0) \ket{0}   \delta\big( \tau - \tau^{(0)}(X) \big)+\text{h.c.} \,.
\end{align}
\end{small}}

For $H\to b\bar b$ the operator basis has only a single operator at $\cO(\lambda)$ (up to helicities and $n\leftrightarrow \bar n$), which was given in \eq{H1_basis}. As said before this operator can not interfere with the leading power one to give rise to an $\cO(\lambda)$ contribution, but it contributes to the cross section at $\cO(\lambda^2)$ when squared.

Since the hard scattering operators at $\cO(\lambda^2)$ must interfere with the leading power operators to give a $\cO(\lambda^2)$ term in the cross section, their contributions are therefore highly constrained. Namely, there must be some intermediate state $\ket{X}$ such that the amplitudes in \eq{xsec_lam2} are nonzero. We will discuss each possible contribution in turn, and the summary of all operators which can contribute to the $\cO(\lambda^2)$ cross section is given in \Tab{tab:summary}. The schematic structure of the jet and soft functions arising from each of the different operator contributions is shown in \Tab{tab:fact_func}. The subleading jet and soft functions enumerated in this table are universal objects that will appear in processes initiated by different Born level amplitudes (such as gluon fusion), unless forbidden by symmetry. In this initial investigation, we content ourselves with only giving the field content of the jet and soft functions. To avoid cumbersome notation, in \Tab{tab:fact_func} we do not write the external vacuum states for the soft functions, or the external proton states for the jet functions, nor do we specify the dependence on the residual space-time variable of the fields. Unlike for the leading power definitions given in \Eqs{eq:soft_func_def}{eq:jet_func_def}, we do not present here the full definitions of the subleading soft and jet functions definitions since it goes hand in hand with presenting the complete factorization theorems for these contributions, which is not the focus of this paper and will be given in future work.

\vspace{0.6cm}
\noindent{\bf{Two Quark-One Gluon Operators:}}

The two quark-one gluon operators, $O_{\cB}^{(1)}$ can contribute to the cross section by interfering with themselves.  They contribute with a leading power gluon channel soft function $S_g^{(0)}$ (Wilson lines in the adjoint), a gluon jet function $J_g^{(0)}$, and a subsubleading power jet function $J_{qq}^{(2)}$ with four quark fields. Here we denote this jet function with subscript $qq$ to indicate that the quarks are massless.

\vspace{0.4cm}
\noindent{\bf{Two Quark-Two Gluon Operators:}}

In the case of the two quark-two gluon operators, all of the operators have one quark field in the $n$-collinear sector and an antiquark field in the $\bar n$-collinear sector. Therefore the fermion number is conserved. However, for operators $O_{\cB1}^{(2)}$ such as $HJ^{\balpha\bt}_{n\bar n0\,}\cB^a_{n-}\cB^b_{\bar n-}$, it needs to interfere with the leading power operator $H(J^\dagger)^{\balpha\bt}_{n\bar n0}$ in order for the amplitude in each collinear sector to transform as a scalar. Then in each collinear sector, the quark field from the leading power operator $H(J^\dagger)^{\balpha\bt}_{n\bar n0}$ and the quark field from the subsubleading power operator $HJ^{\balpha\bt}_{n\bar n0\,}\cB^a_{n-}\cB^b_{\bar n-}$ have opposite chirality. Since the gluon splitting interaction does not change the chirality of the quark, the operator $O_{\cB1}^{(2)}$ can't interfere with the leading power operator. For operators $O_{\cB2}^{(2)}$ and $O_{\cB3}^{(2)}$, the quark field in each collinear sector has the same chirality as the quark field from the leading power operator. Therefore, $O_{\cB2}^{(2)}$ and $O_{\cB3}^{(2)}$ contribute to the cross section at $\cO(\lambda^2)$. $O_{\cB2}^{(2)}$ has two gluon fields in the $n$-collinear sector. This gives a subsubleading power jet function $J_{qgg}^{(2)}$, a quark jet function $J_q^{(0)}$, and a quark soft function $S_q^{(0)}$ (with Wilson lines in the fundamental). The factorization of $O_{\cB3}^{(2)}$ is equivalent to that of $O_{\cB2}^{(2)}$ up to charge conjugation and $n\leftrightarrow \bar n$.

\vspace{0.4cm}
\noindent{\bf{Four Quark Operators:}}

For a four quark operator to interfere with the leading power operator, it must have a bottom quark in one collinear sector and an bottom antiquark in the other collinear sector. This eliminates the operators $O_{bq1}^{(2)}$ and $O_{bq2}^{(2)}$ from contributing to the cross section at $\cO(\lambda^2)$, since for these two operators the bottom quark and the bottom antiquark are in the same collinear sector. The operators $O_{bq3}^{(2)}$ and $O_{bq4}^{(2)}$ contribute to the cross section, and the contributions have a subsubleading power jet function with two quark flavors $J_{q'\!q'\!q}^{(2)}$ (having three massless fermion fields on one side of the measurement, and one on the other), a jet function $J_{q}^{(0)}$, and a soft function $S_q^{(0)}$. For the operators $O_{bb1}^{(2)}$ and $O_{bb2}^{(2)}$ the factorization is similar, but the subsubleading power jet function is $J_{qqq}^{(2)}$ (having three massless fermion fields on one side of the measurement, and one on the other, but all the same flavor).

\vspace{0.4cm}
\noindent{\bf{$\cP_\perp$ Operators:}}

Both the operators involving $\cP_\perp$ insertions have the correct fermion numbers and symmetry properties. Therefore, both $O_{\cP \chi 1}^{(2)}$ and $O_{\cP \chi 2}^{(2)}$ can contribute to the $\cO(\lambda^2)$ cross section. Both contributions have a subsubleading power jet function $J_{qg\cP}^{(2)}$, and a leading power quark jet function $J_q^{(0)}$ and soft function $S_q^{(0)}$. 

\vspace{0.4cm}
\noindent{\bf{Ultrasoft Operators:}}

All ultrasoft operators can contribute to the cross section through interference with the leading power operator due to fermion number conservation and angular momentum conservation. They all have two leading power quark jet functions $J_{q}^{(0)}$. The operators $O^{(2)}_{\cB us(n)}$ and $O^{(2)}_{\cB us(\bn)}$ have a subleading soft function $S_{q\cB}^{(2)}$, and the operators $O^{(2)}_{\partial (us(n))0}$ and $O^{(2)}_{\partial (us(\bar n))\bar 0}$ have a subleading soft function $S_{q\partial}^{(2)}$.

\subsection{Comparison with $e^+e^- \to $ dijet and $ gg \to H$}\label{sec:compare}

The process of Higgs production in quark antiquark annihilation shares some features with both the Drell-yan process, where we have the same $q \bar{q}$ initial state, but a spin-1 particle in the final state, and Higgs production in gluon fusion, where we have the same final state, a color singlet scalar, but different initial states. 
In this section we discuss some interesting differences of the structure of the operator basis, as well as the contributions to the $\cO(\lambda^2)$ cross section, between the basis built in \Sec{sec:basis} and cross section discussed in \Sec{sec:discussion} relative to a process with two collinear sectors initiated by the $\bar q \Gamma_{\!v} q$ vector quark current \cite{Feige:2017zci} and the case of $gg \to H$ as discussed in \cite{Moult:2017rpl}. 
The case of quark antiquark annihilation into a color singlet scalar analyzed in this paper shares more similarities  with the $\bar q \Gamma_{\!v} q$ case rather than with $gg \to H$.

For example the leading power factorization theorems in our case is identical to the $\bar q \Gamma_{\!v} q$ case, while a simple replacement of quark and gluon jet (beam) functions, as well as the color charges of the Wilson lines in the soft functions would give us the leading power factorization theorem for Higgs production in gluon fusion. 
At subleading power this pattern continues, even though there are interesting differences between $q\bar q \to H$ and $\bar q q \to \gamma$ arising from the helicity structure of the currents.

One interesting feature of $gg\to H$ is that the Wilson coefficient for the leading power operator has an explicit dependence on the large label momenta of the collinear gluon fields at tree level. 
This is not the case for both our case (see \Eq{eq:LP_wilson}) and for the $\bar q\, \Gamma q$ current, whose leading power operators have a Wilson coefficient that is independent of the large label momenta at tree level. 
Since the Wilson coefficients of the hard scattering operators involving insertions of $n \cdot \partial$, $\bar n \cdot \partial$, or $\cB_{us(n)0}$ are related to the derivatives of the leading power Wilson coefficients by RPI, as discussed in \Sec{sec:nnlp_soft}, these particular operators vanish at tree level for both $q\bar q \to H$ and $\bar q\, \Gamma q$ current, but are present at tree level for $gg\to H$. 
For the quark scalar and vector currents the power corrections from the ultrasoft sector at LL arise instead only from subleading power Lagrangian insertions. 
Therefore, the nature of power corrections in terms of the organization of the effective theory in the ultrasoft sector is quite different if we consider quark antiquark annihilation or gluon fusion. 

In comparing the operator bases of these three cases we can observe that the  basis constructed here contains the least number of operators. 
The reason is both because the spin-0 constraint allows less helicity configurations than the spin-1 case and that the underlying hard scattering process must come from a pair of quarks in opposite chirality which means that they can't come from gluons. 
This implies strong constraints on operators with gluon building block fields. 
For example, in the scalar current case to ${\cal O}(\lambda^2)$ there are no cross section contributions with gluon initial states (back-to-back incoming gluons), which was not the case for the vector quark current.  
On the other hand since already at the amplitude level we need something very similar to the leading power current, most of the $\cO(\lambda^2)$ operators can interfere with the leading power, therefore the constraints coming from angular momentum conservation don't reduce the cross section contributions with respect to the amplitude level contributions as much as in the case of Higgs production in gluon fusion or the vector quark current case.

From \Tab{tab:fact_func} we can see that ultrasoft operators contribute as an interference of the form $\cO(\lambda^2) \cO(1)$. 
This is guaranteed by the Low-Burnett-Kroll theorem \cite{Low:1958sn,Burnett:1967km} and happens also for the gluon and dijet cases.

As for the vector quark current case, the only operator that enters the cross section at $\cO(\lambda^2)$ without interfering with the LP operator, but  as a squared matrix element is $O_{\cB_\bn}$ in \eq{H1_basis}, a hard scattering operator involving two collinear quarks recoiling against a collinear gluon. 
In the NNLO calculation of power corrections for the $q\Gamma_{\!v} \bar q$ and $gg \to H$ cases \cite{Moult:2016fqy,Moult:2017jsg}, the analogous operators played gave rise to a leading logarithmic divergence not predicted by a naive exponentiation of the one-loop result, and it is expected that the same will be true here.
In our basis there is no operator at $\cO(\lambda)$ with non zero fermion number in each collinear sector, which is present for the $gg\to H$ case. 
It would be interesting to explore in more detail the relation between the leading logarithmic divergences for the quark antiquark initiated processes compared with the gluon fusion ones.

\section{Matching}\label{sec:matching}

In this section we carry out the matching procedure to determine the Wilson coefficient for the operators relevant for the calculation of the $\cO(\lambda^2)$ cross section, which were enumerated in \Sec{sec:contribs} and summarized in \Tab{tab:summary}. As mentioned in \Sec{sec:intro}, we consider the Yukawa interaction after electroweak symmetry breaking in the mass basis
\begin{align}
\cL_m=-m_d^i\bar d^i_Ld^i_R\frac{h}{v}-m_u^i\bar u^i_Lu^i_R\frac{h}{v}\,,
\end{align}
where $i=1,2,3$ is the flavor index, $h$ is the Higgs field, and $v=(\sqrt{2} G_F)^{-1/2}=246$ GeV is the Higgs vacuum expectation value. Therefore, for the process $b\bar b\rightarrow H$ or $H\rightarrow b\bar b$, we should have the factor $-i\frac{m_b}{v}$ for each $Hb\bar b$ full theory vertex. However, since this factor is the same for every full theory diagram, we will simply suppress it and take it to be $1$ throughout the matching calculation, and the dependence on $-i\frac{m_b}{v}$ can be easily reinstated. To be more precise, in what follows, we omit the $-i\frac{m_b}{v}$ factor in every Feynman rule and full theory diagram, and omit the $-\frac{m_b}{v}$ factor in every operator.

In the matching, we take all particles as outgoing. The SCET helicity operators are fully crossing symmetric, see~\cite{Moult:2015aoa}, so amplitudes for incoming particles are easy to obtain. We also restrict to Feynman gauge although we check gauge invariance through relevant Ward identities. For operators involving collinear gluon or quark fields, gauge invariance is guaranteed through the use of gauge-invariant collinear building block, $\cB_{\perp}$ and $\chi$, defined in \eq{chiB}. 

Throughout the matching, we will keep the same diagram conventions as in~\cite{Feige:2017zci} and ~\cite{Moult:2017rpl}. Therefore we will indicate in Feynman diagrams collinear gluons in the effective theory as a spring with a line drawn through them, collinear quarks will be represented by dashed lines, and ultrasoft gluons will be indicated with an explicit ``us". In this way it is clear the distinction between EFT and full theory diagrams for which standard Feynman diagram notation for quarks and gluons is used. Furthermore a purple circle is used to denote a hard scattering operator in the effective theory, while in the full theory diagrams we will use the $\otimes$ symbol to denote the vertex from the Yukawa interaction.

Due to the large number of operators present in our basis, we will express the results of the tree-level matching in the form of the Wilson coefficient multiplying the relevant operator. In order to do so, we define a shorthand notation with a caligraphic {\cal O},
\begin{align}
\cO_X^{(i)} = C_X^{\text{tree}} O_X^{(i)}\,,
\end{align}
where as before, the superscript indicates what is the power suppression of the operator with respect to the leading power, and the subscript is a label that identifies uniquely the operator by denoting its field and helicity content.
We will write results for $\cO_X^{(i)}$ in a form such that it is trivial to separate the tree level Wilson coefficient $C_X^{\text{tree}}$ and the hard scattering operator $O_X^{(i)}$, so that higher order corrections can be added easily as they become available.

\subsection{Leading Power Matching}\label{sec:matching_lp}

The leading power Wilson coefficient for a quark current $\bar q_n \Gamma q_{\bar n}$ is well known in the SCET literature, and is independent of the spin structure $\Gamma$. As explained in \Sec{sec:basis_lp} the unique leading power scalar operator is
\begin{align}
	\cO^{(0)}=\bar \chi_{n} \chi_{\bar n}\,.
\end{align}
It's Wilson coefficient is given to $\cO(\alpha_s)$ by 
\begin{align}\label{eq:LP_wilson}
	C^{(0)}=1+\frac{\alpha_s(\mu) C_F}{4\pi} \left( -\log^2 \left[  \frac{-\omega_1 \omega_2-i0}{\mu^2} \right] +3\log \left[  \frac{-\omega_1 \omega_2-i0}{\mu^2} \right] -8 +\frac{\pi^2}{6} \right) \,.
\end{align}
Throughout this section, we will restrict ourselves to the tree level matching, however, we have given the Wilson coefficient of \Eq{eq:LP_wilson} to one loop, since it can be used with the RPI relations of \Sec{sec:nnlp_soft} for the operators involving ultrasoft insertions, which are first non-trivial at this order. These results can be directly obtained from the LP operator of \cite{Feige:2017zci} with $\Gamma = \id$.

\subsection{Subleading Power Matching}\label{sec:matching_nlp}

We now consider the matching at the subleading power. In \Sec{sec:nlp} we argued that the only $\cO(\lambda)$ operator which can contribute to the cross section at $\cO(\lambda^2)$ is $(b\bar b)_n(g)_\bn$, where the two collinear quark fields are in the same collinear sector and the collinear gluon field is in the other collinear sector. We can therefore perform the matching using this external state. The QCD diagram for the production of one gluon field and two quark fields are
\begin{align}\label{eq:QCD_bbg}
\fd{2.5cm}{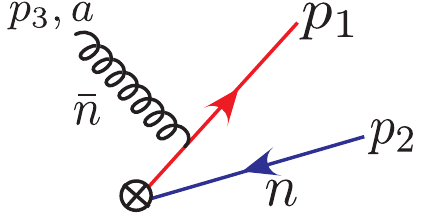} &= \bar u(p_1)(igT^a\Sl{\epsilon}_{3}^{*})\frac{i(\Sl{p}_1+\Sl{p}_3)}{(p_1+p_3)^2} v_(p_2)\, \nn \\
\fd{2.5cm}{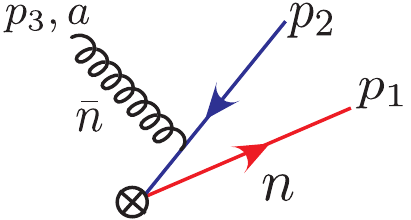} &= \bar u(p_1)\frac{-i(\Sl{p}_2+\Sl{p}_3)}{(p_2+p_3)^2}(igT^a\Sl{\epsilon}_{3}^{*})v_(p_2)\,.
\end{align}
Since all the propagators are far offs-shell, we can use the following kinematics:
\begin{align}
p_1^\mu =\omega_1 \frac{n^\mu}{2}, \qquad  p_2 =\omega_2 \frac{n^\mu}{2}, \qquad p_3 =\omega_3 \frac{\bar n^\mu}{2}\,,
\end{align}
where $\omega_1,~\omega_2,~\omega_3\sim \cO(\lambda^0)$. The polarization vector $\epsilon^*_3$ is also taken to be $\bn$-collinear, which has the scaling $n\cdot\epsilon^*_3\sim \cO(\lambda^{-1}),~\bn\cdot\epsilon^*_3\sim \cO(\lambda),~\epsilon^*_{3\perp}\sim \cO(\lambda^0)$. We denote the projected SCET spinors as
\begin{align}
	u_n(i)=P_nu(p_i),\qquad v_n(i)=P_nv(p_i), \qquad P_n=\frac{\Sl{n}\Sl{\bn}}{4},
\end{align}
where $p_i$ is in the n-collinear sector, and similarly for the $\bn$-collinear case. The relation between full theory and collinear quark spinors are
\begin{align}
	u(p_i)=\left(1+\frac{\Sl{p}_{i\perp}}{\bn\cdot p_i}\frac{\Sl{\bn}}{2}\right)u_n(i), \qquad u(p_i)=\left(1+\frac{\Sl{p}_{i\perp}}{n\cdot p_i}\frac{\Sl{n}}{2}\right)u_\bn(i) \,,
\end{align}
for the $n$-collinear and $\bn$-collinear case respectively, and we have direct analogs for the $v(p_i)$ anti-quark spinors.

Expanding the two QCD diagrams to order $\cO(\lambda)$, we obtain
\begin{align}
	\left. \fd{2.5cm}{figures/matching_subleadingvertex_2.pdf}\right|_{\cO(\lambda)} &= -\frac{gT^a}{\omega_1}\bar u_n(1)\Sl{\epsilon}_{3\perp}^{*}\frac{\Sl{\bn}}{2} v_n(2)\, \nn \\
	\left. \fd{2.5cm}{figures/matching_subleadingvertex_1.pdf}\right|_{\cO(\lambda)} &= \frac{gT^a}{\omega_2}\bar u_n(1)\frac{\Sl{\bn}}{2}\Sl{\epsilon}_{3\perp}^{*} v_n(2)\,.
\end{align}
Therefore, the corresponding hard scattering operator is given by
\begin{align} \label{eq:O1B}
	\cO^{(1)}_{\cB}=\left(\frac{1}{\omega_1}+\frac{1}{\omega_2}\right)g\bar \chi_{n,\omega_1}\frac{\Sl{\bn}}{2}\Sl{\cB}_{\bn\perp,\omega_3}\chi_{n,-\omega_2}H\,,
\end{align}
or in terms of helicity operators
\begin{align}
	\cO^{(1)}_{\cB++}&=\frac{\omega_1+\omega_2}{2\sqrt{2\omega_1\omega_2}}gT^a_{\alpha\bar \bt}(\langle n\bn \rangle)^2\cB_{\bn+}^aJ^{\balpha\bt}_{n+}H\, \nn \\
\cO^{(1)}_{\cB--}&=-\frac{\omega_1+\omega_2}{2\sqrt{2\omega_1\omega_2}}gT^a_{\alpha\bar \bt}([n\bn])^2\cB_{\bn-}^aJ^{\balpha\bt}_{n-}H\,.
\end{align}
For back-to-back vectors $n$ and $\bn$, the spinor factors $\langle n\bn \rangle=2$ and $[n\bn]=-2$. 
The Feynman rule for the operator in \eq{O1B} is given by
\begin{align}
\fd{2.5cm}{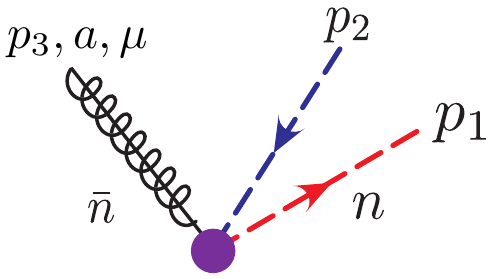}=gT^a\left(\frac{1}{\omega_1}+\frac{1}{\omega_2}\right)\frac{\Sl{\bn}}{2}\left(\gamma^{\mu}_{\perp}-\frac{\Sl{p}_{3\perp}n^{\mu}}{\omega_3}\right)\,.
\end{align}
\subsection{Subsubleading Power Matching}\label{sec:matching_nnlp}

In this section we carry out the matching to the $\cO(\lambda^2)$ operators at tree level, considering only those which contribute at the cross section level at $\cO(\lambda^2)$, as discussed in \Sec{sec:contribs}. Given the number of operators, each with different field content, we find it convenient to consider each case separately.
\subsubsection{$\cP_\perp$ insertion}\label{sec:matching_Pperp}

In \Sec{sec:nnlp_perp}, we show that the allowed subsubleading operators with $\cP_\perp$ insertions all have two collinear quark fields in different collinear sectors and a collinear gluon field. Therefore, the corresponding QCD diagrams are the same as that of the subleading operator, which are shown in \eq{QCD_bbg}. We start with the $(bg\cP_{\perp})_n(\bar b)_\bn$ case, whose allowed helicity configurations are given in \eq{Hbbgpperp_basis1}. To perform the matching, we take the kinematics to be
\begin{align}
p^{\mu}_1=\omega_1\frac{n^\mu}{2}+p^{\mu}_{\perp}+p_{1r}\frac{\bn^{\mu}}{2}, \qquad p^{\mu}_2=\omega_2\frac{\bn^\mu}{2}, \qquad p^\mu_3=\omega_3\frac{n^\mu}{2}-p^\mu_\perp+p_{3r}\frac{\bn^\mu}{2}\,,
\end{align}
with the on-shell conditions $\omega_1p_{1r}+p^2_\perp=0$ and $\omega_3p_{3r}+p^2_\perp=0$.

Expanding the QCD diagrams in \eq{QCD_bbg} with the chosen kinematics, we obtain
\begin{align}
\left. \fd{2.5cm}{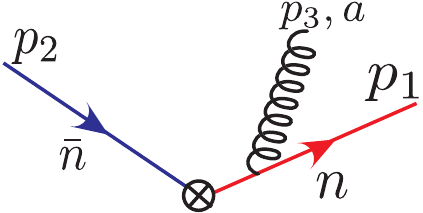}\right|_{\cO(\lambda^2)} &= 0\, \nn \\
\left. \fd{2.5cm}{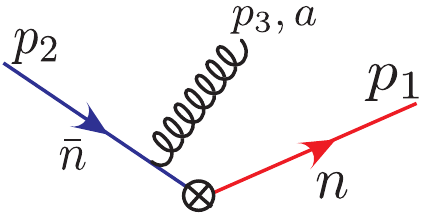}\right|_{\cO(\lambda^2)} &=-gT^a\left(\frac{1}{\omega_1\omega_2}+\frac{1}{\omega_2\omega_3}\right)\bar u_n(1)\Sl{p}_\perp\left(\Sl{\epsilon}^*_{3\perp}+\frac{\bn\cdot\epsilon^*_3\Sl{p}_\perp}{\omega_3}\right)v_\bn(2)\,.
\end{align}
The first diagram has no contribution at $\cO(\lambda^2)$ due to the on-shell propagator, and the amplitude of the second diagram can be recognized as the matrix element of the SCET hard scattering operators using the expansion of the collinear gluon field
\begin{align}\label{eq:gluon_expansion}
	\cB^\mu_{n\perp}=A^{\mu a}_{\perp k}T^a-k^\mu_\perp\frac{\bn\cdot A_{nk}^aT^a}{\bn\cdot k}+...\,,
\end{align}
where the dots represent terms with multiple gluon fields. Note that this expansion is the result of gauge invariance, and it guarantees that the Ward identity is satisfied.

The hard scattering operators are given by
\begin{align}\label{eq:Pqgq_match}
	\cO^{(2)}_{\cP\chi1}=\frac{g}{\omega_2}\left(\frac{1}{\omega_1}+\frac{1}{\omega_3}\right)\bar \chi_{n,\omega_1}[\Sl{\cP}_{\perp}\Sl{\cB}_{n\perp,\omega_3}]\chi_{\bn,-\omega_2}H\,,
\end{align}
or in terms of the helicity operators
\begin{align}\label{eq:Pqgq_hel_match}
	\cO^{(2)}_{\cP\chi1+(0)[-]}&=\frac{\omega_1+\omega_3}{\omega_3\sqrt{\omega_1\omega_2}}gT^a_{\alpha\bar \bt}[n\bn]\cB^a_{n+}\{\cP^{-}_\perp J^{\balpha\bt}_{n\bn 0}\}H\, \nn \\
	\cO^{(2)}_{\cP\chi1-(0^\dagger)[+]}&=\frac{\omega_1+\omega_3}{\omega_3\sqrt{\omega_1\omega_2}}gT^a_{\alpha\bar \bt}\langle n\bn\rangle\cB^a_{n-}\{\cP^{+}_\perp (J^\dagger)^{\balpha\bt}_{n\bn 0}\}H\,,
\end{align}
and the Feynman rule for \eq{Pqgq_match} is given by
\begin{align}\label{eq:Feynmanrule_Pperp}
\fd{2.5cm}{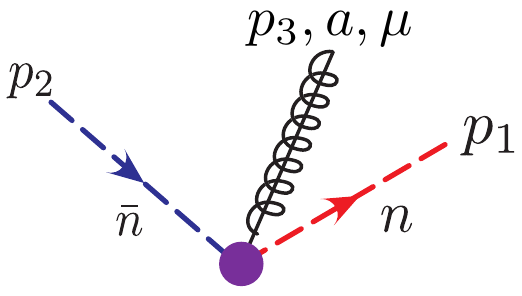}=\frac{gT^a}{\omega_2}\left(\frac{1}{\omega_1}+\frac{1}{\omega_3}\right)\Sl{p}_{3\perp}\left(\gamma^{\mu}_{\perp}-\frac{\Sl{p}_{3\perp}\bn^{\mu}}{\omega_3}\right)\,.
\end{align}
Interestingly, even though there are four allowed helicity configurations in \eq{Hbbgpperp_basis1}, only two of them have non-vanishing Wilson coefficients.

For the $(b)_n(\bar bg\cP_\perp)_\bn$ case, the operator basis is given in \eq{Hbbgpperp_basis2}. We take the kinematics to be
\begin{align}
p^{\mu}_1=\omega_1\frac{n^\mu}{2}, \qquad p^{\mu}_2=\omega_2\frac{\bn^\mu}{2}+p^{\mu}_{\perp}+p_{2r}\frac{n^{\mu}}{2}, \qquad p^\mu_3=\omega_3\frac{\bn^\mu}{2}-p^\mu_\perp+p_{3r}\frac{n^\mu}{2}\,,
\end{align}
Performing a similar matching calculation, or using charge conjugation, we can obtain the corresponding SCET hard scattering operator
\begin{align} \label{eq:O2Pchi2}
\cO^{(2)}_{\cP\chi2}=-\frac{g}{\omega_1}\left(\frac{1}{\omega_2}+\frac{1}{\omega_3}\right)\bar \chi_{n,\omega_1}[\Sl{\cB}_{\bn\perp,\omega_3}\Sl{\cP}_\perp^\dagger]\chi_{\bn,-\omega_2}H\,,
\end{align}
and the helicity operators are given by
\begin{align}
\cO^{(2)}_{\cP\chi2+(0)[-]}&=-\frac{\omega_2+\omega_3}{\omega_3\sqrt{\omega_1\omega_2}}gT^a_{\alpha\bar \bt}[n\bn]\cB^a_{\bn+}\{J^{\balpha\bt}_{n\bn 0}(\cP^{-}_\perp)^\dagger\}H\, \nn \\
\cO^{(2)}_{\cP\chi2-(0^\dagger)[+]}&=-\frac{\omega_2+\omega_3}{\omega_3\sqrt{\omega_1\omega_2}}gT^a_{\alpha\bar \bt}\langle n\bn\rangle\cB^a_{\bn-}\{(J^\dagger)^{\balpha\bt}_{n\bn 0}(\cP^{+}_\perp)^\dagger\}H\,,
\end{align}
with the Feynman rule for \eq{O2Pchi2} being
\begin{align}
\fd{2.5cm}{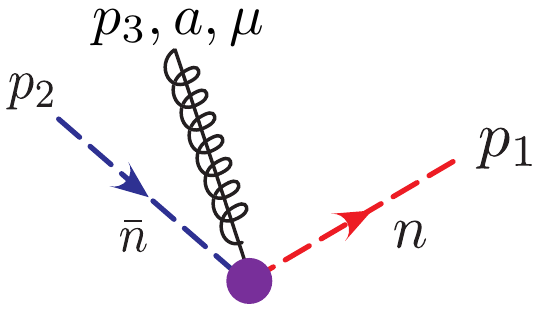}=-\frac{gT^a}{\omega_1}\left(\frac{1}{\omega_2}+\frac{1}{\omega_3}\right)\left(\gamma^{\mu}_{\perp}-\frac{\Sl{p}_{3\perp}n^{\mu}}{\omega_3}\right)\Sl{p}_{3\perp}\,.
\end{align}
Similarly, only two of the four operators in \eq{Hbbgpperp_basis2} have non-vanishing Wilson coefficients.

\subsubsection{bbgg}\label{sec:matching_bbgg}
We now consider the matching for the operators with two collinear quark fields and two collinear gluon fields. By the discussion in \Sec{sec:nnlp_collinear} and \Sec{sec:contribs}, we only have to consider two cases: $(bgg)_n(\bar b)_\bn$ and $(b)_n(\bar bgg)_\bn$. For the production of two quarks and two gluons, there are six QCD diagrams from the quark-gluon vertex:
\begin{align}\label{eq:QCD_bbgg}
\fd{2.5cm}{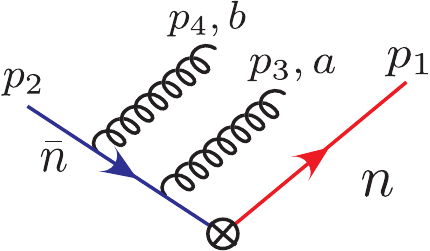}+\fd{2.5cm}{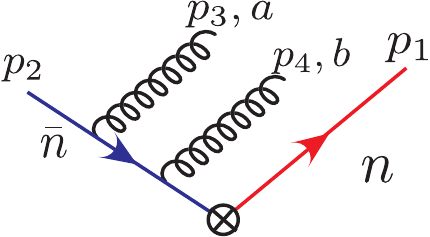}=&\bar u(p_1)\frac{-i(\Sl{p}_2+\Sl{p}_3+\Sl{p}_4)}{(p_2+p_3+p_4)^2}(igT^a\Sl{\epsilon}^*_3)\frac{-i(\Sl{p}_2+\Sl{p}_4)}{(p_2+p_4)^2}(igT^b\Sl{\epsilon}^*_4)v(p_2)\, \nn \\
&+\left((3,a)\leftrightarrow(4,b)\right)\, \nn \\
\fd{2.5cm}{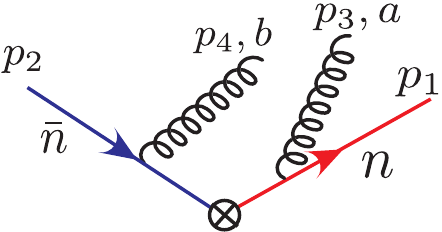}+\fd{2.5cm}{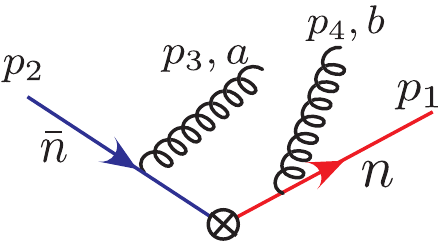}=&\bar u(p_1)(igT^a\Sl{\epsilon}^*_3)\frac{i(\Sl{p}_1+\Sl{p}_3)}{(p_1+p_3)^2}\frac{-i(\Sl{p}_2+\Sl{p}_4)}{(p_2+p_4)^2}(igT^b\Sl{\epsilon}^*_4)v(p_2)\, \nn \\
&+\left((3,a)\leftrightarrow(4,b)\right)\, \nn \\
\fd{2.5cm}{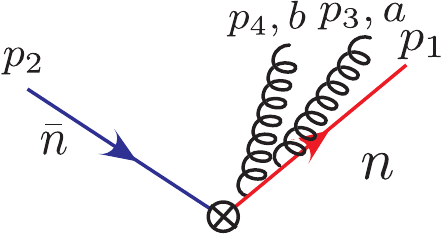}+\fd{2.5cm}{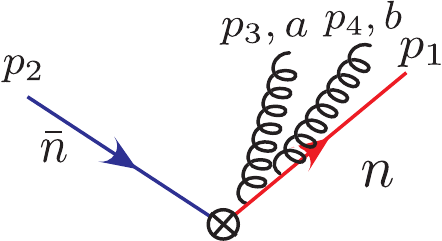}=&\bar u(p_1)(igT^a\Sl{\epsilon}^*_3)\frac{i(\Sl{p}_1+\Sl{p}_3)}{(p_1+p_3)^2}(igT^b\Sl{\epsilon}^*_4)\frac{i(\Sl{p}_1+\Sl{p}_3+\Sl{p}_4)}{(p_1+p_3+p_4)^2}v(p_2)\, \nn \\
&+\left((3,a)\leftrightarrow(4,b)\right)\,,
\end{align}
and two diagrams from the three-gluon vertex:
\begin{align}
\fd{2.5cm}{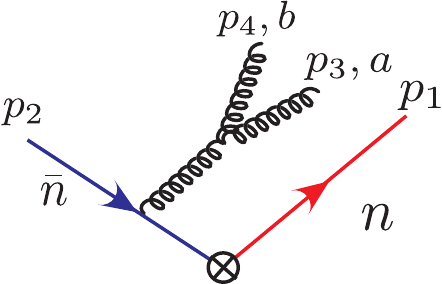}=&\bar u(p_1)\frac{-i(\Sl{p}_2+\Sl{p}_3+\Sl{p}_4)}{(p_2+p_3+p_4)^2}(igT^c\gamma_\lambda)\frac{-i}{(p_3+p_4)^2}gf^{abc}\, \nn \\
&\left[\epsilon^*_3\cdot\epsilon^*_4(-p_3+p_4)^\lambda+\epsilon^*_3\cdot(-p_3-2p_4)\epsilon^{*\lambda}_4+\epsilon^*_4\cdot(2p_3+p_4)\epsilon^{*\lambda}_3\right]v(p_2)\, \nn \\
\fd{2.5cm}{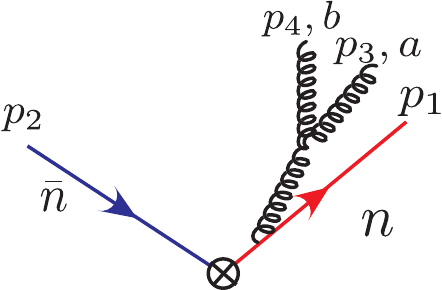}=&\bar u(p_1)(igT^c\gamma_\lambda)\frac{i(\Sl{p}_1+\Sl{p}_3+\Sl{p}_4)}{(p_1+p_3+p_4)^2}\frac{-i}{(p_3+p_4)^2}gf^{abc}\, \nn \\
&\left[\epsilon^*_3\cdot\epsilon^*_4(-p_3+p_4)^\lambda+\epsilon^*_3\cdot(-p_3-2p_4)\epsilon^{*\lambda}_4+\epsilon^*_4\cdot(2p_3+p_4)\epsilon^{*\lambda}_3\right]v(p_2)\,.
\end{align}

For the case $(bgg)_n(\bar b)_\bn$, we take the kinematics to be
\begin{align}\label{eq:kinematics_bbgg}
&p_1^\mu=\omega_1\frac{n^\mu}{2}, \qquad p_2^\mu=\omega_2\frac{\bn^\mu}{2}\, \nn \\
&p_3^\mu=\omega_3\frac{n^\mu}{2}+p^\mu_\perp+p_{3r}\frac{\bn^\mu}{2}, \qquad p_4^\mu=\omega_4\frac{n^\mu}{2}-p^\mu_\perp+p_{4r}\frac{\bn^\mu}{2}\,.
\end{align}
First of all, all the diagrams with on-shell propagators do not have $\cO(\lambda^2)$ term:
\begin{align}
\left.\fd{2.5cm}{figures/matching_2q2g_diagram5_low.pdf}\right|_{\cO(\lambda^2)}=\left.\fd{2.5cm}{figures/matching_2q2g_diagram6_low.pdf}\right|_{\cO(\lambda^2)}=\left.\fd{2.5cm}{figures/matching_2q2g_diagram8_low.pdf}\right|_{\cO(\lambda^2)}=0\,.
\end{align}
Moreover, with our chosen kinematics, the $\cO(\lambda^2)$ term of the other non-abelian diagram also vanishes,
\begin{align}
\left.\fd{2.5cm}{figures/matching_2q2g_diagram7_low.pdf}\right|_{\cO(\lambda^2)}=0\,.
\end{align}
Expanding the other four diagrams, we obtain:
\begin{align}
\left.\left(\fd{2.5cm}{figures/matching_2q2g_diagram2_low.pdf}+\fd{2.5cm}{figures/matching_2q2g_diagram1_low.pdf}\right)\right|_{\cO(\lambda^2)}&=\frac{-g^2T^aT^b}{\omega_2(\omega_3+\omega_4)}\bar u_n(1)\Sl{\epsilon}^*_{3\perp}\Sl{\epsilon}^*_{4\perp}v_\bn(2)\, \nn \\
&+\left((3,a)\leftrightarrow(4,b)\right)\, \nn \\
\left.\left(\fd{2.5cm}{figures/matching_2q2g_diagram4_low.pdf}+\fd{2.5cm}{figures/matching_2q2g_diagram3_low.pdf}\right)\right|_{\cO(\lambda^2)}&=\frac{-g^2T^aT^b}{\omega_1\omega_2\omega_4}(\omega_3+\omega_4)\bar u_n(1)\Sl{\epsilon}^*_{3\perp}\Sl{\epsilon}^*_{4\perp}v_\bn(2)\, \nn \\
&+\left((3,a)\leftrightarrow(4,b)\right)\,,
\end{align}
where we only keep the perpendicular part of the gluon polarization vectors to simplify the matching calculation. The full expression including the longitudinal polarization component can be obtained by gauge invariance and the expansion of the collinear gluon field \eq{gluon_expansion}. As a cross check we carry out the full calculation including the longitudinal polarization vectors in \App{app:polarization_long}.

After we obtain the $\cO(\lambda^2)$ term of the QCD diagrams with two quarks and two gluons, in order to calculate the Wilson coefficient of the hard scattering operator $\cO^{(2)}_{\cB2}$, we have to subtract the contribution from the other operators and SCET Lagrangian insertions. The SCET diagrams that will contribute to the matrix element of the production of two collinear quarks and two collinear gluons are from the operator $\cO^{(2)}_{\cP\chi1}$ and the leading power SCET Lagrangian $\cL^{(0)}$:
\begin{align}
\left.\left(\fd{2.5cm}{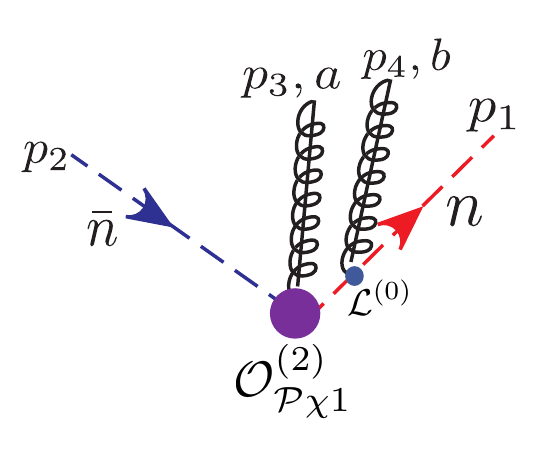}+\fd{2.5cm}{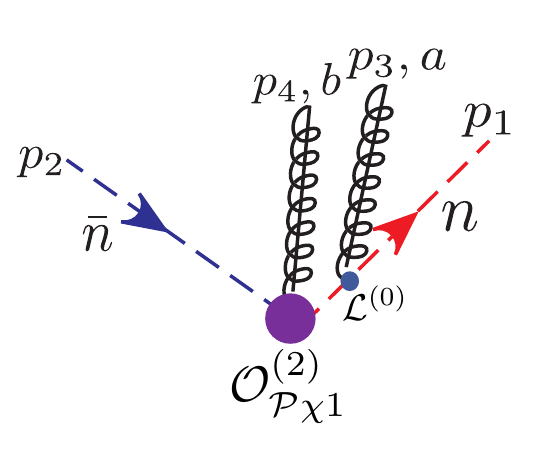}\right)\right|_{\cO(\lambda^2)}=&\frac{-g^2T^aT^b\omega_3(\omega_1+\omega_3+\omega_4)}{\omega_1\omega_2\omega_4(\omega_1+\omega_3)}\bar u_n(1)\Sl{\epsilon}^*_{3\perp}\Sl{\epsilon}^*_{4\perp}v_\bn(2)\, \nn \\
&+\left((3,a)\leftrightarrow(4,b)\right)\,,
\end{align}
and the non-abelian diagram has no contribution at $\cO(\lambda^2)$ since the total perpendicular momentum of the gluon fields is zero.
\begin{align}
\left.\fd{2.5cm}{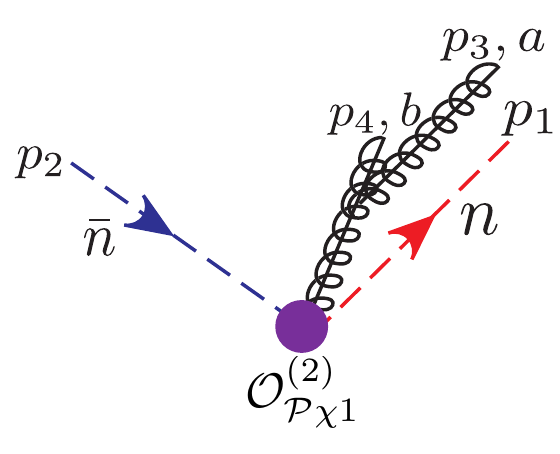}\right|_{\cO(\lambda^2)}=0\,.
\end{align}

Finally, the Wilson coefficient of the operator $\cO^{(2)}_{\cB2}$ can be obtained by taking the $\cO(\lambda^2)$ term of the QCD diagrams and subtract the contribution from the operator $\cO^{(2)}_{\cP\chi1}$ and SCET Lagrangian insertions. The result is given by
\begin{align}\label{eq:matching_bbgg_operator}
\cO^{(2)}_{\cB2}=-\frac{g^2}{\omega_2}\left(\frac{1}{\omega_1+\omega_3}+\frac{1}{\omega_3+\omega_4}\right)\bar\chi_{n,\omega_1}\Sl{\cB}_{n\perp,\omega_3}\Sl{\cB}_{n\perp,\omega_4}\chi_{\bn,-\omega_2}H\,,
\end{align} 
and the helicity operators are
\begin{align}
\cO^{(2)}_{\cB2+-(0)}&=g^2(T^aT^b)_{\alpha\bar\bt}\sqrt{\frac{\omega_1}{\omega_2}}\left(\frac{1}{\omega_1+\omega_3}+\frac{1}{\omega_3+\omega_4}\right)[n\bn]\cB^a_{n-}\cB^b_{n+}J^{\balpha\bt}_{n\bn 0}H\,, \nn \\
\cO^{(2)}_{\cB2+-(0^\dagger)}&=g^2(T^aT^b)_{\alpha\bar\bt}\sqrt{\frac{\omega_1}{\omega_2}}\left(\frac{1}{\omega_1+\omega_3}+\frac{1}{\omega_3+\omega_4}\right)\langle n\bn\rangle\cB^a_{n+}\cB^b_{n-}(J^\dagger)^{\balpha\bt}_{n\bn 0}H\,.
\end{align}
The Feynman rule of the operator in \eq{matching_bbgg_operator} is given by
\begin{align}
\fd{2.5cm}{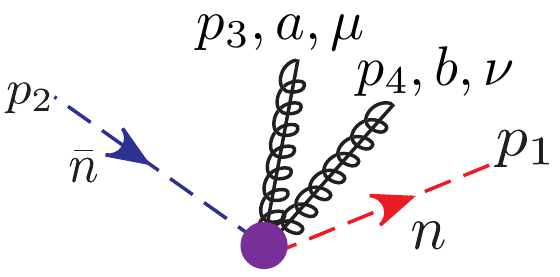}=&-\frac{g^2T^aT^b}{\omega_2}\left(\frac{1}{\omega_1+\omega_3}+\frac{1}{\omega_3+\omega_4}\right)\left(\gamma^\mu_\perp-\frac{\Sl{p}_{3\perp}\bn^\mu}{\omega_3}\right)\left(\gamma^\nu_\perp-\frac{\Sl{p}_{4\perp}\bn^\nu}{\omega_4}\right)\, \nn \\
&+\left((3,a,\mu)\leftrightarrow(4,b,\nu)\right)\,.
\end{align}

For the case $(b)_n(\bar bgg)_\bn$, we can perform a similar calculation or use the charge conjugation symmetry, and the hard scattering operator is given by
\begin{align} \label{eq:O2B3}
\cO^{(2)}_{\cB3}=-\frac{g^2}{\omega_1}\left(\frac{1}{\omega_2+\omega_4}+\frac{1}{\omega_3+\omega_4}\right)\bar\chi_{n,\omega_1}\Sl{\cB}_{\bn\perp,\omega_3}\Sl{\cB}_{\bn\perp,\omega_4}\chi_{\bn,-\omega_2}H\,,
\end{align}
and the helicity operators are
\begin{align}
\cO^{(2)}_{\cB3+-(0)}&=g^2(T^aT^b)_{\alpha\bar\bt}\sqrt{\frac{\omega_2}{\omega_1}}\left(\frac{1}{\omega_2+\omega_4}+\frac{1}{\omega_3+\omega_4}\right)[n\bn]\cB^a_{\bn+}\cB^b_{\bn-}J^{\balpha\bt}_{n\bn 0}H\,, \nn \\
\cO^{(2)}_{\cB3+-(0^\dagger)}&=g^2(T^aT^b)_{\alpha\bar\bt}\sqrt{\frac{\omega_2}{\omega_1}}\left(\frac{1}{\omega_2+\omega_4}+\frac{1}{\omega_3+\omega_4}\right)\langle n\bn\rangle\cB^a_{\bn-}\cB^b_{\bn+}(J^\dagger)^{\balpha\bt}_{n\bn 0}H\,.
\end{align}
The Feynman rule of the operator in \eq{O2B3} is
\begin{align}
\fd{2.5cm}{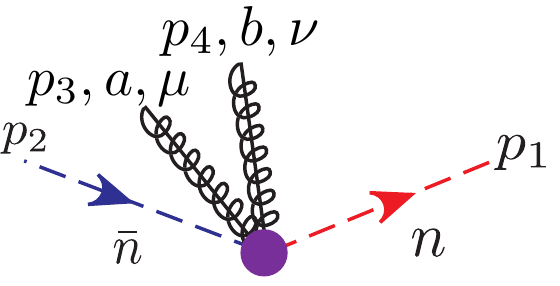}=&-\frac{g^2T^aT^b}{\omega_1}\left(\frac{1}{\omega_2+\omega_4}+\frac{1}{\omega_3+\omega_4}\right)\left(\gamma^\mu_\perp-\frac{\Sl{p}_{3\perp}n^\mu}{\omega_3}\right)\left(\gamma^\nu_\perp-\frac{\Sl{p}_{4\perp}n^\nu}{\omega_4}\right)\, \nn \\
&+\left((3,a,\mu)\leftrightarrow(4,b,\nu)\right)\,.
\end{align}

\subsubsection{bbqq}\label{sec:matching_bbqq}
We now consider the matching of the operators with four collinear quark fields. In this section, we first consider the case when the two quark pairs are of different flavor. As discussed in \Sec{sec:nnlp_collinear}, the bottom quark pair $b\bar b$ should have opposite chirality, and the other quark pair $q\bar q$ should have the same chirality. Therefore, in this case, there are only two QCD diagrams
\begin{align}\label{eq:QCD_bbqq}
\fd{2.5cm}{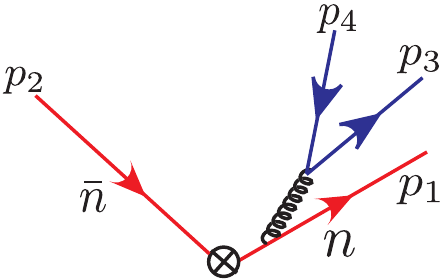}&=\bar u(p_3)(igT^a\gamma_\mu)v(p_4)\frac{-i}{(p_3+p_4)^2}\bar u(p_1)(igT^a\gamma^\mu)\frac{i(\Sl{p}_1+\Sl{p}_3+\Sl{p}_4)}{(p_1+p_3+p_4)^2}v(p_2)\,, \nn \\
\fd{2.5cm}{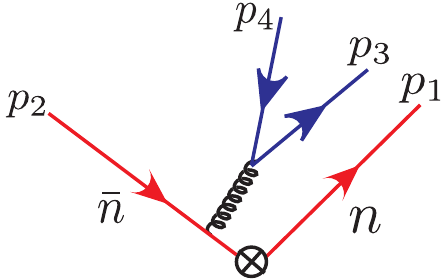}&=\bar u(p_1)\frac{-i(\Sl{p}_2+\Sl{p}_3+\Sl{p}_4)}{(p_2+p_3+p_4)^2}(igT^a\gamma_\mu)v(p_2)\frac{-i}{(p_3+p_4)^2}\bar u(p_3)(igT^a\gamma^\mu)v(p_4)\,.
\end{align}
Moreover, in \Sec{sec:contribs} we show that the operators $\cO^{(2)}_{bq1}$ and $\cO^{(2)}_{bq2}$ in \eqs{Z2_basis_bq_1}{Z2_basis_bq_2} have no contribution to the cross section at $\cO(\lambda^2)$, so we only need to consider the operators $\cO^{(2)}_{bq3}$ and $\cO^{(2)}_{bq4}$ in \eqs{Z2_basis_bq_3}{Z2_basis_bq_4}. For the case $(bq\bar q)_n(\bar b)_\bn$, which has the operator basis \eq{Z2_basis_bq_3}, we take the kinematics to be
\begin{align}\label{eq:kinematics_bbqq}
&p_1^\mu=\omega_1\frac{n^\mu}{2}, \qquad p_2^\mu=\omega_2\frac{\bn^\mu}{2}\, \nn \\
&p_3^\mu=\omega_3\frac{n^\mu}{2}+p^\mu_\perp+p_{3r}\frac{\bn^\mu}{2}, \qquad p_4^\mu=\omega_4\frac{n^\mu}{2}-p^\mu_\perp+p_{4r}\frac{\bn^\mu}{2}\,.
\end{align}
This choice of momentum eliminates all the contributions from the SCET Lagrangian insertions and the other hard scattering operators such as $\cO^{(2)}_{\cP\chi}$ at the $\cO(\lambda^2)$ order.

Expanding the QCD diagrams, we find that the $\cO(\lambda^2)$ terms of the two diagrams both vanish due to the nearly on-shell gluon propagator (which limits the terms that are expanded),
\begin{align}
\left. \fd{2.5cm}{figures/matching_bbqq_diagram1.pdf}\right|_{\cO(\lambda^2)}=\left. \fd{2.5cm}{figures/matching_bbqq_diagram2.pdf}\right|_{\cO(\lambda^2)}=0\,.
\end{align}
A similar calculation can be carried out for the $(b)_n(\bar bq\bar q)_\bn$ case, and the $\cO(\lambda^2)$ terms also vanish. Therefore, the tree level Wilson coefficients in the four quarks operators $\cO^{(2)}_{bq3}$ and $\cO^{(2)}_{bq4}$ are zero.

\subsubsection{bbbb}\label{sec:matching_bbbb}
Now we consider the case when all the quark fields are of the same flavor. We start with the case $(b\bar bb)_n(\bar b)_\bn$, whose operator basis is given by \eq{Z2_basis_bb_1}. All the full theory QCD diagrams and corresponding amplitudes with external state $b\bar bb\bar b$ are given by

\begin{align}\label{eq:QCD_bbbball}
\fd{2.5cm}{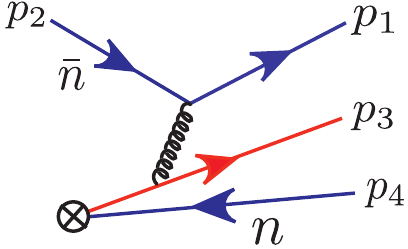}&=\bar u(p_1)(igT^a\gamma_\mu)v(p_2)\frac{-i}{(p_1+p_2)^2}\bar u(p_3)(igT^a\gamma^\mu)\frac{i(\Sl{p}_1+\Sl{p}_2+\Sl{p}_3)}{(p_1+p_2+p_3)^2}v(p_4)\,, \nn \\
\fd{2.5cm}{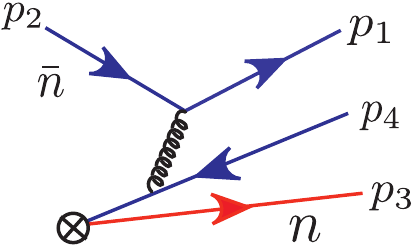}&=\bar u(p_1)(igT^a\gamma_\mu)v(p_2)\frac{-i}{(p_1+p_2)^2}\bar u(p_3)\frac{-i(\Sl{p}_1+\Sl{p}_2+\Sl{p}_4)}{(p_1+p_2+p_4)^2}(igT^a\gamma^\mu)v(p_4)\,, \nn \\
\fd{2.5cm}{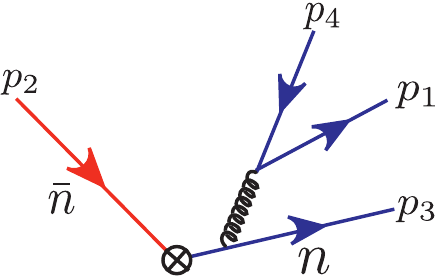}&=\bar u(p_1)(igT^a\gamma_\mu)v(p_4)\frac{-i}{(p_1+p_4)^2}\bar u(p_3)(igT^a\gamma^\mu)\frac{i(\Sl{p}_1+\Sl{p}_3+\Sl{p}_4)}{(p_1+p_3+p_4)^2}v(p_2)\,, \nn \\
\fd{2.5cm}{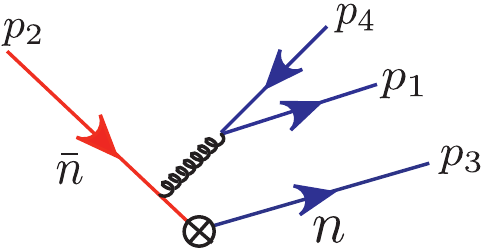}&=\bar u(p_1)(igT^a\gamma_\mu)v(p_4)\frac{-i}{(p_1+p_4)^2}\bar u(p_3)\frac{-i(\Sl{p}_1+\Sl{p}_2+\Sl{p}_4)}{(p_1+p_2+p_4)^2}(igT^a\gamma^\mu)v(p_2)\,.
\end{align}

In the above diagrams, the colors of the lines indicate the helicity of the quarks. Due to the fact that all the quarks are of the same flavor, depending on the different chirality configurations there are different possible QCD diagrams. In particular, for the operators $\cO^{(2)}_{bb1(0,\bar 0)}$, the helicity configuration is $(\underset{+}{b}\underset{-}{\bar b}\underset{-}{b})_n(\underset{-}{\bar b})_\bn$, and for $\cO^{(2)}_{bb1(0^\dagger,0)}$ the helicity configuration is $(\underset{-}{b}\underset{+}{\bar b}\underset{+}{b})_n(\underset{+}{\bar b})_\bn$. In these helicity configurations, the two bottom antiquarks are identical particles, and only the first two diagrams are possible. On the other hand, $\cO^{(2)}_{bb1(0,0)}$ has helicity configuration $(\underset{+}{b}\underset{+}{\bar b}\underset{+}{b})_n(\underset{-}{\bar b})_\bn$ and $\cO^{(2)}_{bb1(0^\dagger,\bar 0)}$ has $(\underset{-}{b}\underset{-}{\bar b}\underset{-}{b})_n(\underset{+}{\bar b})_\bn$. In these two configurations, the two bottom quarks are identical, and they correspond to the third and fourth diagrams.

Now, we apply the kinematics in \eq{kinematics_bbqq}, which will remove the contribution from $\cO^{(2)}_{\cP\chi1}$, and expand the QCD diagrams to $\cO(\lambda^2)$ order:
\begin{align}\label{eq:QCD_bbbb2all}
\left. \fd{2.5cm}{figures/matching_bbbb_diagram3.pdf}\right|_{\cO(\lambda^2)}&=-\frac{g^2}{\omega_1\omega_2(\omega_1+\omega_3)}\bar u_n(1)T^a\gamma_{\perp}^\mu v_\bn(2)\bar u_n(3)T^a\gamma_{\perp\mu}\frac{\Sl{\bn}}{2}v_n(4)\,, \nn \\
\left. \fd{2.5cm}{figures/matching_bbbb_diagram4.pdf}\right|_{\cO(\lambda^2)}&=\frac{g^2}{\omega_1\omega_2(\omega_1+\omega_4)}\bar u_n(1)T^a\gamma^\mu_{\perp} v_\bn(2)\bar u_n(3)\frac{\Sl{\bn}}{2}T^a\gamma_{\perp\mu} v_n(4)\,, \nn \\
\left. \fd{2.5cm}{figures/matching_bbbb_diagram5.pdf}\right|_{\cO(\lambda^2)}&=0\,,\nn \\
\left. \fd{2.5cm}{figures/matching_bbbb_diagram6.pdf}\right|_{\cO(\lambda^2)}&=\frac{g^2}{\omega_1\omega_2}\left(\frac{1}{\omega_1+\omega_4}+\frac{1}{\omega_3}\right)\left(\bar u_n(1)T^a\gamma_{\perp\mu}\frac{\Sl{p}_\perp}{\omega_4}\frac{\Sl{\bn}}{2}v_n(4)\frac{1}{p_{4r}}\bar u_n(3)\Sl{p}_\perp T^a\gamma^\mu_{\perp} v_\bn(2)\right.\,\nn \\
&\left.-\bar u_n(1)T^a\gamma_\mu v_n(4)\frac{1}{p_{4r}}\bar u_n(3)\frac{\Sl{\bn}}{2}\frac{p^2_\perp}{\omega_1+\omega_4} T^a\gamma^\mu v_\bn(2)\right)\,.
\end{align}
The third diagram has no $\cO(\lambda^2)$ term due to the nearly on-shell propagator, and the fourth diagram has a $\cO(\lambda^2)$ term that is nonlocal. In fact, this nonlocal $\cO(\lambda^2)$ term is not produced by the operator $\cO^{(2)}_{bb1}$, but by the operator $\cO^{(2)}_{\cP\chi1}$ and the leading power SCET Lagrangian insertions. This is due to the following SCET diagram, which has to be taken into account only when the two bottom quarks are identical particles:
\begin{align}\label{eq:SCET_bbbb2}
\left. \fd{2.5cm}{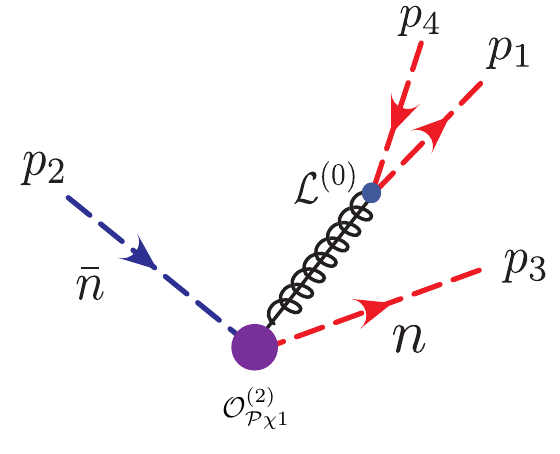}\right|_{\cO(\lambda^2)}&=\frac{g^2}{\omega_1\omega_2}\left(\frac{1}{\omega_1+\omega_4}+\frac{1}{\omega_3}\right)\left(\bar u_n(1)T^a\gamma_{\perp\mu}\frac{\Sl{p}_\perp}{\omega_4}\frac{\Sl{\bn}}{2}v_n(4)\frac{1}{p_{4r}}\bar u_n(3)T^a\Sl{p}_\perp\gamma^\mu_{\perp} v_\bn(2)\right. \, \nn \\
&\left.-\bar u_n(1)T^an_\mu\frac{\Sl{\bn}}{2}v_n(4)\frac{1}{p_{4r}}\bar u_n(3)T^a\frac{\bn^\mu}{\omega_1+\omega_4}p^2_\perp v_\bn(2)\right)\,.
\end{align}
In calculating this amplitude we use the Feynman rule for $\cO^{(2)}_{\cP\chi1}$ in \eq{Feynmanrule_Pperp} and the SCET leading power Lagrangian Feynman rule. One can show that the terms in the fourth diagram in \eq{QCD_bbbb2all} and in \eq{SCET_bbbb2} are exactly the same. Therefore, there is no contribution from the hard scattering operators $\cO^{(2)}_{bb1(0,0)}$ and $\cO^{(2)}_{bb1(0^\dagger,\bar 0)}$, whose helicity configurations have identical bottom quarks. 

As a result, the hard scattering operators corresponding to the external state $b\bar bb\bar b$ only contributes to the first two diagrams in \eq{QCD_bbbb2all}, and is given by
\begin{align} \label{eq:O2bb1}
\cO^{(2)}_{bb1}=-\frac{g^2}{\omega_1\omega_2}\left(\frac{1}{\omega_1+\omega_3}+\frac{1}{\omega_1+\omega_4}\right)\bar \chi_{n,\omega_1}T^a\gamma^\mu_{\perp}\chi_{\bn,-\omega_2}\bar \chi_{n,\omega_3}T^a\gamma_{\perp\mu}\frac{\Sl{\bn}}{2}\chi_{n,-\omega_4}\,.
\end{align}
After projecting to the helicity operators, we obtain
\begin{align}
\cO^{(2)}_{bb1}=-g^2\sqrt{\frac{\omega_3\omega_4}{\omega_1\omega_2}}\left(\frac{1}{\omega_1+\omega_3}+\frac{1}{\omega_1+\omega_4}\right)T^a_{\alpha\bar\bt}T^a_{\gamma\bar\delta}\left(\langle n\bn\rangle J^{\balpha\bt}_{n\bn-}J^{\bar\gamma\delta}_{n+}H+[n\bn]J^{\balpha\bt}_{n\bn+}J^{\bar\gamma\delta}_{n-}H\right)\,.
\end{align}
Using the Fierz identity for the $SU(3)$ generators
\begin{align}
T^a_{\alpha\bar\bt}T^a_{\gamma\bar\delta}=\frac{1}{2}\left(\delta_{\alpha\bar\delta}\delta_{\gamma\bar\bt}-\frac{1}{3}\delta_{\alpha\bar\bt}\delta_{\gamma\bar\delta}\right)\,,
\end{align}
and the Fierz identity for the spinors to change the order of the spinors, the helicity operators written in terms of the color basis in \eq{qqqq_color} and the helicity basis in \eq{Z2_basis_bb_1} are
\begin{align}
\cO^{(2)}_{bb1(0,\bar 0)}&=\frac{g^2}{16}\sqrt{\frac{\omega_3\omega_4}{\omega_1\omega_2}}\left(\frac{1}{\omega_1+\omega_3}+\frac{1}{\omega_1+\omega_4}\right)\left(\delta_{\alpha\bar\delta}\delta_{\gamma\bar\bt}-\frac{1}{3}\delta_{\alpha\bar\bt}\delta_{\gamma\bar\delta}\right)[n\bn]J^{\bar\gamma\bt}_{n\bn0}J^{\bar\alpha\delta}_{n\bar 0}H\,, \nn \\
\cO^{(2)}_{bb1(0^\dagger,0)}&=\frac{g^2}{16}\sqrt{\frac{\omega_3\omega_4}{\omega_1\omega_2}}\left(\frac{1}{\omega_1+\omega_3}+\frac{1}{\omega_1+\omega_4}\right)\left(\delta_{\alpha\bar\delta}\delta_{\gamma\bar\bt}-\frac{1}{3}\delta_{\alpha\bar\bt}\delta_{\gamma\bar\delta}\right)\langle n\bn\rangle (J^\dagger)^{\bar\gamma\bt}_{n\bn0}J^{\bar\alpha\delta}_{n0}H\,.
\end{align}
The other two possible operators in \eq{Z2_basis_bb_1}, although allowed by helicity constraints, have zero tree level Wilson coefficients.
Finally, the Feynman rule for the operator in \eq{O2bb1} is given by
\begin{align}
\fd{2.5cm}{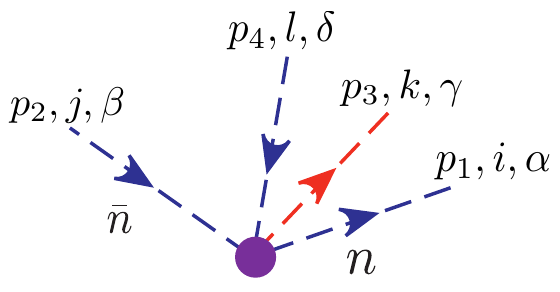}&=-\frac{g^2}{2\omega_1\omega_2}\left(\frac{1}{\omega_1+\omega_3}+\frac{1}{\omega_1+\omega_4}\right)\left(\delta_{\alpha\bar\delta}\delta_{\gamma\bar\bt}-\frac{1}{3}\delta_{\alpha\bar\bt}\delta_{\gamma\bar\delta}\right)\gamma^\mu_{\perp ij}\left(\gamma_{\perp\mu}\frac{\Sl{\bn}}{2}\right)_{kl}\,,
\end{align}
where the colors of the collinear quark propagators represent different chirality, $\alpha,\bt,\gamma,\delta$ are color indices, and $i,j,k,l$ are spinor indices.

The calculation is similar for the case $(b)_n(b\bar b\bar b)_\bn$. Here the two bottom antiquarks are identical, and a direct calculation or charge conjugation both imply that the Wilson coefficient of the operators $\cO^{(2)}_{bb2(0,\bar 0)}$ and $\cO^{(2)}_{bb2(0^\dagger,0)}$ are zero. The non-vanishing contribution comes from the 
\begin{align}  \label{eq:O2bb2}
\cO^{(2)}_{bb2}=-\frac{g^2}{\omega_1\omega_2}\left(\frac{1}{\omega_2+\omega_3}+\frac{1}{\omega_2+\omega_4}\right)\bar\chi_{n,\omega_1}T^a\gamma_{\perp\mu}\chi_{\bn,-\omega_2}\bar\chi_{\bn,\omega_3}T^a\gamma^\mu_{\perp}\frac{\Sl{n}}{2}\chi_{\bn,-\omega_4}H,
\end{align}
which in terms of helicity operators is
\begin{align}
\cO^{(2)}_{bb2(0,0)}&=-\frac{g^2}{16}\sqrt{\frac{\omega_3\omega_4}{\omega_1\omega_2}}\left(\frac{1}{\omega_2+\omega_3}+\frac{1}{\omega_2+\omega_4}\right)\left(\delta_{\alpha\bar\delta}\delta_{\gamma\bar\bt}-\frac{1}{3}\delta_{\alpha\bar\bt}\delta_{\gamma\bar\delta}\right)[n\bn]J^{\bar\gamma\bt}_{n\bn0}J^{\bar\alpha\delta}_{\bn0}H\,, \nn \\
\cO^{(2)}_{bb2(0^\dagger,\bar 0)}&=-\frac{g^2}{16}\sqrt{\frac{\omega_3\omega_4}{\omega_1\omega_2}}\left(\frac{1}{\omega_2+\omega_3}+\frac{1}{\omega_2+\omega_4}\right)\left(\delta_{\alpha\bar\delta}\delta_{\gamma\bar\bt}-\frac{1}{3}\delta_{\alpha\bar\bt}\delta_{\gamma\bar\delta}\right)\langle n\bn\rangle (J^\dagger)^{\bar\gamma\bt}_{n\bn0}J^{\bar\alpha\delta}_{\bn\bar 0}H\,.
\end{align}
The Feynman rule of the operator in \eq{O2bb2} is given by
\begin{align}
\fd{2.5cm}{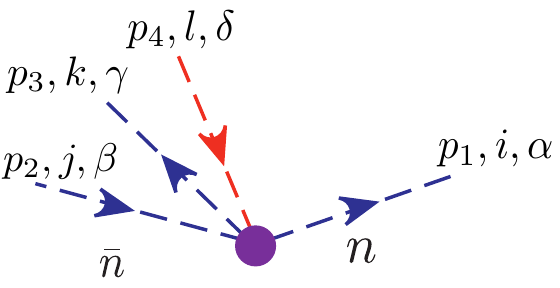}&=-\frac{g^2}{2\omega_1\omega_2}\left(\frac{1}{\omega_2+\omega_3}+\frac{1}{\omega_2+\omega_4}\right)\left(\delta_{\alpha\bar\delta}\delta_{\gamma\bar\bt}-\frac{1}{3}\delta_{\alpha\bar\bt}\delta_{\gamma\bar\delta}\right)\gamma^\mu_{\perp ij}\left(\gamma_{\perp\mu}\frac{\Sl{n}}{2}\right)_{kl}\,,
\end{align}
where the colors of the collinear quark propagators represent different chirality, $\alpha,\bt,\gamma,\delta$ are color indices, and $i,j,k,l$ are spinor indices.

\subsubsection{Ultrasoft Gluon and Derivatives}\label{sec:matching_gus}
Finally, we consider the matching for the operators with ultrasoft gluon or derivative insertions. Because of the RPI relation in \eq{usRPIrelation} and \eq{usRPIrelationb}, and the fact that the Wilson coefficient of the leading power operator is $1$ at tree level, the Wilson coefficient of both the operators with ultrasoft gluon and the ultrasoft derivative insertion should vanish at tree level. Here we'll give an explicit verification of this statement for the case of an ultrasoft gluon emission.

In \Sec{sec:nnlp_soft}, we show that every operator with ultrasoft gluon insertion has a quark in the $n$-collinear sector, an antiquark in the $\bn$-collinear sector, and a gluon with ultrasoft momentum. Therefore, we should consider the QCD diagrams in \eq{QCD_bbg} with the kinematics
\begin{align}
p^{\mu}_1=\omega_1\frac{n^\mu}{2}, \qquad p^{\mu}_2=\omega_2\frac{\bn^\mu}{2}, \qquad p^\mu_3=\bn\cdot p_3\frac{n^\mu}{2}+p^\mu_{3\perp}+n\cdot p_{3}\frac{\bn^\mu}{2}\,,
\end{align}
where all the components of the gluon momentum $p_3$ and polarization vector $\epsilon^*_3$ scale as $\sim\cO(\lambda^2)$.
Expanding the diagrams with the given kinematics, we obtain
\begin{align}
\left. \fd{2.5cm}{figures/matching_subleadingvertex_2_low.pdf}\right|_{\cO(\lambda^2)} &= -\frac{gT^a}{\omega_1}\bar u_n(1)\frac{1}{n\cdot p_3}\left(\Sl{\epsilon}^*_{3\perp}\Sl{p}_{3\perp}+n\cdot\epsilon^*_3\bn\cdot p_3\right)v_\bn(2)\, \nn \\
\left. \fd{2.5cm}{figures/matching_subleadingvertex_1_low.pdf}\right|_{\cO(\lambda^2)} &=\frac{gT^a}{\omega_2}\bar u_n(1)\frac{1}{\bn\cdot p_3}\left(\Sl{p}_{3\perp}\Sl{\epsilon}^*_{3\perp}+\bn\cdot\epsilon^*_3n\cdot p_3\right)v_\bn(2)\,.
\end{align}
For these two diagrams, there are also contributions from the subsubleading SCET Lagrangian $\cL^{(2)}$. The Feynman rules for the SCET Lagrangian at $\cO(\lambda^2)$ order involving a $n$-collinear quark and a ultrasoft gluon are given by
\begin{align}\label{eq:feynman_rule_subsubleading_prop}
\fd{2.5cm}{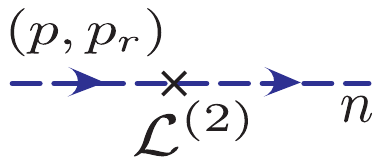}=i\frac{\Sl{\bn}}{2}\frac{p^2_{r\perp}}{\bn\cdot p}, \qquad
\fd{2.5cm}{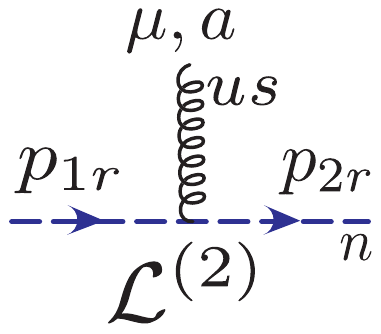}=igT^a\left(\frac{\gamma^\mu_\perp\Sl{p}_{1r\perp}}{\bn\cdot p}+\frac{\Sl{p}_{2r\perp}\gamma^\mu_\perp}{\bn\cdot p}\right)\frac{\Sl{\bn}}{2}\,.
\end{align}
Therefore, using the Feynman rules in \Eq{eq:feynman_rule_subsubleading_prop} we can compute the matrix element due to the SCET Lagrangian insertions. They are given by the following SCET diagrams
\begin{align}
\left.\left(\fd{2.5cm}{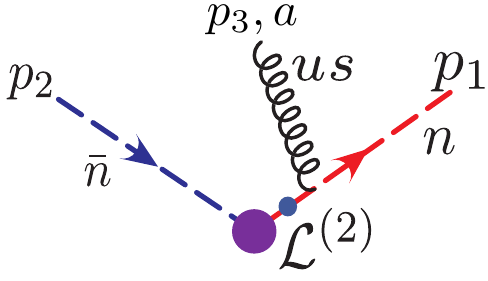}+\fd{2.5cm}{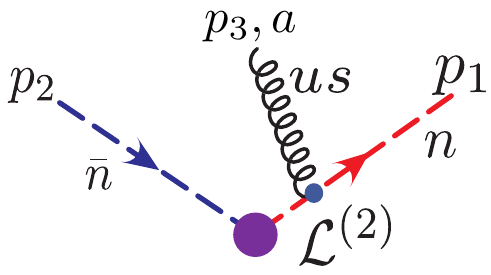}\right)\right|_{\cO(\lambda^2)}&=-\frac{gT^a}{\omega_1}\bar u_n(1)\frac{1}{n\cdot p_3}\left(\Sl{\epsilon}^*_{3\perp}\Sl{p}_{3\perp}+n\cdot\epsilon^*_s\bn\cdot p_s\right)\, \nn \\
\left.\left(\fd{2.5cm}{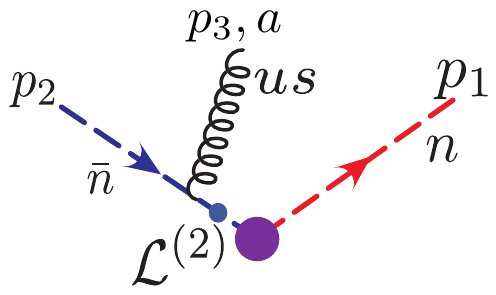}+\fd{2.5cm}{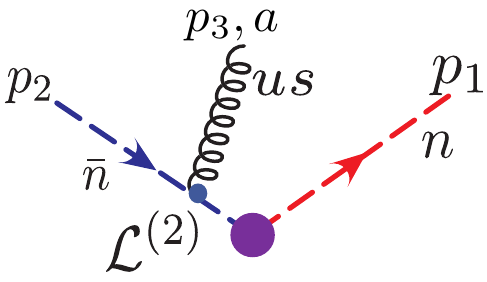}\right)\right|_{\cO(\lambda^2)}&=\frac{gT^a}{\omega_2}\bar u_n(1)\frac{1}{\bn\cdot p_3}\left(\Sl{p}_{3\perp}\Sl{\epsilon}^*_{3\perp}+\bn\cdot\epsilon^*_sn\cdot p_s\right)\,,
\end{align}
where for the two diagrams on the left the ultrasoft gluons are produced by the leading power SCET Lagrangian $\cL^{(0)}$, and for the diagrams on the right the ultrasoft gluons are from the subsubleading Lagrangian $\cL^{(2)}$. One can see that the above SCET diagrams exactly reproduce the matrix element calculated from the QCD diagrams. Therefore, we explicitly show that the Wilson coefficients for the hard scattering operators with ultrasoft gluon insertions are zero at tree level.

From the 1-loop matching of the LP operator of \eq{LP_wilson} we can derive the 1-loop Wilson coefficients of ultrasoft operators just by using the RPI relation in \eq{usRPIrelation} and \eq{usRPIrelationb}. This gives us 
\begin{align}\label{eq:us_matching}
	C^{(2)}_{\partial(us)n}&= C^{(2)}_{\cB(us)n}=-\frac{\partial C^{(0)} }{\partial \omega_1} = \frac{\alpha_s(\mu) C_F}{4\pi} \frac{1}{\omega_1} \left[2 \ln\left(\frac{-\omega_1\omega_2 -i 0}{\mu^2}\right) - 3\right] + \cO(\alpha_s^2)
\end{align}
Therefore the Feynman rules for the ultrasoft operators are
\begin{align}
	\fd{2.5cm}{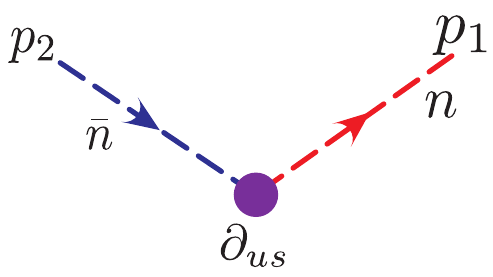}&=\frac{\alpha_s(\mu) C_F}{4\pi} \frac{\bn \cdot p_{1r}}{\omega_1} \left[2 \ln\left(\frac{-\omega_1\omega_2 -i 0}{\mu^2}\right) - 3\right]
  \,, \nn \\
	\fd{2.5cm}{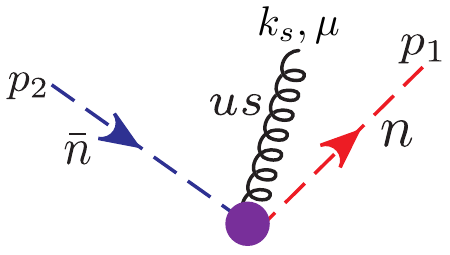}&=gT^a\frac{\alpha_s(\mu) C_F}{4\pi} \frac{1}{\omega_1} \left[2 \ln\left(\frac{-\omega_1\omega_2 -i 0}{\mu^2}\right) - 3\right]\left( \bn^\mu - \bn \cdot k_{s}\frac{n^\mu}{n \cdot k_{s}} \right)\,.
\end{align}
In terms of helicity operators we have
\begin{alignat}{2} \label{eq:soft_derivative_matching}
&\cO_{\partial(us(n))0:(0)}^{(2)\,\balpha\bt}
=  \frac{\alpha_s(\mu) C_F}{4\pi} \left[2 \ln\left(\frac{-\omega_1\omega_2 -i 0}{\mu^2}\right) - 3\right]\sqrt{\frac{\omega_2}{\omega_1}}\frac{[n\bn]}{2}\{i\partial_{us(n)0}\, J_{n\bar n\,0}^{\balpha\bt}\}\, H
\,,\nn \\
&\cO_{\partial(us(n))0:(0^\dagger)}^{(2)\,\balpha\bt}
=  \frac{\alpha_s(\mu) C_F}{4\pi} \left[2 \ln\left(\frac{-\omega_1\omega_2 -i 0}{\mu^2}\right) - 3\right]\sqrt{\frac{\omega_2}{\omega_1}}\frac{\langle n\bn\rangle}{2}\{i\partial_{us(n)0} \, (J^\dagger)_{n\bar n\,0}^{\balpha\bt}\}\, H\,,
\end{alignat}
and
\begin{alignat}{2} \label{eq:soft_insert_matching}
&O_{\cB(us(n))0:(0)}^{(2)a\,\balpha\bt}
= \frac{\alpha_s(\mu) C_F}{4\pi} \left[2 \ln\left(\frac{-\omega_1\omega_2 -i 0}{\mu^2}\right) - 3\right]\sqrt{\frac{\omega_2}{\omega_1}}\frac{[n\bn]}{2}g\cB^a_{us(n)0}J_{n\bar n\,0}^{\balpha\bt}\, H
\, \nn \\
&O_{\cB(us(n))0:(0^\dagger)}^{(2)a\,\balpha\bt}
=\frac{\alpha_s(\mu) C_F}{4\pi} \left[2 \ln\left(\frac{-\omega_1\omega_2 -i 0}{\mu^2}\right) - 3\right]\sqrt{\frac{\omega_2}{\omega_1}}\frac{\langle n\bn\rangle}{2}g\cB^a_{us(n)0}(J^\dagger)_{n\bar n\,0}^{\balpha\bt}\, H
\,.
\end{alignat}

\section{Conclusion}\label{sec:conclusions}

In this paper we presented a complete basis for power suppressed hard scattering operators describing the quark antiquark initiated production (or exclusive decay) of a color singlet scalar in \SCETi. 
This basis includes all the operators up to $\cO(\lambda^2)$ in the SCET expansion and it is summarized in \Tab{tab:summary}.  
Given the scalar nature of the color singlet operator the helicity selection rules for the hard scattering operators are particularly constraining. 
Therefore, the basis has been constructed using SCET helicity building blocks which both guarantee a gauge invariant definition of the hard scattering operators and easily allow to enforce helicity selection rules and rule out a large number of operators by angular momentum conservation.

Starting from our basis of operators we have analyzed the subset of them that contributes to cross section up to $\cO(\lambda^2)$. 
The constraints coming from helicity selection rules helped also in this case, allowing us to rule out some of the terms coming from the interference of the leading power operator with sub-subleading operators. 
Given the minimal basis of operators entering the cross section at $\cO(\lambda^2)$, we have determined in \Tab{tab:fact_func}, a schematic form of the factorization theorems up to $\cO(\lambda^2)$ and the field content of the soft and beam functions appearing in them.

For the operators that enter the cross section at $\cO(\lambda^2)$ we have also carried out the tree level calculation of their Wilson coefficients. After  subtracting contributions arising from the T-product of hard scattering operators and SCET Lagrangians all the Wilson coefficients are free of $\cO(\lambda^2)$ non localities as expected. 
Since some of these T-products involve subleading hard scattering operators, this gives a cross-check on the consistency of our calculations. 
RPI symmetry also relates the Wilson coefficient of ultrasoft operators to the derivative of the leading power one, resulting in the vanishing of the ultrasoft operators at tree level since at that order the leading power Wilson coefficient is a constant. 
We have verified this relation in the case of an ultrasoft gluon emission with an explicit calculation, showing that the T-product of the leading power operator and sub-subleading power lagrangians completely reproduces the full theory diagrams. 
Our results for the tree level Wilson coefficients will allow for a study of the power corrections at one-loop and for the study of the leading (in $\alpha_s$) logarithmic renormalization group structure at subleading power (in $\lambda$). 
Future directions therefore include both studies at fixed order, relevant for N-jettiness subtractions, studies of resummation for subleading power cross sections, and studies of factorization, including the universality of results with different underlying hard scattering processes.

\begin{acknowledgments}
We thank Ian Moult for discussions. This work was supported in part by the Office of Nuclear Physics of the U.S. Department of Energy under the Grant No.~DE-SC0011090. I.S. was also supported by the Simons Foundation through the Investigator grant 327942.
\end{acknowledgments}

\appendix
\section{Generalized Basis with $\cP_{\perp n},~\cP_{\perp \bar n}\neq 0$ and Mass Insertion Helicity Flip}\label{app:gen_pt}
In the main text, we presented a complete basis of operators to $\cO(\lambda^2)$ in a frame where the total $\cP_\perp$ in each collinear sector is restricted to be zero and the quark masses are taken to be zero. 
In this section we extend the basis, giving the additional operators present when the individual collinear sectors have non-vanishing $\cP_\perp$ and when the quarks are massive. 
We then perform a tree level matching calculation to those operators which can contribute to the cross section at $\cO(\lambda^2)$.

\subsection{Operators contributing when $\cP_{\perp n},~\cP_{\perp \bar n}\neq 0$}

While the coefficients of the operators discussed in this section could in principal be  fixed by RPI, we choose to find their coefficients by simply performing the tree level matching with more general kinematics.

We first consider the operators with two collinear quark fields and $\cP_\perp$ insertions.   Since the two quarks have opposite chirality and are in different collinear sectors, they must form a current with zero helicity. Therefore, there must be at least two $\cP_\perp$ for the operator to have zero total helicity. When both $\cP_\perp$ operators act on the $n$-collinear sector the full operator basis is given by
\begin{align}
&   \boldsymbol{(\cP_\perp\cP_\perp b)_n (\bar b)_{\bn}:} {\vcenter{\includegraphics[width=0.18\columnwidth]{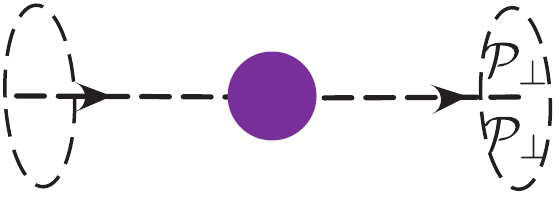}}}  \nn
\end{align}
\vspace{-0.4cm}
\begin{alignat}{2}\label{eq:Hbbpperp_basis1}
&O_{\cP \cP1(0)[+-]}^{(2)\,\balpha\bt}
= \big\{\cP_{\perp}^{+}\cP_{\perp}^{-} J_{n\bar n\, 0}^{\balpha\bt} \big\}\,  H
\,,\qquad &
&O_{\cP \cP1(0^\dagger)[+-]}^{(2)\,\balpha\bt}
= \big\{\cP_{\perp}^{+}\cP_{\perp}^{-} (J^\dagger)_{n\bar n\, 0}^{\balpha\bt} \big\}\,  H \,,
\end{alignat}
and when both $\cP_\perp$ operators act on the $\bn$-collinear sector we have
\begin{align}
&   \boldsymbol{(b)_n (\cP_\perp\cP_\perp \bar b)_{\bn}:} {\vcenter{\includegraphics[width=0.18\columnwidth]{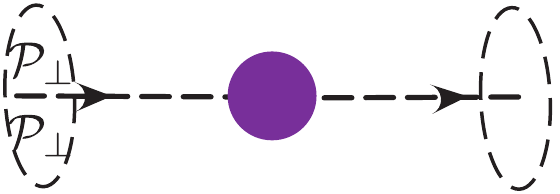}}}  \nn
\end{align}
\vspace{-0.4cm}
\begin{alignat}{2}\label{eq:Hbbpperp_basis2}
&O_{\cP \cP2(0)[+-]}^{(2)\,\balpha\bt}
= \big\{J_{n\bar n\, 0}^{\balpha\bt}(\cP_{\perp}^{+})^\dagger(\cP_{\perp}^{-})^\dagger  \big\}\,  H
\,,\qquad &
&O_{\cP \cP2(0^\dagger)[+-]}^{(2)\,\balpha\bt}
= \big\{(J^\dagger)_{n\bar n\, 0}^{\balpha\bt}(\cP_{\perp}^{+})^\dagger(\cP_{\perp}^{-})^\dagger  \big\}\,  H \,,
\end{alignat}
whereas when one $\cP_\perp$ operator acts on each sector we have
\begin{align}
&   \boldsymbol{(\cP_\perp b)_n (\cP_\perp\bar b)_{\bn}:} {\vcenter{\includegraphics[width=0.18\columnwidth]{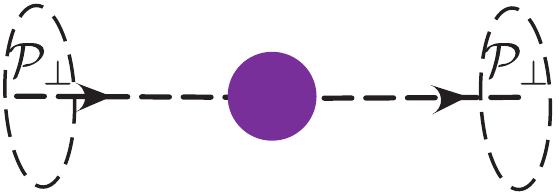}}}  \nn
\end{align}
\vspace{-0.4cm}
\begin{alignat}{2}\label{eq:Hbbpperp_basis3}
&O_{\cP \cP3(0)[++]}^{(2)\,\balpha\bt}
= \big\{\cP_{\perp}^{+}J_{n\bar n\, 0}^{\balpha\bt}(\cP_{\perp}^{+})^\dagger\big\}\,  H
\,,\qquad &
&O_{\cP \cP3(0^\dagger)[++]}^{(2)\,\balpha\bt}
= \big\{\cP_{\perp}^{+}(J^\dagger)_{n\bar n\, 0}^{\balpha\bt}(\cP_{\perp}^{+})^\dagger\big\}\, H \,, \\
&O_{\cP \cP3(0)[--]}^{(2)\,\balpha\bt}
= \big\{\cP_{\perp}^{-}J_{n\bar n\, 0}^{\balpha\bt}(\cP_{\perp}^{-})^\dagger\big\}\,  H
\,,\qquad &
&O_{\cP \cP3(0^\dagger)[--]}^{(2)\,\balpha\bt}
= \big\{\cP_{\perp}^{-}(J^\dagger)_{n\bar n\, 0}^{\balpha\bt}(\cP_{\perp}^{-})^\dagger\big\}\, H\,. \nn
\end{alignat}
Here we have used integration by parts to avoid including a $\cP_\perp$ operator acting on the Higgs $H$. The color structure of the above operators are given by \eq{leading_color}.

Next we consider the operators with two quark fields, one gluon field and a $\cP_\perp$ insertion. The operator basis assuming $\cP_\perp$ to be zero in each sector is given in \eqs{Hbbgpperp_basis1}{Hbbgpperp_basis2}. However, in the case that $\cP_\perp$ in each sector is nonzero, the basis becomes more complicated.  For the case that the gluon field is in the $n$-collinear sector, the basis when the $\cP_\perp$ operator acts on the $n$-collinear sector is 
\begin{align}
&   \boldsymbol{(bg\cP_\perp)_n (\bar b)_{\bn}:} {\vcenter{\includegraphics[width=0.18\columnwidth]{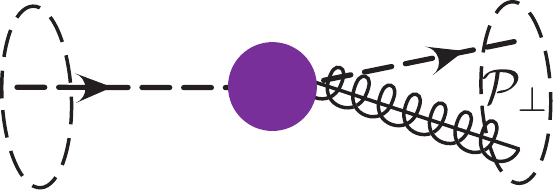}}}  \nn
\end{align}
\vspace{-0.4cm}
\begin{alignat}{2}\label{eq:Hbbgpperp_extension_basis1}
&O_{\cP\chi1-(0)[+]}^{(2)\,a\balpha\bt}
= \cB^a_{n-}\big\{\cP_{\perp}^{+}J_{n\bar n\, 0}^{\balpha\bt}\big\}\,  H
\,,\qquad &
&O_{\cP\chi1-(0^\dagger)[+]}^{(2)\,a\balpha\bt}
= \cB^a_{n-}\big\{\cP_{\perp}^{+}(J^\dagger)_{n\bar n\, 0}^{\balpha\bt}\big\}\,  H \,, \\
&O_{\cP\chi1+(0)[-]}^{(2)\,a\balpha\bt}
= \cB^a_{n+}\big\{\cP_{\perp}^{-}J_{n\bar n\, 0}^{\balpha\bt}\big\}\,  H
\,,\qquad &
&O_{\cP\chi1+(0^\dagger)[-]}^{(2)\,a\balpha\bt}
= \cB^a_{n+}\big\{\cP_{\perp}^{-}(J^\dagger)_{n\bar n\, 0}^{\balpha\bt}\big\}\,  H\,, \nn \\
&O_{\cP\cB1-(0)[+]}^{(2)\,a\balpha\bt}
= [\cP_\perp^{+}\cB^a_{n-}]J_{n\bar n\, 0}^{\balpha\bt}\,  H
\,,\qquad &
&O_{\cP\cB1-(0^\dagger)[+]}^{(2)\,a\balpha\bt}
= [\cP_\perp^{+}\cB^a_{n-}](J^\dagger)_{n\bar n\, 0}^{\balpha\bt}\,  H \,, \nn \\
&O_{\cP\cB1+(0)[-]}^{(2)\,a\balpha\bt}
= [\cP_\perp^{-}\cB^a_{n+}]J_{n\bar n\, 0}^{\balpha\bt}\,  H
\,,\qquad &
&O_{\cP\cB1+(0^\dagger)[-]}^{(2)\,a\balpha\bt}
= [\cP_\perp^{-}\cB^a_{n+}](J^\dagger)_{n\bar n\, 0}^{\balpha\bt}\,  H \,, \nn
\end{alignat}
and when the $\cP_\perp$ operator acts on the $\bn$-collinear sector we have
\begin{align}
&   \boldsymbol{(bg)_n (\bar b\cP_\perp)_{\bn}:} {\vcenter{\includegraphics[width=0.18\columnwidth]{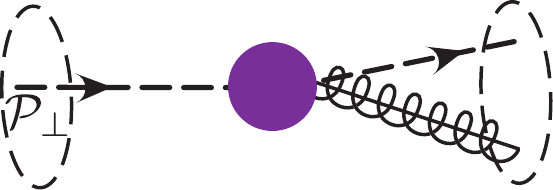}}}  \nn
\end{align}
\vspace{-0.4cm}
\begin{alignat}{2}\label{eq:Hbbgpperp_extension_basis2}
&O_{\cP\chi1-(0)[-]}^{(2)\,a\balpha\bt}
= \cB^a_{n-}\big\{J_{n\bar n\, 0}^{\balpha\bt}(\cP_{\perp}^{-})^\dagger\big\}\,  H
\,,\qquad &
&O_{\cP\chi1-(0^\dagger)[-]}^{(2)\,a\balpha\bt}
= \cB^a_{n-}\big\{(J^\dagger)_{n\bar n\, 0}^{\balpha\bt}(\cP_{\perp}^{-})^\dagger\big\}\,  H \,, \\
&O_{\cP\chi1+(0)[+]}^{(2)\,a\balpha\bt}
= \cB^a_{n+}\big\{J_{n\bar n\, 0}^{\balpha\bt}(\cP_{\perp}^{+})^\dagger\big\}\,  H
\,,\qquad &
&O_{\cP\chi1+(0^\dagger)[+]}^{(2)\,a\balpha\bt}
= \cB^a_{n+}\big\{(J^\dagger)_{n\bar n\, 0}^{\balpha\bt}(\cP_{\perp}^{+})^\dagger\big\}\,  H\,. \nn
\end{alignat}
The color structure of these operators are the same as \eq{nnlp_color_quark_perpBPS1}.

For the case that the gluon is in the $\bn$-collinear sector, the basis when the $\cP_\perp$ operator acts on the $\bn$-collinear sector is
\begin{align}
&   \boldsymbol{(b)_n (\bar bg\cP_\perp)_{\bn}:} {\vcenter{\includegraphics[width=0.18\columnwidth]{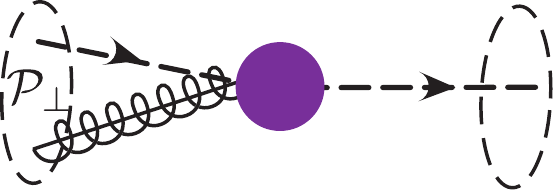}}}  \nn
\end{align}
\vspace{-0.4cm}
\begin{alignat}{2}\label{eq:Hbbgpperp_extension_basis3}
&O_{\cP\chi2-(0)[+]}^{(2)\,a\balpha\bt}
= \cB^a_{\bn-}\big\{J_{n\bar n\, 0}^{\balpha\bt}(\cP_{\perp}^{+})^\dagger\big\}\,  H
\,,\qquad &
&O_{\cP\chi2-(0^\dagger)[+]}^{(2)\,a\balpha\bt}
= \cB^a_{\bn-}\big\{(J^\dagger)_{n\bar n\, 0}^{\balpha\bt}(\cP_{\perp}^{+})^\dagger\big\}\,  H \,, \\
&O_{\cP\chi2+(0)[-]}^{(2)\,a\balpha\bt}
= \cB^a_{\bn+}\big\{J_{n\bar n\, 0}^{\balpha\bt}(\cP_{\perp}^{-})^\dagger\big\}\,  H
\,,\qquad &
&O_{\cP\chi2+(0^\dagger)[-]}^{(2)\,a\balpha\bt}
= \cB^a_{\bn+}\big\{(J^\dagger)_{n\bar n\, 0}^{\balpha\bt}(\cP_{\perp}^{-})^\dagger\big\}\,  H\,, \nn \\
&O_{\cP\cB2-(0)[+]}^{(2)\,a\balpha\bt}
= [\cP_\perp^{+}\cB^a_{\bn-}]J_{n\bar n\, 0}^{\balpha\bt}\,  H
\,,\qquad &
&O_{\cP\cB2-(0^\dagger)[+]}^{(2)\,a\balpha\bt}
= [\cP_\perp^{+}\cB^a_{\bn-}](J^\dagger)_{n\bar n\, 0}^{\balpha\bt}\,  H \,, \nn \\
&O_{\cP\cB2+(0)[-]}^{(2)\,a\balpha\bt}
= [\cP_\perp^{-}\cB^a_{\bn+}]J_{n\bar n\, 0}^{\balpha\bt}\,  H
\,,\qquad &
&O_{\cP\cB2+(0^\dagger)[-]}^{(2)\,a\balpha\bt}
= [\cP_\perp^{-}\cB^a_{\bn+}](J^\dagger)_{n\bar n\, 0}^{\balpha\bt}\,  H \,, \nn
\end{alignat}
and when the $\cP_\perp$ operator acts on the $n$-collinear sector,
\begin{align}
&   \boldsymbol{(b\cP_\perp)_n (\bar bg)_{\bn}:} {\vcenter{\includegraphics[width=0.18\columnwidth]{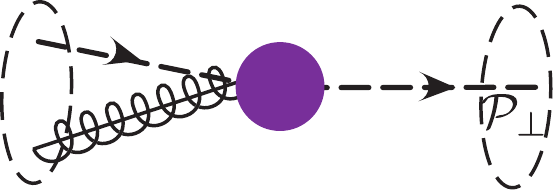}}}  \nn
\end{align}
\vspace{-0.4cm}
\begin{alignat}{2}\label{eq:Hbbgpperp_extension_basis4}
&O_{\cP\chi2-(0)[-]}^{(2)\,a\balpha\bt}
= \cB^a_{\bn-}\big\{\cP_{\perp}^{-}J_{n\bar n\, 0}^{\balpha\bt}\big\}\,  H
\,,\qquad &
&O_{\cP\chi2-(0^\dagger)[-]}^{(2)\,a\balpha\bt}
= \cB^a_{\bn-}\big\{\cP_{\perp}^{-}(J^\dagger)_{n\bar n\, 0}^{\balpha\bt}\big\}\,  H \,, \\
&O_{\cP\chi2+(0)[+]}^{(2)\,a\balpha\bt}
= \cB^a_{\bn+}\big\{\cP_{\perp}^{+}J_{n\bar n\, 0}^{\balpha\bt}\big\}\,  H
\,,\qquad &
&O_{\cP\chi2+(0^\dagger)[+]}^{(2)\,a\balpha\bt}
= \cB^a_{\bn+}\big\{\cP_{\perp}^{+}(J^\dagger)_{n\bar n\, 0}^{\balpha\bt}\big\}\,  H\,. \nn
\end{alignat}
The color structure of the operators are the same as \eq{nnlp_color_quark_perpBPS2}.

The matching for the operator with two quark fields and two $\cP_\perp$ insertions can be done by expanding the matrix element with $b\bar b$ external state. We take the kinematics to be
\begin{align}
p_1^\mu=\omega_1\frac{n^\mu}{2}+p_{1\perp}^\mu+p_{1r}\frac{\bn^\mu}{2}, \qquad p_2^\mu=\omega_2\frac{\bn^\mu}{2}+p_{2\perp}^\mu+p_{2r}\frac{n^\mu}{2}\,.
\end{align}
Then we obtain
\begin{align}
\left.\bar u(p_1)v(p_2)\right|_{\cO(\lambda^2)}=-\frac{1}{\omega_1\omega_2}\bar u_n(1)\Sl{p}_{1\perp}\Sl{p}_{2\perp}v_\bn(2)\,
\end{align}
which corresponds to the operator
\begin{align} \label{eq:O2PP3}
\cO^{(2)}_{\cP\cP3}=-\frac{1}{\omega_1\omega_2}[\bar\chi_{n,\omega_1}\Sl{\cP}_{1\perp}^\dagger][\Sl{\cP}_{2\perp}\chi_{\bn,-\omega_2}]H\,,
\end{align}
or in terms of the helicity operator
\begin{align} \label{eq:O2PP3h}
\cO^{(2)}_{\cP\cP3(0)[--]}&=\frac{1}{\sqrt{\omega_1\omega_2}}\delta_{\alpha\bar\bt}[n\bn]\big\{\cP_\perp^{-}J^{\balpha\bt}_{n\bn 0}(\cP_\perp^{-})^\dagger\big\}H\, \nn \\
\cO^{(2)}_{\cP\cP3(0^\dagger)[++]}&=\frac{1}{\sqrt{\omega_1\omega_2}}\delta_{\alpha\bar\bt}\langle n\bn\rangle\big\{\cP_\perp^{+}(J^\dagger)^{\balpha\bt}_{n\bn 0}(\cP_\perp^{+})^\dagger\big\}H\,,
\end{align}
and the Feynman rule for the operator in \eq{O2PP3} is 
\begin{align}
\fd{2.5cm}{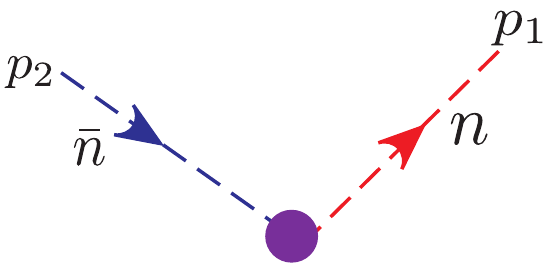}=-\frac{1}{\omega_1\omega_2}\Sl{p}_{1\perp}\Sl{p}_{2\perp}\,.
\end{align}
Except for the two operators in \eq{O2PP3h}, the Wilson coefficients of all the other operators in \eqss{Hbbpperp_basis1}{Hbbpperp_basis2}{Hbbpperp_basis3} are zero. 

For the operators with two quark fields, one gluon field, and one $\cP_\perp$ insertion, we have to consider the QCD diagrams with $b\bar bg$ external state, which are given in \eq{QCD_bbg}. In the case that the gluon is in the $n$-collinear sector, we use the following kinematics:
\begin{align}
p_1^\mu=\omega_1\frac{n^\mu}{2}+p_{1\perp}^\mu+p_{1r}\frac{\bn^\mu}{2}, \qquad p_2^\mu=\omega_2\frac{\bn^\mu}{2}+p_{2\perp}^\mu+p_{2r}\frac{n^\mu}{2}, \qquad p_3^\mu=\omega_3\frac{n^\mu}{2}+p_{3\perp}^\mu+p_{3r}\frac{\bn^\mu}{2}\,.
\end{align}
To do the matching for the operators with $\cP_\perp$ insertion in the $n$-collinear sector, which are given in \eq{Hbbgpperp_extension_basis1}, we first take $p_{2\perp}=0$ and expand the QCD diagram. The $\cO(\lambda^2)$ term is given by
\begin{align}
\left. \fd{2.5cm}{figures/matching_subleadingvertex_2_low.pdf}\right|_{\cO(\lambda^2)} = & \, 0\,, \nn \\
\left. \fd{2.5cm}{figures/matching_subleadingvertex_1_low.pdf}\right|_{\cO(\lambda^2)} =&\frac{gT^a}{\omega_2\omega_3}\bar u_n(1)\Sl{p}_{3\perp}\left(\Sl{\epsilon}^*_{3\perp}-\frac{\bn\cdot\epsilon^*_3\Sl{p}_{3\perp}}{\omega_3}\right)v_\bn(2)\, \nn \\
&-\frac{gT^a}{\omega_1\omega_2}\bar u_n(1)\Sl{p}_{1\perp}\left(\Sl{\epsilon}^*_{3\perp}-\frac{\bn\cdot\epsilon^*_3\Sl{p}_{3\perp}}{\omega_3}\right)v_\bn(2) \,.
\end{align}
Therefore the operators are
\begin{align} \label{eq:O2PchiB1}
\cO^{(2)}_{\cP\chi1}&=-\frac{g}{\omega_1\omega_2}\left[\bar\chi_{n,\omega_1}\Sl{\cP}_{1\perp}^\dagger\right]\Sl{\cB}_{n\perp,\omega_3}\chi_{\bn,-\omega_2}\,, \nn \\
\cO^{(2)}_{\cP\cB1}&=\frac{g}{\omega_2\omega_3}\bar\chi_{n,\omega_1}\left[\Sl{\cP}_{3\perp}\Sl{\cB}_{n\perp,\omega_3}\right]\chi_{\bn,-\omega_2}\,.
\end{align}
The helicity operators are
\begin{align}
\cO^{(2)}_{\cP\chi1+(0)[-]}&=\frac{g}{\sqrt{\omega_1\omega_2}}T^a_{\alpha\bar\bt}[n\bn]\cB^a_{n+}\big\{\cP^{-}_{\perp}J_{n\bn 0}^{\balpha\bt}\big\}H\,, \nn \\
\cO^{(2)}_{\cP\chi1-(0^\dagger)[+]}&=\frac{g}{\sqrt{\omega_1\omega_2}}T^a_{\alpha\bar\bt}\langle n\bn\rangle\cB^a_{n-}\big\{\cP^{+}_{\perp}(J^\dagger)_{n\bn 0}^{\balpha\bt}\big\}H\,, \nn \\
\cO^{(2)}_{\cP\cB1+(0)[-]}&=-\frac{g}{\omega_3}\sqrt{\frac{\omega_1}{\omega_2}}T^a_{\alpha\bar\bt}[n\bn][\cP^{-}_{\perp}\cB^a_{n+}]J_{n\bn 0}^{\balpha\bt}H\,, \nn \\
\cO^{(2)}_{\cP\cB1-(0^\dagger)[+]}&=-\frac{g}{\omega_3}\sqrt{\frac{\omega_1}{\omega_2}}T^a_{\alpha\bar\bt}\langle n\bn\rangle[\cP^{+}_{\perp}\cB^a_{n-}](J^\dagger)_{n\bn 0}^{\balpha\bt}H\,.
\end{align}
The Feynman rule for the operators in \eq{O2PchiB1} is given by
\begin{align}
\fd{2.5cm}{figures/matching_scetPperp.pdf}=-\frac{gT^a}{\omega_1\omega_2}\Sl{p}_{1\perp}\left(\gamma^\mu_\perp-\frac{\Sl{p}_{3\perp}\bn^\mu}{\omega_3}\right)+\frac{gT^a}{\omega_2\omega_3}\Sl{p}_{3\perp}\left(\gamma^\mu_\perp-\frac{\Sl{p}_{3\perp}\bn^\mu}{\omega_3}\right)\,.
\end{align}

For the case that the $\cP_\perp$ insertion is in the $\bn$-collinear sector, we take $p_{1\perp}+p_{3\perp}=0$ and expand the diagrams. The $\cO(\lambda^2)$ term proportional to $\Sl{p}_{2\perp}$ is given by
\begin{align}
\left.\left(\fd{2.5cm}{figures/matching_subleadingvertex_2_low.pdf}+\fd{2.5cm}{figures/matching_subleadingvertex_1_low.pdf}\right)\right|_{\cO(\lambda^2)}&\, \nn \\
=\frac{gT^a}{\omega_2\omega_3}\bar u_n(1)\Sl{p}_{2\perp}\left(\Sl{\epsilon}^*_{3\perp}-\frac{\bn\cdot\epsilon^*_3\Sl{p}_{3\perp}}{\omega_3}\right)v_\bn(2)&\, \nn \\
+\frac{gT^a}{\omega_2}\left(\frac{1}{\omega_3}+\frac{1}{\omega_1+\omega_3}\right)&\bar u_n(1)\left(\Sl{\epsilon}^*_{3\perp}-\frac{\bn\cdot\epsilon^*_3\Sl{p}_{3\perp}}{\omega_3}\right)\Sl{p}_{2\perp}v_\bn(2)\,.
\end{align}
From this we obtain
\begin{align} \label{eq:O2Pchi1}
\cO^{(2)}_{\cP\chi1}=\frac{g}{\omega_2\omega_3}\bar\chi_{n,\omega_1}\Sl{\cP}_{2\perp}\Sl{\cB}_{n\perp,\omega_3}\chi_{\bn,-\omega_2}H+\frac{g}{\omega_2}\left(\frac{1}{\omega_3}+\frac{1}{\omega_1+\omega_3}\right)\bar\chi_{n,\omega_1}\Sl{\cB}_{n\perp,\omega_3}\left[\Sl{\cP}_{2\perp}\chi_{\bn,-\omega_2}\right]H\,,
\end{align}
where $\cP_{2\perp}$ acts on the $\bn$-collinear field $\chi_{\bn,-\omega_2}$. The helicity operators are then given by
\begin{align}
\cO^{(2)}_{\cP\chi1-(0)[-]}&=-g\sqrt{\frac{\omega_1}{\omega_2}}\left(\frac{1}{\omega_3}+\frac{1}{\omega_1+\omega_3}\right)T^a_{\alpha\bar\bt}[n\bn]\cB^a_{n-}\big\{J^{\balpha\bt}_{n\bn 0}(\cP_\perp^{-})^\dagger\big\}H\,, \nn \\
\cO^{(2)}_{\cP\chi1+(0^\dagger)[+]}&=-g\sqrt{\frac{\omega_1}{\omega_2}}\left(\frac{1}{\omega_3}+\frac{1}{\omega_1+\omega_3}\right)T^a_{\alpha\bar\bt}\langle n\bn\rangle\cB^a_{n+}\big\{(J^\dagger)^{\balpha\bt}_{n\bn 0}(\cP_\perp^{+})^\dagger\big\}H\,, \nn \\
\cO^{(2)}_{\cP\chi1+(0)[+]}&=-\frac{g}{\omega_3}\sqrt{\frac{\omega_1}{\omega_2}}T^a_{\alpha\bar\bt}[n\bn]\cB^a_{n+}\big\{J^{\balpha\bt}_{n\bn 0}(\cP_\perp^{+})^\dagger\big\}H\,, \nn \\
\cO^{(2)}_{\cP\chi1-(0^\dagger)[-]}&=-\frac{g}{\omega_3}\sqrt{\frac{\omega_1}{\omega_2}}T^a_{\alpha\bar\bt}\langle n\bn\rangle\cB^a_{n-}\big\{(J^\dagger)^{\balpha\bt}_{n\bn 0}(\cP_\perp^{-})^\dagger\big\}H\,,
\end{align}
and the Feynman rule for \eq{O2Pchi1} is
\begin{align}
\fd{2.5cm}{figures/matching_scetPperp.pdf}=\frac{gT^a}{\omega_2\omega_3}\Sl{p}_{2\perp}\left(\gamma^\mu_\perp-\frac{\Sl{p}_{3\perp}\bn^\mu}{\omega_3}\right)+\frac{gT^a}{\omega_2}\left(\frac{1}{\omega_3}+\frac{1}{\omega_1+\omega_3}\right)\left(\gamma^\mu_\perp-\frac{\Sl{p}_{3\perp}\bn^\mu}{\omega_3}\right)\Sl{p}_{2\perp}\,.
\end{align}

For the other case that the gluon field is in the $\bn$-collinear sector, the matching calculation is similar. When $\cP_\perp$ is in the $\bn$-collinear sector, the operators are
\begin{align} \label{eq:O2PchiB2}
\cO^{(2)}_{\cP\chi2}&=\frac{g}{\omega_1\omega_2}\bar\chi_{n,\omega_1}\Sl{\cB}_{\bn\perp,\omega_3}\left[\Sl{\cP}_{2\perp}^\dagger\chi_{\bn,-\omega_2}\right]\,, \nn \\
\cO^{(2)}_{\cP\cB2}&=-\frac{g}{\omega_1\omega_3}\bar\chi_{n,\omega_1}\left[\Sl{\cB}_{\bn\perp,\omega_3}(\Sl{\cP}_{3\perp})^\dagger\right]\chi_{\bn,-\omega_2}\,,
\end{align}
and the helicity operators are given by
\begin{align}
\cO^{(2)}_{\cP\chi2+(0)[-]}&=-\frac{g}{\sqrt{\omega_1\omega_2}}T^a_{\alpha\bar\bt}[n\bn]\cB^a_{\bn+}\big\{J_{n\bn 0}^{\balpha\bt}(\cP^{-}_{\perp})^\dagger\big\}H\,, \nn \\
\cO^{(2)}_{\cP\chi2-(0^\dagger)[+]}&=-\frac{g}{\sqrt{\omega_1\omega_2}}T^a_{\alpha\bar\bt}\langle n\bn\rangle\cB^a_{\bn-}\big\{(J^\dagger)_{n\bn 0}^{\balpha\bt}(\cP^{+}_{\perp})^\dagger\big\}H\,, \nn \\
\cO^{(2)}_{\cP\cB2+(0)[-]}&=\frac{g}{\omega_3}\sqrt{\frac{\omega_2}{\omega_1}}T^a_{\alpha\bar\bt}[n\bn][\cP^{-}_{\perp}\cB^a_{\bn+}]J_{n\bn 0}^{\balpha\bt}H\,, \nn \\
\cO^{(2)}_{\cP\cB2-(0^\dagger)[+]}&=\frac{g}{\omega_3}\sqrt{\frac{\omega_2}{\omega_1}}T^a_{\alpha\bar\bt}\langle n\bn\rangle[\cP^{+}_{\perp}\cB^a_{\bn-}](J^\dagger)_{n\bn 0}^{\balpha\bt}H\,,
\end{align}
with the Feynman rule for the operators in \eq{O2PchiB2} being
\begin{align}
\fd{2.5cm}{figures/matching_scetPperp2.pdf}=\frac{gT^a}{\omega_1\omega_2}\left(\gamma^\mu_\perp-\frac{\Sl{p}_{3\perp}n^\mu}{\omega_3}\right)\Sl{p}_{2\perp}-\frac{gT^a}{\omega_1\omega_3}\left(\gamma^\mu_\perp-\frac{\Sl{p}_{3\perp}n^\mu}{\omega_3}\right)\Sl{p}_{3\perp}\,.
\end{align}
When $\cP_\perp$ is in the $n$-collinear sector, we obtain
\begin{align} \label{eq:O2Pchi2}
\cO^{(2)}_{\cP\chi2}=-\frac{g}{\omega_1\omega_3}\bar\chi_{n,\omega_1}\Sl{\cB}_{n\perp,\omega_3}\Sl{\cP}_{1\perp}\chi_{\bn,-\omega_2}H-\frac{g}{\omega_1}\left(\frac{1}{\omega_3}+\frac{1}{\omega_2+\omega_3}\right)\left[\bar\chi_{n,\omega_1}\Sl{\cP}_{1\perp}^\dagger\right]\Sl{\cB}_{n\perp,\omega_3}\chi_{\bn,-\omega_2}H\,,
\end{align}
and the helicity operators are
\begin{align}
\cO^{(2)}_{\cP\chi2-(0)[-]}&=g\sqrt{\frac{\omega_2}{\omega_1}}\left(\frac{1}{\omega_3}+\frac{1}{\omega_2+\omega_3}\right)T^a_{\alpha\bar\bt}[n\bn]\cB^a_{\bn-}\big\{\cP_\perp^{-}J^{\balpha\bt}_{n\bn 0}\big\}H\, \nn \\
\cO^{(2)}_{\cP\chi2+(0^\dagger)[+]}&=g\sqrt{\frac{\omega_2}{\omega_1}}\left(\frac{1}{\omega_3}+\frac{1}{\omega_2+\omega_3}\right)T^a_{\alpha\bar\bt}\langle n\bn\rangle\cB^a_{\bn+}\big\{\cP_\perp^{+}(J^\dagger)^{\balpha\bt}_{n\bn 0}\big\}H\, \nn \\
\cO^{(2)}_{\cP\chi2+(0)[+]}&=\frac{g}{\omega_3}\sqrt{\frac{\omega_2}{\omega_1}}T^a_{\alpha\bar\bt}[n\bn]\cB^a_{\bn+}\big\{\cP_\perp^{+}J^{\balpha\bt}_{n\bn 0}\big\}H\, \nn \\
\cO^{(2)}_{\cP\chi2-(0^\dagger)[-]}&=\frac{g}{\omega_3}\sqrt{\frac{\omega_2}{\omega_1}}T^a_{\alpha\bar\bt}\langle n\bn\rangle\cB^a_{\bn-}\big\{\cP_\perp^{-}(J^\dagger)^{\balpha\bt}_{n\bn 0}\big\}H\,.
\end{align}
The Feynman rule for the operator in \eq{O2Pchi2} is
\begin{align}
\fd{2.5cm}{figures/matching_scetPperp2.pdf}=-\frac{gT^a}{\omega_1\omega_3}\left(\gamma^\mu_\perp-\frac{\Sl{p}_{3\perp}n^\mu}{\omega_3}\right)\Sl{p}_{1\perp}-\frac{gT^a}{\omega_1}\left(\frac{1}{\omega_3}+\frac{1}{\omega_2+\omega_3}\right)\Sl{p}_{1\perp}\left(\gamma^\mu_\perp-\frac{\Sl{p}_{3\perp}n^\mu}{\omega_3}\right)\,.
\end{align}

\subsection{Helicity flip operators}

We now consider the operator basis when we allow quark mass insertions. The quark mass term in the Lagrangian looks like $-m(\psi_L^\dagger\psi_R+\psi_R^\dagger\psi_L)$ and it couples quarks of different chirality. Of course the Yukawa coupling \Eq{eq:yukawaint} between the Higgs and the quark-antiquark pair is also proportional to the quark mass. Therefore the basis presented in \Sec{sec:basis} and the Wilson coefficients computed in \Sec{sec:matching} should be thought of as the results at the first non-trivial order in the mass expansion. In this appendix we consider how the basis is extended when working to second order in the mass expansion, which allows an additional quark helicity flip.

For collinear particles with energy that is much larger than their mass, we can relate the expansion in mass to the expansion in the power counting parameter $\lambda$ as follows. For a collinear field we have $p^2 = p^+ p^- + p_\perp^2 \sim\lambda^2$, and by imposing the on-shell condition $p^2=m^2$, so we see that the mass $m$ scales as $\lambda$. Therefore, for collinear quark fields, each helicity flip is $\sim\lambda$ in the power counting, and we relax the constraint of the chiral conserving gluon interaction in one location. For each helicity flip, there is an additional factor of the mass $m$.  From these considerations we can then construct the operator basis involving helicity flips. 

At subleading power $\cO(\lambda)$, we can have two collinear fields and a helicity flip. The possible outcomes are quark-antiquark pair and gluon-gluon pair. However, for the quark-antiquark pair case, the two quarks have the same helicity after one helicity flip, so the current does not have spin 0 and is thus ruled out by conservation of angular momentum. Therefore, the operator basis only has the gluons
\begin{align}
 \boldsymbol{g_n \bar g_{\bn}:}   {\vcenter{\includegraphics[width=0.18\columnwidth]{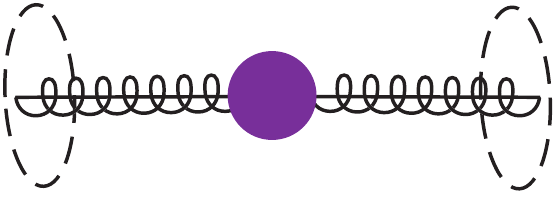}}} \nn
\end{align}
\vspace{-0.4cm}
\begin{alignat}{2}\label{eq:Hgg_flip}
 &O_{m\cB++}^{(1)ab}
=\cB^a_{n+}\cB^b_{\bn+}\,H
\,, \qquad &&
O_{m\cB--}^{(1)a}
=\cB^a_{n-}\cB^b_{\bn-}\,H\,,
\end{alignat}
with the color sturcture
\begin{align} \label{eq:subleading_color_flip}
 \vT_{\BPS}^{ab} = \bigl(\cY_{n}^T \cY_{\bn} \bigr)^{ab}\,.
\end{align}
In the full theory diagram, the two gluons are produced from the fermionic loop formed by the quark fields $b\bar b$ with the same chirality. The one-loop matching calculation for this operator will not be considered here.

At the subsubleading power $\cO(\lambda^2)$, the possible operator can have three collinear fields and one helicity flip. The operators with the outcoming particles being three gluons are again ruled out due to the fact that they cannot form a combination of zero helicity. Therefore the only possible situation is when the outcoming particles are $b\bar bg$, and $b$ and $\bar b$ have the same chirality. For the case that the gluon is in the $n$-collinear sector, the operator basis is given by
\begin{align}
\boldsymbol{(bg)_n \bar b_{\bn}:}   {\vcenter{\includegraphics[width=0.18\columnwidth]{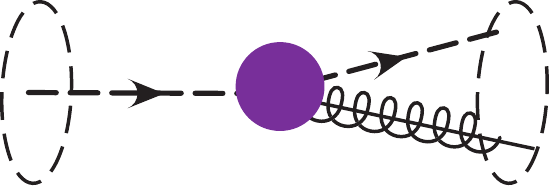}}} \nn
\end{align}
\vspace{-0.4cm}
\begin{alignat}{2}\label{eq:Hbbg_flip}
 &O_{m\cB n+(-)}^{(2)a\balpha\bt}
=\cB^a_{n+}J^{\balpha\bt}_{n\bn-}\,H
\,, \qquad 
&& O_{m\cB n-(+)}^{(2)a\balpha\bt}
=\cB^a_{n-}J^{\balpha\bt}_{n\bn+}\,H\,,
\end{alignat}
and for the case that the gluon is in the $\bn$-collinear sector, the operator basis is
\begin{align}
\boldsymbol{(b)_n (\bar bg)_{\bn}:}   {\vcenter{\includegraphics[width=0.18\columnwidth]{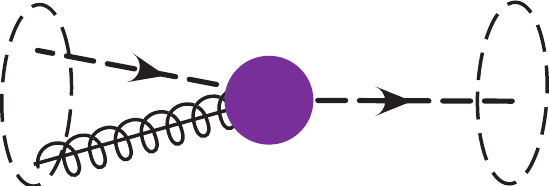}}} \nn
\end{align}
\vspace{-0.4cm}
\begin{alignat}{2}\label{eq:Hbbg_flip_2}
 &O_{m\cB\bn+(+)}^{(2)a\balpha\bt}
=\cB^a_{\bn+}J^{\balpha\bt}_{n\bn+}\,H
\,, \qquad 
&&O_{m\cB\bn-(-)}^{(2)a\balpha\bt}
=\cB^a_{\bn-}J^{\balpha\bt}_{n\bn-}\,H\,.
\end{alignat}
The color structure for the above operators are
\begin{align} \label{eq:subleading_color_flip}
&\vT_{\BPS}^{a\alpha\bar \bt} = \bigl(T^aY_{n}^\dagger Y_{\bn} \bigr)^{\alpha\bar\bt}\,, \qquad \bigl(Y_{n}^\dagger Y_{\bn}T^a \bigr)^{\alpha\bar\bt}\,,
\end{align}
for \eqs{Hbbg_flip}{Hbbg_flip_2} respectively.

In this case, the matching can be done at tree level. We take the full theory QCD diagram with $b\bar bg$ being the external state, which is given in \eq{QCD_bbg}, and rewrite the amplitude with bottom quark mass $m\neq0$:
\begin{align}\label{eq:QCD_bbg_flip}
\fd{2.5cm}{figures/matching_subleadingvertex_2_low.pdf} &= \bar u(p_1)(igT^a\Sl{\epsilon}_{3}^{*})\frac{i(\Sl{p}_1+\Sl{p}_3+m)}{(p_1+p_3)^2+m^2} v(p_2)\,, \nn \\
\fd{2.5cm}{figures/matching_subleadingvertex_1_low.pdf} &= \bar u(p_1)\frac{-i(\Sl{p}_2+\Sl{p}_3-m)}{(p_2+p_3)^2+m^2}(igT^a\Sl{\epsilon}_{3}^{*})v(p_2)\,.
\end{align}
Then we take the kinematics to be
\begin{align}
p_1^\mu=\omega_1\frac{n^\mu}{2}+p_{1r}\frac{\bn^\mu}{2}, \qquad p_2^\mu=\omega_2\frac{\bn^\mu}{2}+p_{2r}\frac{n^\mu}{2}, \qquad p_3^\mu=\omega_3\frac{n^\mu}{2}\,,
\end{align}
where $\omega_1p_{1r}=\omega_2p_{2r}=m^2$. Expanding the diagrams to $\cO(\lambda^2)$, we obtain
\begin{align}\label{eq:QCD_bbg_flip}
\left.\fd{2.5cm}{figures/matching_subleadingvertex_2_low.pdf}\right|_{\cO(\lambda^2)} &=0\,, \nn \\
\left.\fd{2.5cm}{figures/matching_subleadingvertex_1_low.pdf}\right|_{\cO(\lambda^2)} &= -m\frac{gT^a}{\omega_2\omega_3}\bar u_n(1)\Sl{\epsilon}_{3\perp}^*v_\bn(2)\,,
\end{align}
which implies that the hard scattering operator is given by
\begin{align} \label{eq:O2mBn}
\cO^{(2)}_{m\cB n}=-m\frac{g}{\omega_2\omega_3}\bar\chi_{n,\omega_1}\Sl{\cB}_{n\perp,\omega_3}\chi_{\bn,-\omega_2}\,,
\end{align}
and the helicity operators are 
\begin{align}
\cO^{(2)}_{m\cB n+(-)}&=m\frac{g}{\omega_3}\sqrt{\frac{\omega_1}{2\omega_2}}T^a_{\alpha\bar\bt}[n\bn]\cB_{n+}^aJ^{\balpha\bt}_{n\bn-}\,, \nn \\
\cO^{(2)}_{m\cB n-(+)}&=-m\frac{g}{\omega_3}\sqrt{\frac{\omega_1}{2\omega_2}}T^a_{\alpha\bar\bt}\langle n\bn\rangle\cB_{n-}^aJ^{\balpha\bt}_{n\bn+}\,.
\end{align}
As expected, the Wilson coefficient is proportional to the bottom quark mass $m$ due to the helicity flip. The Feynman rule of the operator in \eq{O2mBn} is given by
\begin{align}
\fd{2.5cm}{figures/matching_scetPperp.pdf}=-m\frac{gT^a}{\omega_2\omega_3}\gamma^\mu_\perp\,.
\end{align}

The Wilson coefficient for operator with the gluon being in the $\bn$-collinear sector can be obtained by charge conjugation or a similar calculation, and the hard scattering operator is 
\begin{align} \label{eq:O2mBnb}
\cO^{(2)}_{m\cB \bn}=-m\frac{g}{\omega_1\omega_3}\bar\chi_{n,\omega_1}\Sl{\cB}_{\bn\perp,\omega_3}\chi_{\bn,-\omega_2}\,,
\end{align}
or in terms of the helicity operators
\begin{align}
\cO^{(2)}_{m\cB n+(+)}&=-m\frac{g}{\omega_3}\sqrt{\frac{\omega_2}{2\omega_1}}T^a_{\alpha\bar\bt}\langle n\bn\rangle\cB_{\bn+}^aJ^{\balpha\bt}_{n\bn+}\, \nn \\
\cO^{(2)}_{m\cB n-(-)}&=m\frac{g}{\omega_3}\sqrt{\frac{\omega_2}{2\omega_1}}T^a_{\alpha\bar\bt}[n\bn]\cB_{\bn-}^aJ^{\balpha\bt}_{n\bn-}\,,
\end{align}
and the Feynman rule for \eq{O2mBnb} is
\begin{align}
\fd{2.5cm}{figures/matching_scetPperp2.pdf}=-m\frac{gT^a}{\omega_1\omega_3}\gamma^\mu_\perp\,.
\end{align}

\section{Projection onto Helicities}\label{app:helicity}

In \Sec{sec:matching} we have given the result for the Wilson coefficients for the operators in SCET and their projections onto the helicity building blocks of \Sec{sec:review}. In this appendix we want to give a detailed example of how to get the projection of an operator onto helicities and how to generalize the result in the case of generic $n^\mu$ and $\bn^\mu$ axis.

First of all let's note that throughout the text, to slightly simplify the expressions, we chose to define our spinors with respect to the jet axis $n$, which we take to be in the $z$ direction, $n^\mu = (1, 0, 0, 1)$. In this case we could simplify the spinor products between $n$ and $\bn$ using
\be 
	[n \bn] = -2 \,, \qquad  \langle n\bn\rangle = 2 \,,
\ee
though we have often left these factors explicit. 

To give a detailed example of how to do the projection of a subleading hard scattering operator onto helicities we choose the projection of the sub-subleading operator $\cO^{(2)}_{\cP\chi1}$ defined in \eq{Hbbgpperp_basis1}. The following are the steps of how to get from \eq{Pqgq_match} to \eq{Pqgq_hel_match}. First of all, we insert an identity matrix $1=\left(\frac{1+\gamma_5}{2}\right)^2+\left(\frac{1-\gamma_5}{2}\right)^2$ in the spinor product
\begin{align}
\cO^{(2)}_{\cP\chi1}&=\frac{g}{\omega_2}\left(\frac{1}{\omega_1}+\frac{1}{\omega_3}\right)\bar \chi_{n,\omega_1}\left(\left(\frac{1+\gamma_5}{2}\right)^2+\left(\frac{1-\gamma_5}{2}\right)^2\right)[\Sl{\cP}_{\perp}\Sl{\cB}_{n\perp,\omega_3}]\chi_{\bn,-\omega_2}H\,\nn \\
&=\frac{g}{\omega_2}\left(\frac{1}{\omega_1}+\frac{1}{\omega_3}\right)\left(\bar \chi_{n+}[\Sl{\cP}_{\perp}\Sl{\cB}_{n\perp,\omega_3}]\chi_{\bn,-}+\bar \chi_{n-}[\Sl{\cP}_{\perp}\Sl{\cB}_{n\perp,\omega_3}]\chi_{\bn,+}\right)H\,,
\end{align}
where we use the definition of the helicity quark fields in \eq{quarkhel_def} and $\{\gamma_\mu,\gamma_5\}=0$.
Secondly, note that the gluon polarization vectors satisfy the following identity:
\begin{align}
g_{\mu\nu}=\frac{n_\mu\bn_\nu+n_\nu\bn_\mu}{2}-(\epsilon_\mu^+(n,\bn)\epsilon_\nu^-(n,\bn)+\epsilon_\mu^-(n,\bn)\epsilon_\nu^+(n,\bn))\,.
\end{align}
With \eqs{cBpm_def}{Pperppm} and the above identity, the term $\Sl{\cP}_{\perp}\Sl{\cB}_{n\perp,\omega_3}$ can be further simplified to
\begin{align}
\Sl{\cP}_{\perp}\Sl{\cB}_{n\perp,\omega_3}=(\Sl{\epsilon}^+\cP^+_\perp+\Sl{\epsilon}^-\cP^-_\perp)(\Sl{\epsilon}^+\cB_{n+}+\Sl{\epsilon}^-\cB_{n-})\,.
\end{align}
Simply plugging in the above expression of $\Sl{\cP}_{\perp}\Sl{\cB}_{n\perp,\omega_3}$ one will obtain 8 different helicity operators. However, the gluon polarization vectors also satisfy
\begin{align}
\Sl{\epsilon}_{-}(n,\bn)|n\rangle=\Sl{\epsilon}_{+}(n,\bn)|n]=0, \qquad [n|\Sl{\epsilon}_{+}(n,\bn)=\langle n|\Sl{\epsilon}_{-}(n,\bn)=0\,.
\end{align}
Therefore, only 2 of the 8 helicity operators are nonzero (i.e., satisfy the helicity constraint), and we obtain
\begin{align}
\cO^{(2)}_{\cP\chi1}&=-\frac{2g}{\omega_2}\left(\frac{1}{\omega_1}+\frac{1}{\omega_3}\right)(\bar\chi_{n+}[\cP^-_\perp\cB_{n+}]\chi_{\bn-}+\bar\chi_{n-}[\cP^+_\perp\cB_{n-}]\chi_{\bn+})\,.
\end{align}
Using the definition of the helicity currents in \eqs{jpm_back_to_bacjdef}{coll_subl} and integration by parts, we then obtain the two helicity operators given in \eq{Pqgq_hel_match}.

\section{Matching Calculation with Longitudinal Polarizations}\label{app:polarization_long}

In \Sec{sec:matching_bbgg}, we only kept the perpendicular component of the polarization vectors when doing the matching for the two quark two gluon operators $\cO_{\cB2}^{(2)}$ and $\cO_{\cB3}^{(2)}$. Here we will give the full matching calculation of $\cO_{\cB2}^{(2)}$ that keeps also the longitudinal polarization vectors. Due to gauge invariance and the collinear gluon field expansion \eq{gluon_expansion}, we expect to find the matrix element of $\cO_{\cB2}^{(2)}$ which using \eq{matching_bbgg_operator} as input, should be
\begin{align}\label{eq:matching_bbgg_final_long}
-\frac{g^2T^aT^b}{\omega_2}\left(\frac{1}{\omega_1+\omega_3}+\frac{1}{\omega_3+\omega_4}\right)\bar u_n(1)\left(\Sl{\epsilon}_{3\perp}^*-\frac{\bn\cdot\epsilon_3^*\Sl{p}_{\perp}}{\omega_3}\right)\left(\Sl{\epsilon}_{4\perp}^*+\frac{\bn\cdot\epsilon_4^*\Sl{p}_{\perp}}{\omega_4}\right)v_\bn(2)\,\nn \\
+((3,a)\leftrightarrow(4,b))\,,
\end{align}
where the kinematics are taken to be as in \eq{kinematics_bbgg}. Here we show that we can recover this result for the $\cO_{\cB2}^{(2)}$ matrix element by expanding the full QCD diagrams and subtracting the contributions from other hard scattering operators. 

First of all, the QCD diagrams with nonzero $\cO(\lambda^2)$ terms are
\begin{align}\label{eq:matching_bbgg_QCD_long}
&\left.\left(\fd{2.5cm}{figures/matching_2q2g_diagram2_low.pdf}+\fd{2.5cm}{figures/matching_2q2g_diagram1_low.pdf}\right)\right|_{\cO(\lambda^2)}\, \nn \\
&=\frac{-g^2T^aT^b}{\omega_2(\omega_3+\omega_4)}\bar u_n(1)\left(\Sl{\epsilon}^*_{3\perp}+\frac{\bn\cdot\epsilon^*_3\Sl{p}_\perp}{\omega_4}\right)\left(\Sl{\epsilon}^*_{4\perp}+\frac{\bn\cdot\epsilon^*_4\Sl{p}_\perp}{\omega_4}\right)v_\bn(2)+\left((3,a)\leftrightarrow(4,b)\right)\, \nn \\
&\left.\left(\fd{2.5cm}{figures/matching_2q2g_diagram4_low.pdf}+\fd{2.5cm}{figures/matching_2q2g_diagram3_low.pdf}\right)\right|_{\cO(\lambda^2)}\, \nn \\
&=\frac{-g^2T^aT^b}{\omega_1\omega_2\omega_4}(\omega_3+\omega_4)\bar u_n(1)\Sl{\epsilon}^*_{3\perp}\left(\Sl{\epsilon}^*_{4\perp}+\frac{\bn\cdot\epsilon^*_4\Sl{p}_\perp}{\omega_4}\right)v_\bn(2)\, \nn \\
&+\frac{g^2T^aT^b}{\omega_1\omega_2\omega_4p_3^r}(\omega_1+\omega_3+\omega_4)\bar u_n(1)n\cdot\epsilon^*_3\Sl{p}_\perp\left(\Sl{\epsilon}^*_{4\perp}+\frac{\bn\cdot\epsilon^*_4\Sl{p}_\perp}{\omega_4}\right)v_\bn(2)+\left((3,a)\leftrightarrow(4,b)\right)\,,
\end{align}
where now the longitudinal polarizations are also included. We see that there is an additional nonlocal term due to the longitudinal polarizations. Also, in the $\cO(\lambda^2)$ term of the first two diagrams, the expression $\left(\Sl{\epsilon}^*_{3\perp}+\frac{\bn\cdot\epsilon^*_3\Sl{p}_\perp}{\omega_4}\right)$ is not the desired form. This is due to the fact that the $\cO(\lambda^2)$ terms of the QCD diagrams also contain contributions from the hard scattering operator $\cO^{(2)}_{\cP\chi1}$, whose Feynman rule is given by \eq{Feynmanrule_Pperp}. Therefore, this operator can contribute to the $bbgg$ QCD diagrams through a leading power SCET Lagrangian insertion that produces an additional collinear gluon. The SCET diagrams and the corresponding amplitudes are
\begin{align}\label{eq:matching_bbgg_SCET_long}
&\left.\left(\fd{2.5cm}{figures/matching_bbgg_SCET_contribution.pdf}+\fd{2.5cm}{figures/matching_bbgg_SCET_contribution2.pdf}\right)\right|_{\cO(\lambda^2)}\, \nn \\
&=\frac{-g^2T^aT^b\omega_3(\omega_1+\omega_3+\omega_4)}{\omega_1\omega_2\omega_4(\omega_1+\omega_3)}\bar u_n(1)\Sl{\epsilon}^*_{3\perp}\Sl{\epsilon}^*_{4\perp}v_\bn(2)\, \nn \\
&+\frac{g^2T^aT^b}{\omega_1\omega_2\omega_4p_3^r}(\omega_1+\omega_3+\omega_4)\bar u_n(1)n\cdot\epsilon^*_3\Sl{p}_\perp\left(\Sl{\epsilon}^*_{4\perp}+\frac{\bn\cdot\epsilon^*_4\Sl{p}_\perp}{\omega_4}\right)v_\bn(2)+\left((3,a)\leftrightarrow(4,b)\right)\,,
\end{align}
and the non-abelian diagram involving a three collinear gluon vertex has no contribution due to our choice of kinematics. One can see that the nonlocal terms of the QCD diagrams are exactly from the operator $\cO^{(2)}_{\cP\chi1}$ and a SCET Lagrangian insertion. Also, after subtracting \eq{matching_bbgg_SCET_long} from \eq{matching_bbgg_QCD_long}, we obtain
\begin{align}\label{eq:matching_bbgg_intermediate_long}
&-\frac{g^2T^aT^b}{\omega_2}\left(\frac{1}{\omega_1+\omega_3}+\frac{1}{\omega_3+\omega_4}\right)\bar u_n(1)\Sl{\epsilon}_{3\perp}^*\left(\Sl{\epsilon}_{4\perp}^*+\frac{\bn\cdot\epsilon_4^*\Sl{p}_{\perp}}{\omega_4}\right)v_\bn(2)\, \nn \\
&-\frac{g^2T^aT^b}{\omega_2\omega_4(\omega_3+\omega_4)}\bar u_n(1)\bn\cdot\epsilon^*_3\Sl{p}_\perp\left(\Sl{\epsilon}_{4\perp}^*+\frac{\bn\cdot\epsilon_4^*\Sl{p}_{\perp}}{\omega_4}\right)v_\bn(2)+((3,a)\leftrightarrow(4,b))\,,
\end{align}
One can see that the coefficient for the perpendicular component of the polarization vectors is correct, as given in \eq{matching_bbgg_final_long}. However, the longitudinal components do not yet have the correct coefficient.

This is because, besides the leading power Lagrangian insertion, the additional longitudinal polarized collinear gluon can also be produced from the Wilson line attached to the collinear quark field. From the definition of the collinear quark field and the collinear Wilson line \eqs{chiB}{Wn}, we can see that the collinear quark field  has the expansion
\begin{align}
\bar\chi_{n,\omega}=\bar\xi_{n,\omega}-\frac{g}{\bn\cdot k}\bar\xi_{n,\omega-\bn\cdot k}\bn\cdot A_{nk}^aT^a+...\,,
\end{align}
where the dots denote terms with multiple gluon fields, and $k$ is the momentum of the additional gluon field. Therefore, starting from the original expression \eq{Pqgq_match}, we can expand the expression of the operator $\cO^{(2)}_{\cP\chi1}$ in the following way:
\begin{align}
\cO^{(2)}_{\cP\chi1}=\frac{g}{\omega_2}\left(\frac{1}{\omega_1}+\frac{1}{\omega_3}\right)\left(\bar\xi_{n,\omega_1}-\frac{g}{\bn\cdot k}\bar\xi_{n,\omega_1-\bn\cdot k}\bn\cdot A_{nk}^aT^a+...\right)[\Sl{\cP}_\perp\Sl{\cB}_{\bn\perp,\omega_3}]\chi_{\bn,-\omega_2}\,.
\end{align}
The second term in the parenthesis then gives a matrix element of $bbgg$ external state at power $\cO(\lambda^2)$. Using our kinematics $(\omega_1\rightarrow\omega_1+\omega_3,\omega_2\rightarrow\omega_2,\omega_3\rightarrow\omega_4,k\rightarrow p_3)$, the SCET diagram and the amplitude is given by
\begin{align}
&\left.\left(\fd{2.5cm}{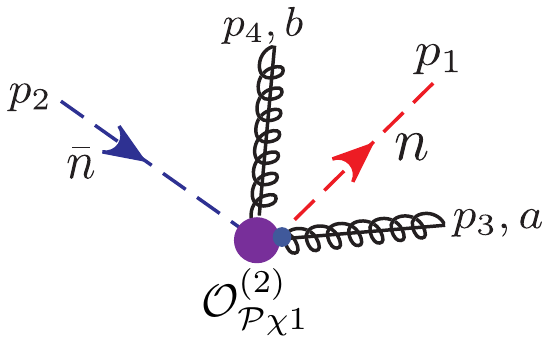}+\fd{2.5cm}{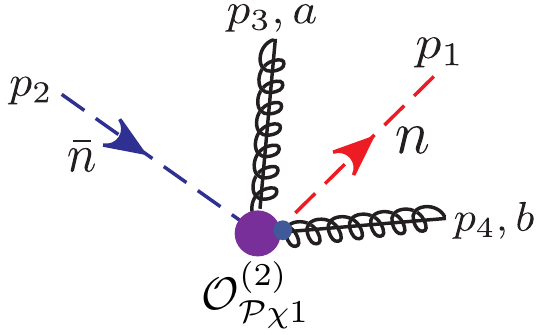}\right)\right|_{\cO(\lambda^2)}\, \nn \\
&=\frac{g^2T^aT^b}{\omega_2\omega_3}\left(\frac{1}{\omega_1+\omega_3}+\frac{1}{\omega_4}\right)\bar u_n(1)\bn\cdot\epsilon_3^*\Sl{p}_\perp\left(\Sl{\epsilon}^*_{4\perp}+\frac{\bn\cdot\epsilon^*_4\Sl{p}_\perp}{\omega_4}\right)v_\bn(2)+\left((3,a)\leftrightarrow(4,b)\right)\,
\end{align}
where the small blue dots on the diagrams denote the collinear gluon emission from the Wilson lines. Finally, we subtract this amplitude from \eq{matching_bbgg_intermediate_long}, and the result is then exactly the same as \eq{matching_bbgg_final_long} as expected. Therefore, by carefully considering all the relevant full theory diagrams and EFT diagrams, we have explicitly carried out the matching calculation with all polarization directions for the hard scattering operator $\cO^{(2)}_{\cB2}$.

\bibliography{bibliography}{}

\providecommand{\href}[2]{#2}\begingroup\raggedright\begin{thebibliography}{10}

\bibitem{Collins:1989gx}
J.~C. Collins, D.~E. Soper, and G.~F. Sterman, {\it {Factorization of Hard
  Processes in QCD}},  {\em Adv. Ser. Direct. High Energy Phys.} {\bf 5} (1989)
  1--91, [\href{http://arxiv.org/abs/hep-ph/0409313}{{\tt hep-ph/0409313}}].

\bibitem{Bauer:2000ew}
C.~W. Bauer, S.~Fleming, and M.~E. Luke, {\it {Summing Sudakov logarithms in $B
  \to X_s \gamma$ in effective field theory}},  {\em Phys. Rev.} {\bf D63}
  (2000) 014006, [\href{http://arxiv.org/abs/hep-ph/0005275}{{\tt
  hep-ph/0005275}}].

\bibitem{Bauer:2000yr}
C.~W. Bauer, S.~Fleming, D.~Pirjol, and I.~W. Stewart, {\it {An Effective field
  theory for collinear and soft gluons: Heavy to light decays}},  {\em Phys.
  Rev.} {\bf D63} (2001) 114020,
  [\href{http://arxiv.org/abs/hep-ph/0011336}{{\tt hep-ph/0011336}}].

\bibitem{Bauer:2001ct}
C.~W. Bauer and I.~W. Stewart, {\it {Invariant operators in collinear effective
  theory}},  {\em Phys. Lett.} {\bf B516} (2001) 134--142,
  [\href{http://arxiv.org/abs/hep-ph/0107001}{{\tt hep-ph/0107001}}].

\bibitem{Bauer:2001yt}
C.~W. Bauer, D.~Pirjol, and I.~W. Stewart, {\it {Soft collinear factorization
  in effective field theory}},  {\em Phys. Rev.} {\bf D65} (2002) 054022,
  [\href{http://arxiv.org/abs/hep-ph/0109045}{{\tt hep-ph/0109045}}].

\bibitem{Bauer:2002nz}
C.~W. Bauer, S.~Fleming, D.~Pirjol, I.~Z. Rothstein, and I.~W. Stewart, {\it
  {Hard scattering factorization from effective field theory}},  {\em Phys.
  Rev.} {\bf D66} (2002) 014017,
  [\href{http://arxiv.org/abs/hep-ph/0202088}{{\tt hep-ph/0202088}}].

\bibitem{Larkoski:2014bxa}
A.~J. Larkoski, D.~Neill, and I.~W. Stewart, {\it {Soft Theorems from Effective
  Field Theory}},  {\em JHEP} {\bf 06} (2015) 077,
  [\href{http://arxiv.org/abs/1412.3108}{{\tt arXiv:1412.3108}}].

\bibitem{Lee:2004ja}
K.~S.~M. Lee and I.~W. Stewart, {\it {Factorization for power corrections to $B
  \to X_s \gamma$ and $B \to X_u \ell \bar\nu_\ell$}},  {\em Nucl. Phys.} {\bf
  B721} (2005) 325--406, [\href{http://arxiv.org/abs/hep-ph/0409045}{{\tt
  hep-ph/0409045}}].

\bibitem{Beneke:2004in}
M.~Beneke, F.~Campanario, T.~Mannel, and B.~D. Pecjak, {\it {Power corrections
  to $\bar B \to X_u \ell \bar\nu$ ($X_s\gamma$) decay spectra in the
  'shape-function' region}},  {\em JHEP} {\bf 06} (2005) 071,
  [\href{http://arxiv.org/abs/hep-ph/0411395}{{\tt hep-ph/0411395}}].

\bibitem{Hill:2004if}
R.~J. Hill, T.~Becher, S.~J. Lee, and M.~Neubert, {\it {Sudakov resummation for
  subleading SCET currents and heavy-to-light form-factors}},  {\em JHEP} {\bf
  07} (2004) 081, [\href{http://arxiv.org/abs/hep-ph/0404217}{{\tt
  hep-ph/0404217}}].

\bibitem{Bosch:2004cb}
S.~W. Bosch, M.~Neubert, and G.~Paz, {\it {Subleading shape functions in
  inclusive B decays}},  {\em JHEP} {\bf 11} (2004) 073,
  [\href{http://arxiv.org/abs/hep-ph/0409115}{{\tt hep-ph/0409115}}].

\bibitem{Beneke:2004rc}
M.~Beneke, Y.~Kiyo, and D.~s. Yang, {\it {Loop corrections to subleading heavy
  quark currents in SCET}},  {\em Nucl. Phys.} {\bf B692} (2004) 232--248,
  [\href{http://arxiv.org/abs/hep-ph/0402241}{{\tt hep-ph/0402241}}].

\bibitem{Paz:2009ut}
G.~Paz, {\it {Subleading Jet Functions in Inclusive B Decays}},  {\em JHEP}
  {\bf 06} (2009) 083, [\href{http://arxiv.org/abs/0903.3377}{{\tt
  arXiv:0903.3377}}].

\bibitem{Benzke:2010js}
M.~Benzke, S.~J. Lee, M.~Neubert, and G.~Paz, {\it {Factorization at Subleading
  Power and Irreducible Uncertainties in $\bar B\to X_s\gamma$ Decay}},  {\em
  JHEP} {\bf 08} (2010) 099, [\href{http://arxiv.org/abs/1003.5012}{{\tt
  arXiv:1003.5012}}].

\bibitem{Freedman:2013vya}
S.~M. Freedman, {\it {Subleading Corrections To Thrust Using Effective Field
  Theory}},  \href{http://arxiv.org/abs/1303.1558}{{\tt arXiv:1303.1558}}.

\bibitem{Freedman:2014uta}
S.~M. Freedman and R.~Goerke, {\it {Renormalization of Subleading Dijet
  Operators in Soft-Collinear Effective Theory}},  {\em Phys. Rev.} {\bf D90}
  (2014), no.~11 114010, [\href{http://arxiv.org/abs/1408.6240}{{\tt
  arXiv:1408.6240}}].

\bibitem{Kolodrubetz:2016uim}
D.~W. Kolodrubetz, I.~Moult, and I.~W. Stewart, {\it {Building Blocks for
  Subleading Helicity Operators}},  {\em JHEP} {\bf 05} (2016) 139,
  [\href{http://arxiv.org/abs/1601.02607}{{\tt arXiv:1601.02607}}].

\bibitem{Moult:2016fqy}
I.~Moult, L.~Rothen, I.~W. Stewart, F.~J. Tackmann, and H.~X. Zhu, {\it
  {Subleading Power Corrections for N-Jettiness Subtractions}},
  \href{http://arxiv.org/abs/1612.00450}{{\tt arXiv:1612.00450}}.

\bibitem{Feige:2017zci}
I.~Feige, D.~W. Kolodrubetz, I.~Moult, and I.~W. Stewart, {\it {A Complete
  Basis of Helicity Operators for Subleading Factorization}},
  \href{http://arxiv.org/abs/1703.03411}{{\tt arXiv:1703.03411}}.

\bibitem{Goerke:2017lei}
R.~Goerke and M.~Inglis-Whalen, {\it {Renormalization of Dijet Operators at
  Order $1/Q^2$ in Soft-Collinear Effective Theory}},
  \href{http://arxiv.org/abs/1711.09147}{{\tt arXiv:1711.09147}}.

\bibitem{Moult:2017rpl}
I.~Moult, I.~W. Stewart, and G.~Vita, {\it {A Subleading Operator Basis and
  Matching for $gg \to H$}},  \href{http://arxiv.org/abs/1703.03408}{{\tt
  arXiv:1703.03408}}.

\bibitem{Laenen:2008gt}
E.~Laenen, G.~Stavenga, and C.~D. White, {\it {Path integral approach to
  eikonal and next-to-eikonal exponentiation}},  {\em JHEP} {\bf 03} (2009)
  054, [\href{http://arxiv.org/abs/0811.2067}{{\tt arXiv:0811.2067}}].

\bibitem{Laenen:2008ux}
E.~Laenen, L.~Magnea, and G.~Stavenga, {\it {On next-to-eikonal corrections to
  threshold resummation for the Drell-Yan and DIS cross sections}},  {\em Phys.
  Lett.} {\bf B669} (2008) 173--179,
  [\href{http://arxiv.org/abs/0807.4412}{{\tt arXiv:0807.4412}}].

\bibitem{Bonocore:2015esa}
D.~Bonocore, E.~Laenen, L.~Magnea, S.~Melville, L.~Vernazza, and C.~D. White,
  {\it {A factorization approach to next-to-leading-power threshold
  logarithms}},  {\em JHEP} {\bf 06} (2015) 008,
  [\href{http://arxiv.org/abs/1503.05156}{{\tt arXiv:1503.05156}}].

\bibitem{Moult:2015aoa}
I.~Moult, I.~W. Stewart, F.~J. Tackmann, and W.~J. Waalewijn, {\it {Employing
  Helicity Amplitudes for Resummation}},  {\em Phys. Rev.} {\bf D93} (2016),
  no.~9 094003, [\href{http://arxiv.org/abs/1508.02397}{{\tt
  arXiv:1508.02397}}].

\bibitem{Stewart:2009yx}
I.~W. Stewart, F.~J. Tackmann, and W.~J. Waalewijn, {\it {Factorization at the
  LHC: From PDFs to Initial State Jets}},  {\em Phys. Rev.} {\bf D81} (2010)
  094035, [\href{http://arxiv.org/abs/0910.0467}{{\tt arXiv:0910.0467}}].

\bibitem{Stewart:2010tn}
I.~W. Stewart, F.~J. Tackmann, and W.~J. Waalewijn, {\it {N-Jettiness: An
  Inclusive Event Shape to Veto Jets}},  {\em Phys. Rev. Lett.} {\bf 105}
  (2010) 092002, [\href{http://arxiv.org/abs/1004.2489}{{\tt
  arXiv:1004.2489}}].

\bibitem{Gaunt:2014ska}
J.~R. Gaunt, {\it {Glauber Gluons and Multiple Parton Interactions}},  {\em
  JHEP} {\bf 07} (2014) 110, [\href{http://arxiv.org/abs/1405.2080}{{\tt
  arXiv:1405.2080}}].

\bibitem{Rothstein:2016bsq}
I.~Z. Rothstein and I.~W. Stewart, {\it {An Effective Field Theory for Forward
  Scattering and Factorization Violation}},  {\em JHEP} {\bf 08} (2016) 025,
  [\href{http://arxiv.org/abs/1601.04695}{{\tt arXiv:1601.04695}}].

\bibitem{Zeng:2015iba}
M.~Zeng, {\it {Drell-Yan process with jet vetoes: breaking of generalized
  factorization}},  {\em JHEP} {\bf 10} (2015) 189,
  [\href{http://arxiv.org/abs/1507.01652}{{\tt arXiv:1507.01652}}].

\bibitem{Fleming:2006cd}
S.~Fleming, A.~K. Leibovich, and T.~Mehen, {\it {Resummation of Large Endpoint
  Corrections to Color-Octet $J/\psi$ Photoproduction}},  {\em Phys. Rev.} {\bf
  D74} (2006) 114004, [\href{http://arxiv.org/abs/hep-ph/0607121}{{\tt
  hep-ph/0607121}}].

\bibitem{Manohar:2002fd}
A.~V. Manohar, T.~Mehen, D.~Pirjol, and I.~W. Stewart, {\it {Reparameterization
  invariance for collinear operators}},  {\em Phys. Lett.} {\bf B539} (2002)
  59--66, [\href{http://arxiv.org/abs/hep-ph/0204229}{{\tt hep-ph/0204229}}].

\bibitem{Chay:2002vy}
J.~Chay and C.~Kim, {\it {Collinear effective theory at subleading order and
  its application to heavy - light currents}},  {\em Phys. Rev.} {\bf D65}
  (2002) 114016, [\href{http://arxiv.org/abs/hep-ph/0201197}{{\tt
  hep-ph/0201197}}].

\bibitem{Beneke:2002ni}
M.~Beneke and T.~Feldmann, {\it {Multipole expanded soft collinear effective
  theory with nonAbelian gauge symmetry}},  {\em Phys. Lett.} {\bf B553} (2003)
  267--276, [\href{http://arxiv.org/abs/hep-ph/0211358}{{\tt hep-ph/0211358}}].

\bibitem{Beneke:2002ph}
M.~Beneke, A.~P. Chapovsky, M.~Diehl, and T.~Feldmann, {\it {Soft collinear
  effective theory and heavy to light currents beyond leading power}},  {\em
  Nucl. Phys.} {\bf B643} (2002) 431--476,
  [\href{http://arxiv.org/abs/hep-ph/0206152}{{\tt hep-ph/0206152}}].

\bibitem{Pirjol:2002km}
D.~Pirjol and I.~W. Stewart, {\it {A Complete basis for power suppressed
  collinear ultrasoft operators}},  {\em Phys. Rev.} {\bf D67} (2003) 094005,
  [\href{http://arxiv.org/abs/hep-ph/0211251}{{\tt hep-ph/0211251}}]. [Erratum:
  Phys. Rev.D69,019903(2004)].

\bibitem{Bauer:2003mga}
C.~W. Bauer, D.~Pirjol, and I.~W. Stewart, {\it {On Power suppressed operators
  and gauge invariance in SCET}},  {\em Phys. Rev.} {\bf D68} (2003) 034021,
  [\href{http://arxiv.org/abs/hep-ph/0303156}{{\tt hep-ph/0303156}}].

\bibitem{Catani:2007vq}
S.~Catani and M.~Grazzini, {\it {An NNLO subtraction formalism in hadron
  collisions and its application to Higgs boson production at the LHC}},  {\em
  Phys. Rev. Lett.} {\bf 98} (2007) 222002,
  [\href{http://arxiv.org/abs/hep-ph/0703012}{{\tt hep-ph/0703012}}].

\bibitem{Boughezal:2015aha}
R.~Boughezal, C.~Focke, W.~Giele, X.~Liu, and F.~Petriello, {\it {Higgs boson
  production in association with a jet at NNLO using jettiness subtraction}},
  {\em Phys. Lett.} {\bf B748} (2015) 5--8,
  [\href{http://arxiv.org/abs/1505.03893}{{\tt arXiv:1505.03893}}].

\bibitem{Gaunt:2015pea}
J.~Gaunt, M.~Stahlhofen, F.~J. Tackmann, and J.~R. Walsh, {\it {N-jettiness
  Subtractions for NNLO QCD Calculations}},  {\em JHEP} {\bf 09} (2015) 058,
  [\href{http://arxiv.org/abs/1505.04794}{{\tt arXiv:1505.04794}}].

\bibitem{Catani:2009sm}
S.~Catani, L.~Cieri, G.~Ferrera, D.~de~Florian, and M.~Grazzini, {\it {Vector
  boson production at hadron colliders: a fully exclusive QCD calculation at
  NNLO}},  {\em Phys. Rev. Lett.} {\bf 103} (2009) 082001,
  [\href{http://arxiv.org/abs/0903.2120}{{\tt arXiv:0903.2120}}].

\bibitem{Ferrera:2011bk}
G.~Ferrera, M.~Grazzini, and F.~Tramontano, {\it {Associated WH production at
  hadron colliders: a fully exclusive QCD calculation at NNLO}},  {\em Phys.
  Rev. Lett.} {\bf 107} (2011) 152003,
  [\href{http://arxiv.org/abs/1107.1164}{{\tt arXiv:1107.1164}}].

\bibitem{Catani:2011qz}
S.~Catani, L.~Cieri, D.~de~Florian, G.~Ferrera, and M.~Grazzini, {\it {Diphoton
  production at hadron colliders: a fully-differential QCD calculation at
  NNLO}},  {\em Phys. Rev. Lett.} {\bf 108} (2012) 072001,
  [\href{http://arxiv.org/abs/1110.2375}{{\tt arXiv:1110.2375}}]. [Erratum:
  Phys. Rev. Lett.117,no.8,089901(2016)].

\bibitem{Grazzini:2013bna}
M.~Grazzini, S.~Kallweit, D.~Rathlev, and A.~Torre, {\it {$Z\gamma$ production
  at hadron colliders in NNLO QCD}},  {\em Phys. Lett.} {\bf B731} (2014)
  204--207, [\href{http://arxiv.org/abs/1309.7000}{{\tt arXiv:1309.7000}}].

\bibitem{Cascioli:2014yka}
F.~Cascioli, T.~Gehrmann, M.~Grazzini, S.~Kallweit, P.~Maierh{\"o}fer, A.~von
  Manteuffel, S.~Pozzorini, D.~Rathlev, L.~Tancredi, and E.~Weihs, {\it {ZZ
  production at hadron colliders in NNLO QCD}},  {\em Phys. Lett.} {\bf B735}
  (2014) 311--313, [\href{http://arxiv.org/abs/1405.2219}{{\tt
  arXiv:1405.2219}}].

\bibitem{Ferrera:2014lca}
G.~Ferrera, M.~Grazzini, and F.~Tramontano, {\it {Associated ZH production at
  hadron colliders: the fully differential NNLO QCD calculation}},  {\em Phys.
  Lett.} {\bf B740} (2015) 51--55, [\href{http://arxiv.org/abs/1407.4747}{{\tt
  arXiv:1407.4747}}].

\bibitem{Gehrmann:2014fva}
T.~Gehrmann, M.~Grazzini, S.~Kallweit, P.~Maierh{\"o}fer, A.~von Manteuffel,
  S.~Pozzorini, D.~Rathlev, and L.~Tancredi, {\it {$W^+W^-$ Production at
  Hadron Colliders in Next to Next to Leading Order QCD}},  {\em Phys. Rev.
  Lett.} {\bf 113} (2014), no.~21 212001,
  [\href{http://arxiv.org/abs/1408.5243}{{\tt arXiv:1408.5243}}].

\bibitem{Grazzini:2015nwa}
M.~Grazzini, S.~Kallweit, and D.~Rathlev, {\it {$W\gamma$ and $Z\gamma$
  production at the LHC in NNLO QCD}},  {\em JHEP} {\bf 07} (2015) 085,
  [\href{http://arxiv.org/abs/1504.01330}{{\tt arXiv:1504.01330}}].

\bibitem{Grazzini:2015hta}
M.~Grazzini, S.~Kallweit, and D.~Rathlev, {\it {ZZ production at the LHC:
  fiducial cross sections and distributions in NNLO QCD}},  {\em Phys. Lett.}
  {\bf B750} (2015) 407--410, [\href{http://arxiv.org/abs/1507.06257}{{\tt
  arXiv:1507.06257}}].

\bibitem{Campbell:2016yrh}
J.~M. Campbell, R.~K. Ellis, Y.~Li, and C.~Williams, {\it {Predictions for
  diphoton production at the LHC through NNLO in QCD}},  {\em JHEP} {\bf 07}
  (2016) 148, [\href{http://arxiv.org/abs/1603.02663}{{\tt arXiv:1603.02663}}].

\bibitem{Boughezal:2016wmq}
R.~Boughezal, J.~M. Campbell, R.~K. Ellis, C.~Focke, W.~Giele, X.~Liu,
  F.~Petriello, and C.~Williams, {\it {Color Singlet Production at NNLO in
  MCFM}},  {\em Eur. Phys. J.} {\bf C77} (2017), no.~1 7,
  [\href{http://arxiv.org/abs/1605.08011}{{\tt arXiv:1605.08011}}].

\bibitem{Boughezal:2015dva}
R.~Boughezal, C.~Focke, X.~Liu, and F.~Petriello, {\it {$W$-boson production in
  association with a jet at next-to-next-to-leading order in perturbative
  QCD}},  {\em Phys. Rev. Lett.} {\bf 115} (2015), no.~6 062002,
  [\href{http://arxiv.org/abs/1504.02131}{{\tt arXiv:1504.02131}}].

\bibitem{Boughezal:2016isb}
R.~Boughezal, X.~Liu, and F.~Petriello, {\it {Phenomenology of the Z-boson plus
  jet process at NNLO}},  {\em Phys. Rev.} {\bf D94} (2016), no.~7 074015,
  [\href{http://arxiv.org/abs/1602.08140}{{\tt arXiv:1602.08140}}].

\bibitem{Boughezal:2016dtm}
R.~Boughezal, X.~Liu, and F.~Petriello, {\it {W-boson plus jet differential
  distributions at NNLO in QCD}},  {\em Phys. Rev.} {\bf D94} (2016), no.~11
  113009, [\href{http://arxiv.org/abs/1602.06965}{{\tt arXiv:1602.06965}}].

\bibitem{Campbell:2016lzl}
J.~M. Campbell, R.~K. Ellis, and C.~Williams, {\it {Direct photon production at
  next-to-next-to-leading order}},  \href{http://arxiv.org/abs/1612.04333}{{\tt
  arXiv:1612.04333}}.

\bibitem{Grazzini:2017mhc}
M.~Grazzini, S.~Kallweit, and M.~Wiesemann, {\it {Fully differential NNLO
  computations with MATRIX}},  \href{http://arxiv.org/abs/1711.06631}{{\tt
  arXiv:1711.06631}}.

\bibitem{Boughezal:2016zws}
R.~Boughezal, X.~Liu, and F.~Petriello, {\it {Power Corrections in the
  N-jettiness Subtraction Scheme}},
  \href{http://arxiv.org/abs/1612.02911}{{\tt arXiv:1612.02911}}.

\bibitem{Moult:2017jsg}
I.~Moult, L.~Rothen, I.~W. Stewart, F.~J. Tackmann, and H.~X. Zhu, {\it
  {N-Jettiness Subtractions for $gg\to H$ at Subleading Power}},
  \href{http://arxiv.org/abs/1710.03227}{{\tt arXiv:1710.03227}}.

\bibitem{iain_notes}
I.~W. Stewart and C.~W. Bauer, ``Lectures on the soft-collinear effective
  theory.''
  \url{http://ocw.mit.edu/courses/physics/8-851-effective-field-theory-spring-2013/lecture-notes/MIT8_851S13_scetnotes.pdf}.

\bibitem{Becher:2014oda}
T.~Becher, A.~Broggio, and A.~Ferroglia, {\it {Introduction to Soft-Collinear
  Effective Theory}},  {\em Lect. Notes Phys.} {\bf 896} (2015) pp.1--206,
  [\href{http://arxiv.org/abs/1410.1892}{{\tt arXiv:1410.1892}}].

\bibitem{Bauer:2002aj}
C.~W. Bauer, D.~Pirjol, and I.~W. Stewart, {\it {Factorization and endpoint
  singularities in heavy to light decays}},  {\em Phys. Rev.} {\bf D67} (2003)
  071502, [\href{http://arxiv.org/abs/hep-ph/0211069}{{\tt hep-ph/0211069}}].

\bibitem{Marcantonini:2008qn}
C.~Marcantonini and I.~W. Stewart, {\it {Reparameterization Invariant Collinear
  Operators}},  {\em Phys. Rev.} {\bf D79} (2009) 065028,
  [\href{http://arxiv.org/abs/0809.1093}{{\tt arXiv:0809.1093}}].

\bibitem{Dixon:1996wi}
L.~J. Dixon, {\it {Calculating scattering amplitudes efficiently}},  in {\em
  {QCD and beyond. Proceedings, Theoretical Advanced Study Institute in
  Elementary Particle Physics, TASI-95, Boulder, USA, June 4-30, 1995}},
  pp.~539--584, 1996.
\newblock \href{http://arxiv.org/abs/hep-ph/9601359}{{\tt hep-ph/9601359}}.

\bibitem{Chay:2004zn}
J.~Chay, C.~Kim, Y.~G. Kim, and J.-P. Lee, {\it {Soft Wilson lines in
  soft-collinear effective theory}},  {\em Phys. Rev.} {\bf D71} (2005) 056001,
  [\href{http://arxiv.org/abs/hep-ph/0412110}{{\tt hep-ph/0412110}}].

\bibitem{Arnesen:2005nk}
C.~M. Arnesen, J.~Kundu, and I.~W. Stewart, {\it {Constraint equations for
  heavy-to-light currents in SCET}},  {\em Phys. Rev.} {\bf D72} (2005) 114002,
  [\href{http://arxiv.org/abs/hep-ph/0508214}{{\tt hep-ph/0508214}}].

\bibitem{Buras:1989xd}
A.~J. Buras and P.~H. Weisz, {\it {QCD Nonleading Corrections to Weak Decays in
  Dimensional Regularization and 't Hooft-Veltman Schemes}},  {\em Nucl. Phys.}
  {\bf B333} (1990) 66--99.

\bibitem{Dugan:1990df}
M.~J. Dugan and B.~Grinstein, {\it {On the vanishing of evanescent operators}},
   {\em Phys. Lett.} {\bf B256} (1991) 239--244.

\bibitem{Herrlich:1994kh}
S.~Herrlich and U.~Nierste, {\it {Evanescent operators, scheme dependences and
  double insertions}},  {\em Nucl. Phys.} {\bf B455} (1995) 39--58,
  [\href{http://arxiv.org/abs/hep-ph/9412375}{{\tt hep-ph/9412375}}].

\bibitem{Low:1958sn}
F.~E. Low, {\it {Bremsstrahlung of very low-energy quanta in elementary
  particle collisions}},  {\em Phys. Rev.} {\bf 110} (1958) 974--977.

\bibitem{Burnett:1967km}
T.~H. Burnett and N.~M. Kroll, {\it {Extension of the low soft photon
  theorem}},  {\em Phys. Rev. Lett.} {\bf 20} (1968) 86.

\end{thebibliography}\endgroup
\bibliographystyle{jhep}

\end{document}